\documentclass[onecolumn,superscriptaddress,amsmath,amssymb,aps,floatfix]{revtex4-1}

\usepackage{graphicx}
\usepackage{slashed}
\usepackage{epstopdf}
\usepackage{bm}
\usepackage{hyperref}
\usepackage{graphicx,color}
\usepackage{multirow}
\usepackage{amsmath,amstext,array}
\usepackage[normalem]{ulem}

\newcolumntype{C}{>{$}c<{$}}

\begin{document}
\title{ 
A Complete Set of 4-Point Amplitudes in the Constructive Standard Model}
\author{Neil Christensen}
\email{nchris3@ilstu.edu}
\affiliation{Department of Physics, Illinois State University, Normal, IL 61790}

\date{\today}

\begin{abstract}
We present a complete set of 4-point amplitudes in the constructive Standard Model at tree level.  Any 4-point amplitude can be obtained from the results presented here by a suitable choice of masses, a permutation of the particles (by crossing symmetry), and a reversal of the momenta of the outgoing particles.  We have validated all of these amplitudes by comparing with Feynman diagrams for a variety of masses, scattering energy and angles, and helicities of the photons and gluons, when they are in the initial states.  The standard constructive techniques work for these amplitudes without the need for any contact terms and indeed, contact terms are not allowed.  Only three 4-point vertices are used (allowed), involving the Higgs boson and the $W$ and $Z$ bosons.  When external photons or gluons are present, the amplitude simplifies to a single spinor-product structure, present in the numerator.  In a few cases, however, the propagator structure is more complex, with different terms depending on the charge or color structure.  In the case of internal photons or gluons, we find that the massless limit of a massive photon or gluon diagram gives the correct result in all cases.  We have additionally found that using the $x$ factor directly gives the correct result in all cases and agrees with the massless limit calculation.
\end{abstract}

\maketitle

\tableofcontents

\section{Introduction}

Ever since the breath-taking result of Parke and Taylor \cite{Parke:1986gb}, where the maximally-helicity-violating gluon amplitude could be written in a single mathematical expression no matter how many thousands or millions of Feynman diagrams were required, it has been realized that Feynman diagrams are not the most economical or the most efficient method to calculate scattering amplitudes.  As research continued, this view was further solidified, culminating in the discovery of a complete recursion algorithm for calculating any pure gluon scattering amplitude \cite{Gastmans:1990xh,Dixon:1996wi,Britto:2005fq,Elvang:2013cua}.  Another milestone for this constructive method was the generalization to any mass and any spin \cite{Arkani-Hamed:2017jhn}.  With this development, it was possible, in principle, to calculate any scattering amplitude using recursion relations and bypassing field theory and Feynman diagrams all together.  The full set of 3-point vertices of the constructive SM (CSM) was released \cite{Christensen:2018zcq} and some 4- and 5-point amplitudes that did not involve photons or gluons were calculated \cite{Christensen:2019mch}, as well as many other calculations \cite{Ochirov:2018uyq,Aoude:2019tzn,Durieux:2019eor,Franken:2019wqr,Bachu:2019ehv,Balkin:2021dko,Baratella:2020lzz,Durieux:2020gip,Alves:2021rjc,Bertuzzo:2023slg,Liu:2023jbq,Bachu:2023fjn,Liu:2022alx}.  

At this point, we intended to analyze perturbative unitarity in the CSM as a way of discovering the 4-point vertices and then calculate the full set of 4-point amplitudes in the CSM.  However, an issue with photons and gluons along internal lines (such as in the processes $W \bar{W} \to W \bar{W}$ and $t \bar{t} \to W \bar{W}$) created a roadblock for these calculations \cite{Christensen:2022nja}.  A workaround was found in that paper by using a massive photon (or gluon) and then taking the massless limit at the end.  Soon afterwards, \cite{Lai:2023upa} showed that the $x$ factor \cite{Arkani-Hamed:2017jhn} had a term that vanished on-shell, that had been missed in earlier calculations, which allowed the correct amplitude for the process $e \bar{e}\to \mu \bar{\mu}$ to be obtained using the $x$ factor,  and \cite{Ema:2024vww} found the correct momentum shift and clarified further why the constructive method works for this process, finalizing the resolution of the challenge.

With this resolution of the perceived issue with the internal photons and gluons, it became possible again to analyze perturbative unitarity and determine the complete set of 4-point vertices in the CSM.  We did this in a companion to this paper \cite{Christensen:2024C}, where we showed that the CSM is perturbatively unitary and that only three 4-point vertices are required.  In that paper, we give a complete set of 4-point vertices in the CSM and note that this set is smaller than the set of 4-point vertices in Feynman diagrams.  Not only do we not need the 4-gluon vertex (which was already known), but we also do not need a 4-point vertex for $\gamma Z W \bar{W}$, $\gamma \gamma W \bar{W}$, $Z Z \bar{W} W$ or $W W \bar{W} \bar{W}$.  Indeed, they are not allowed.  

We further show in that paper that any process that involves a 4-point vertex involving a massive vector boson has a significant rearrangement of contributions to the scattering amplitude relative to Feynman diagrams.  That is, even if there is a superficial resemblance of the Feynman diagrams and the constructive diagrams, these diagrams do not contribute identically to the final amplitude.  The constructive diagrams are not always equal to Feynman diagrams, written in spinor notation.  Often they are, but not when there is a 4-point vertex involving a massive vector boson.  We emphasize that we found this to be true even in amplitudes that do not involve any photons or gluons at all ($Z Z \to W \bar{W}$, $W W \to W W$, $h h \to Z Z$ and $h h \to W \bar{W}$), so that, in these cases, this has nothing to do massless helicity-$\pm1$ particles in the diagram.  

With a complete set of 3- and 4-point vertices, we have carried out a calculation of a complete set of 4-point amplitudes in the CSM, and present our results in this paper.  By complete, we mean that any tree-level 4-point amplitude in the CSM can be obtained from the results presented here by a suitable change of masses, rearrangement of particle numbers (by crossing symmetry) and a reversal of the momenta of any outgoing particles.  (All the amplitudes presented in this paper have all momenta ingoing.)  We note that we use the convention that the numbers in the spinor products and momenta refer to the order of the particles in the amplitude.  Furthermore, a spinor with a bold number represents a massive spinor, while a spinor with a non-bold number represents a massless spinor.

Additionally, we have validated all these amplitudes by comparing with Feynman diagrams in the following way.  We realized early in this project that comparing the analytic expression for the squared amplitudes was increasingly too complicated.  Therefore, we set out to create a numerical computational package that could calculate any constructive scattering amplitude at any phase-space point.  This work culminated in the package SPINAS, which we publish in another companion to this paper \cite{Christensen:2024B}.  Using this package, we implemented every amplitude described in this paper for a variety of masses, scattering energies, and angles.  Further, we tested at least two amplitudes related by crossing symmetry for each expression.  And, whenever a photon or gluon was in the initial state, we also validated the amplitude for their individual helicities.  All together, we validated the amplitude for 137 processes.  Moreover, any interested person can download the SPINAS code, which comes with this complete set of implemented 4-point CSM amplitudes and their validations, and run them independently.  They can also use these as a starting point in the implementation of their own constructive amplitudes.

As we calculated and validated all the 4-point amplitudes in the CSM, we found that the amplitudes resulted from the direct application of the constructive rules, with two caveats to be described in the next paragraphs.  Some diagrams required a complicated set of simplification procedures involving the application of mass identities, Schouten identities, rearrangements or momenta and the application of conservation of energy, but in the end they could be reduced to a form that agreed with Feynman diagrams.  This further solidified our claim that no contact terms beyond the small set of (expected) 4-point vertices described in \cite{Christensen:2024C} were needed.  In deed, we find that adding contact terms would ruin the agreement with Feynman diagrams at 4-point tree level.  Furthermore, we conjecture that the 3-point vertices and the 4-point vertices described in \cite{Christensen:2018zcq,Christensen:2024C} are all that are needed for any multiplicity and any loop in the CSM and that a similar statement would apply to any renormalizable theory.  

For the few diagrams with internal photons (or gluons) but massive external particles, we initially used a massive photon and took the massless limit, as we described in \cite{Christensen:2022nja}.  This worked for all amplitudes in the CSM, including $f \bar{f} \bar{f} f$, $f \bar{f} \bar{W} W$ and $W W \bar{W} \bar{W}$.  Afterwards, we also calculated these amplitudes using the insight of \cite{Lai:2023upa} and found that we could obtain agreement with the validated expressions directly using the $x$ factor as well.  We have reviewed the use of the $x$ factor for $f \bar{f} \bar{f} f$ and shown how to use it for $f \bar{f} \bar{W} W$ and $W W \bar{W} \bar{W}$ in App.~\ref{app:internal photons}.

Although the standard methods were sufficient to find the right spinor amplitude structures otherwise, we did find that in some cases involving external photons or gluons, obtaining the correct propagator structure required some further details.  If there were only two diagrams that could potentially contribute to the amplitude, we found that they both gave identical results.  However, if three diagrams potentially contributed, we found that only two propagator denominators contributed per term and that their coefficients depended on the electric charge or color structure.  To be more precise, if the photon or gluon interacts with different particles in the different diagrams that have different electric charge or color structure, then the coefficient of each propagator term will depend on those charges or color structures.  

One example of this is the amplitude for  $f_1 \bar{f}_2 \gamma W$, where $f_1$ and $f_2$ are fermions (found in Sec. \ref{sec:ffAW}), where the photon interacts with $f_1$ in one diagram (with charge $Q_1$), $f_2$ in another diagram (with charge $Q_2$) and with the $W$ boson in the final diagram (with charge $Q_W$).  All together, there is only one numerator including all the spinor products for all three diagrams, but the propagator structure has two terms (in its simplest form), with coefficients that depend on the charges.  Another example is the amplitude for $q \bar{q} g g$ for a quark-anti-quark pair and two gluons (found in Sec.~\ref{sec:ffAA}).  The interaction of the two gluons and the quark in the T- and U-channel diagrams have the same color structure, but the S-channel diagram has a triple-gluon vertex and a single gluon-quark vertex.  This last diagram has a different color structure.  All together, once again, the numerator has the same structure for all three diagrams (including all the spinor products), but the propagator structure is more complicated.  Further details can be found in that section.  

As a final comment about the general structure of the amplitudes, we note that every amplitude involving an external photon or gluon simplified to a single numerator term (containing all the spinor products).  Often the propagator structure was simple and only had a single term, while a few had more complicated propagator structures, as described above.  All of these amplitudes are very obviously simpler than their Feynman counterparts and are a significant rearrangement of contributions to the amplitude, again compared to their Feynman diagram alternatives.  In deed, each of these amplitudes is trivially gauge invariant since they do not have a gauge parameter and there are no unphysical degrees of freedom to cancel that would require a gauge symmetry. Each of these single simple expressions are physically meaningful.  All these amplitudes are described in Sec.~\ref{sec:A and g}.  
Otherwise, all the amplitudes that do not have an external photon or gluon, require the same number of diagrams as Feynman diagrams, even if in some cases the contributions from the diagrams is rearranged, as we discussed previously.

The structure of this paper is as follows: Section~\ref{sec:4-fermion} addresses 4-fermion amplitudes. In Section~\ref{sec:A and g}, we present amplitudes with external photons or gluons. Section~\ref{sec:other with no 4-point vertex} details the remaining 4-point amplitudes without 4-point vertices, while Section~\ref{sec:4-point vertex} discusses those with 4-point vertices. In Section~\ref{sec:conclusions}, we conclude.  We consider the $x$ factor in the amplitude for $f \bar{f} \bar{f} f$, $f \bar{f} \bar{W} W$ and $W W \bar{W} \bar{W}$ in App.~\ref{app:internal photons}.

\section{\label{sec:4-fermion}4-Fermion Amplitudes}
In this section, we present the 4-point amplitudes with four fermions.  They all have a similar set of diagrams, so we will first give a set of diagram contributions and then we will describe which diagrams contribute to each amplitude.  

\subsection{\label{sec:4-fermion:f1f1f2f2}$\mathbf{f_1 \bar{f}_1 \bar{f}_2 f_2}$}
We begin with two fermion, anti-fermion pairs with a neutral boson connecting the pairs.  If there is a photon or a gluon connecting them, the S-channel contribution is \cite{Christensen:2022nja,Lai:2023upa}
\begin{align}
    \mathcal{M}_{(\gamma/g) S} &=
    \frac{-2Q_1 Q_2 e^2}{s}
    \big(
    \lbrack\mathbf{14}\rbrack \langle\mathbf{23}\rangle +\lbrack\mathbf{13}\rbrack \langle\mathbf{24}\rangle
    +\langle\mathbf{14}\rangle \lbrack\mathbf{23}\rbrack +\langle\mathbf{13}\rangle \lbrack\mathbf{24}\rbrack 
    \big) ,
    \label{eq:4-f:A S}
\end{align}
where $Q$ is the charge in units of $e$.  If this is a gluon contribution, then replace $Q_1Q_2e^2$ with $g_s^2$ and include a QCD matrix $T_a$ for each vertex, namely $\sum_a T_{a\ i_2}^{\ i_1} T_{a\ i_3}^{\ i_4}$,
where $i_1, i_2, i_3$ and $i_4$ are the colors of the quarks.  
If $f_2$ is the same as $f_1$, then a photon or gluon can connect them in the T channel as well, giving the contribution
\begin{align}
    \mathcal{M}_{(\gamma/g) T} &=
    \frac{2Q_1 Q_2 e^2}{t}
    \big(
    -\lbrack\mathbf{14}\rbrack \langle\mathbf{2}\mathbf{3}\rangle 
    +\lbrack\mathbf{12}\rbrack \langle\mathbf{3}\mathbf{4}\rangle 
    -\langle\mathbf{14}\rangle \lbrack\mathbf{2}\mathbf{3}\rbrack 
    +\langle\mathbf{12}\rangle \lbrack\mathbf{3}\mathbf{4}\rbrack 
    \big) ,
\end{align}
where the minus sign for interchanging identical fermions is already taken into account.
If this is a gluon diagram then, again, replace $Q_1Q_2e^2$ with $g_s^2$ and include $\sum_a T_{a\ i_3}^{\ i_1} T_{a\ i_2}^{\ i_4}$.

There is also a Higgs contribution in the S channel, given by
\begin{align}
    \mathcal{M}_{hS} &=
    \frac{e^2m_1m_2}{4M_W^2s_W^2}
    \frac{
    \left(\langle\mathbf{12}\rangle +\lbrack\mathbf{12}\rbrack \right)
    \left(\langle\mathbf{3}\mathbf{4}\rangle +\lbrack\mathbf{3}\mathbf{4}\rbrack \right)
    }{s-m_h^2},
\end{align}
where $s_W=\sin(\theta_W)$ and $\theta_W$ is the Weinberg angle.
If $f_2=f_1$, then the Higgs also contributes in the T channel, given by
\begin{align}
    \mathcal{M}_{hT} &=
    \frac{-e^2m_1m_2}{4M_W^2s_W^2}
    \frac{\left(\langle\mathbf{13}\rangle +\lbrack\mathbf{13}\rbrack \right)
    \left(\langle\mathbf{2}\mathbf{4}\rangle +\lbrack\mathbf{2}\mathbf{4}\rbrack \right)
    }{t-m_h^2},
\end{align}
where, once again, the minus sign for interchanging identical fermions is already taken into account.

We next consider the Z-boson contribution.  In the S channel, the contribution is
\begin{align}
    \mathcal{M}_{ZS} &=
    \frac{-e^2 m_1 m_2 (g_{L1}-g_{R1}) (g_{L2}-g_{R2})}{4 M_W^2 s_W^2}
    \frac{\left(\langle\mathbf{12}\rangle -\lbrack\mathbf{12}\rbrack \right)
    \left(\langle\mathbf{3}\mathbf{4}\rangle -\lbrack\mathbf{3}\mathbf{4}\rbrack \right)}
    {s-M_Z^2}
    \nonumber\\
    &-\frac{e^2}{2c_W^2s_W^2}
    \frac{\left(
    g_{R1}g_{R2}\lbrack\mathbf{14}\rbrack \langle\mathbf{2}\mathbf{3}\rangle 
    +g_{L2}g_{R1}\lbrack\mathbf{13}\rbrack \langle\mathbf{2}\mathbf{4}\rangle 
    +g_{L1}g_{R2}\langle\mathbf{13}\rangle \lbrack\mathbf{2}\mathbf{4}\rbrack 
    +g_{L1}g_{L2}\langle\mathbf{14}\rangle \lbrack\mathbf{2}\mathbf{3}\rbrack 
    \right)}
    {s-M_Z^2} ,
\end{align}
where $g_L=2T_3-2Qs_W^2$, $g_R=-2Qs_W^2$ and $c_W=\cos(\theta_W)$.
A Z-boson contribution in the T channel, if $f_2=f_1$, is given by
\begin{align}
    \mathcal{M}_{ZT} &=
    \frac{e^2 m_1 m_2 (g_{L1}-g_{R1})^2 }{4 M_W^2 s_W^2}
    \frac{\left(\langle\mathbf{13}\rangle -\lbrack\mathbf{13}\rbrack \right)
    \left(\langle\mathbf{2}\mathbf{4}\rangle -\lbrack\mathbf{2}\mathbf{4}\rbrack \right)}{t-M_Z^2}
    \nonumber\\
    &+\frac{e^2}{2c_W^2s_W^2}
    \frac{\left(
    -g_{R1}^{2} \lbrack\mathbf{14}\rbrack \langle\mathbf{2}\mathbf{3}\rangle 
    +g_{L1}g_{R1}\left(\lbrack\mathbf{12}\rbrack \langle\mathbf{3}\mathbf{4}\rangle +\langle\mathbf{12}\rangle \lbrack\mathbf{3}\mathbf{4}\rbrack 
    \right)
    -g_{L1}^{2} \langle\mathbf{14}\rangle \lbrack\mathbf{2}\mathbf{3}\rbrack 
    \right)}{t-M_Z^2} ,
\end{align}
where, as before, the minus sign for interchanging identical fermions is already taken into account.

Finally, in some cases, when $f_2$ is the isospin partner of $f_1$, we have a $W$ boson contribute in the T channel.  It's contribution is
\begin{align}
    \mathcal{M}_{WT} &=
    \frac{e^2}{2M_W^2s_W^2}
    \frac{\left(
    2M_W^{2} \langle\mathbf{14}\rangle \lbrack\mathbf{2}\mathbf{3}\rbrack 
    +\left(
        m_2\langle\mathbf{13}\rangle -m_1\lbrack\mathbf{13}\rbrack
    \right)
    \left(
        m_2\lbrack\mathbf{2}\mathbf{4}\rbrack
        -m_1\langle\mathbf{2}\mathbf{4}\rangle
    \right)
    \right)}
    {t-M_W^2} .
\end{align}

We are now prepared to consider specific cases of 4-fermion amplitudes of this type.  We will begin with the simplest cases with four neutrinos and build our way towards the more complicated amplitudes with four quarks.  We begin with two neutrino pairs of different generations.  For this amplitude, we only have a contribution from the Z boson in the S channel and $g_{L\nu}=1$ and $g_{R\nu}=0$.  In this case, since the expression is so simple, we give the explicit formula.  The amplitude is
\begin{align}
    \mathcal{M}_{\nu_1\bar{\nu}_1\bar{\nu}_2\nu_2} &=
    -\frac{e^2}{2c_W^2s_W^2}
    \frac{\langle14\rangle\lbrack23\rbrack}
    {(s-M_Z^2)},
\end{align}
where $\nu_1\neq \nu_2$.  We note that this is the only possible spinor producta allowed (without an intermediate momentum), consdering the helicities of the neutrinos.  We have validated this amplitude against Feynman diagrams in the processes $\nu_e \bar{\nu}_e \to \nu_\mu \bar{\nu}_\mu$ and $\nu_e \nu_\mu \to \nu_e \nu_\mu$ in SPINAS for a variety of masses (values of $M_Z$ in this case) and collider energies.  On the other hand, if the neutrinos are from the same generation, we have a contribution from the Z in both the S and T channels.  Our amplitude is
\begin{align}
    \mathcal{M}_{\nu_1\bar{\nu}_1\bar{\nu}_1\nu_1} &=
    -\frac{e^2\langle14\rangle\lbrack23\rbrack}{2c_W^2s_W^2}
    \left(\frac{1}{s-M_Z^2}+\frac{1}{t-M_Z^2}\right).
\end{align}
We have validated this amplitude against Feynman diagrams in the processes $\nu_e \bar{\nu}_e \to \nu_e \bar{\nu}_e$ and $\nu_e \nu_e \to \nu_e \nu_e$ in SPINAS for a variety of masses and collider energies.

We next consider two fermions and two neutrinos, where the two fermions are either charged leptons of a different generation or any quarks.  We still only have contributions from the Z boson in the S channel, giving us
\begin{align}
    \mathcal{M}_{f_1\bar{f}_1\bar{\nu}_2\nu_2} &=
    -\frac{e^2}{2c_W^2s_W^2}
    \frac{\left(g_{Rf}\langle\mathbf{2}4\rangle\lbrack\mathbf{1}3\rbrack +g_{Lf}\langle\mathbf{1}4\rangle\lbrack\mathbf{2}3\rbrack \right)}{s-M_Z^2},
\end{align}
where $g_{Le}=-1+2s_W^2$, $g_{Re}=2s_W^2$, $g_{Lu}=1-4s_W^2/3$, $g_{Ru}=-4s_W^2/3$, $g_{Ld}=-1+2s_W^2/3$, $g_{Rd}=2s_W^2/3$ and the quark colors are the same if $f$ is a quark (there is a $\delta^{i_1}_{i_2})$.  We have validated this amplitude against Feynman diagrams in the processes $e \bar{e} \to \nu_\mu \bar{\nu}_\mu$, $e \nu_\mu \to e \nu_\mu$,        , $u \bar{u} \to \nu_e \bar{\nu}_e$, $u  \nu_\mu \to \nu_\mu u$, $d \bar{d} \to \nu_e \bar{\nu}_e$, $d \nu_\mu \to \nu_\mu d$ in SPINAS for a variety of masses and collider energies.

On the other hand, if the charged lepton and neutrinos are from the same generation, we have a contribution from the $Z$ in the S channel as well as a contribution from the $W$ in the T channel.  All together, we have
\begin{align}
    \mathcal{M}_{l_1\bar{l}_1\bar{\nu}_1\nu_1} &=
    -\frac{e^2}{2c_W^2s_W^2}
    \frac{\left(g_{Re}\langle\mathbf{2}4\rangle\lbrack\mathbf{1}3\rbrack +g_{Le}\langle\mathbf{1}4\rangle\lbrack\mathbf{2}3\rbrack \right)}{s-M_Z^2}
    -\frac{e^2}{2M_W^2s_W^2}
    \frac{\left(
    M_l^{2} \langle\mathbf{2}4\rangle\lbrack\mathbf{1}3\rbrack 
    +2M_W^{2} \langle\mathbf{1}4\rangle\lbrack\mathbf{2}3\rbrack 
    \right)}
    {t-M_W^2}.
\end{align}
We have validated this amplitude against Feynman diagrams in the processes $e \bar{e} \to \nu_e \bar{\nu}_e$, $e \nu_e \to \nu_e e$ in SPINAS for a variety of masses and collider energies.

We next consider two charged leptons and two fermions that are not neutrinos and not from the same generation if charged leptons.  The amplitude is
\begin{align}
    \mathcal{M}_{l_1\bar{l}_1\bar{f}_2f_2} &=
    \mathcal{M}_{\gamma S} + \mathcal{M}_{h S} + \mathcal{M}_{Z S},
\end{align}
where $Q_l=-1$, $Q_u=2/3$, $Q_d=-1/3$ and the quark colors are the same (if $f$ is a quark).
We have validated this amplitude against Feynman diagrams in the processes $e \bar{e} \to \mu \bar{\mu}$, $e \mu \to e \mu$, $u \bar{u} \to e \bar{e}$, $u e \to u e$, $d \bar{d} \to e \bar{e}$ and $d e \to e d$ in SPINAS for a variety of masses and collider energies.

On the other hand, if all the charged leptons are from the same generation, we have
\begin{align}
    \mathcal{M}_{l\bar{l}\bar{l}l} &=
    \mathcal{M}_{\gamma S} + \mathcal{M}_{h S} + \mathcal{M}_{Z S} + \mathcal{M}_{\gamma T} + \mathcal{M}_{h T} + \mathcal{M}_{Z T}.
\end{align}
We have validated this amplitude against Feynman diagrams in the processes $e \bar{e} \to e \bar{e}$ and $e e\to e e$ in SPINAS for a variety of masses and collider energies.

We next consider two pairs of quarks from different generations.  We have
\begin{align}
    \mathcal{M}_{q_1\bar{q}_1\bar{q}_2q_2} &=
    \left(\mathcal{M}_{\gamma S} + \mathcal{M}_{h S} + \mathcal{M}_{Z S}\right)\delta^{i_1}_{i_2}\delta^{i_4}_{i_3} + \mathcal{M}_{g S}\sum_a T_{a\ i_2}^{\ i_1} T_{a\ i_3}^{\ i_4},
\end{align}
where $i_1, i_2, i_3$ and $i_4$ are the colors of the quarks.  We have validated this amplitude against Feynman diagrams in the processes $u \bar{u} \to c \bar{c}$, $u c \to u c$, $u \bar{u} \to s \bar{s}$, $u s \to u s$, $d \bar{d} \to s \bar{s}$ and $d s\to d s$  in SPINAS for a variety of masses and collider energies.

If all four quarks are of the same type, we obtain
\begin{align}
    \mathcal{M}_{q\bar{q}\bar{q}q} &=
    \left(\mathcal{M}_{\gamma S} + \mathcal{M}_{h S} + \mathcal{M}_{Z S}\right)\delta^{i_1}_{i_2}\delta^{i_4}_{i_3} 
    + \left(\mathcal{M}_{\gamma T} + \mathcal{M}_{h T} + \mathcal{M}_{Z T}\right) \delta^{i_1}_{i_3}\delta^{i_4}_{i_2} 
    + \mathcal{M}_{g S}\sum_a T_{a\ i_2}^{\ i_1} T_{a\ i_3}^{\ i_4} 
    + \mathcal{M}_{g T}\sum_a T_{a\ i_3}^{\ i_1} T_{a\ i_2}^{\ i_4} .
\end{align}
We have validated this amplitude against Feynman diagrams in the processes $u \bar{u} \to u \bar{u}$, $u u \to u u$, $d \bar{d} \to d \bar{d}$ and $d d \to d d$  in SPINAS for a variety of masses and collider energies.

If we have a quark and its isospin partner, we have
\begin{align}
    \mathcal{M}_{q_1\bar{q}_1\bar{q}_2q_2} &=
    \left(\mathcal{M}_{\gamma S} + \mathcal{M}_{h S} + \mathcal{M}_{Z S}\right)\delta^{i_1}_{i_2}\delta^{i_4}_{i_3} 
    + \mathcal{M}_{W T}\delta^{i_1}_{i_3}\delta^{i_4}_{i_2} 
    + \mathcal{M}_{g S}\sum_a T_{a\ i_2}^{\ i_1} T_{a\ i_3}^{\ i_4}.
\end{align}
We have validated this amplitude against Feynman diagrams in the processes $u \bar{u} \to d \bar{d}$ and $u d \to u d$ in SPINAS for a variety of masses and collider energies.

\subsection{\label{sec:4-fermion:f1f2f3f4}$\mathbf{f_1\bar{f}_2\bar{f}_3f_4}$}
In this subsection, we consider 4-fermion amplitudes with a charged S channel.  We also found them in \cite{Christensen:2019mch}.  There is one contribution, from a $W$ boson in the S channel.  The amplitude is
\begin{align}
    \mathcal{M}_{W S} &=
    \frac{e^2}{M_W^2s_W^2}
    \frac{\left(
    2M_W^{2} \langle\mathbf{1}\mathbf{4}\rangle \lbrack\mathbf{2}\mathbf{3}\rbrack 
    +m_2m_3\langle\mathbf{1}\mathbf{2}\rangle \langle\mathbf{3}\mathbf{4}\rangle 
    -m_1m_3\lbrack\mathbf{1}\mathbf{2}\rbrack\langle\mathbf{3}\mathbf{4}\rangle  
    -m_2m_4\langle\mathbf{1}\mathbf{2}\rangle \lbrack\mathbf{3}\mathbf{4}\rbrack 
    +m_1m_4\lbrack\mathbf{1}\mathbf{2}\rbrack \lbrack\mathbf{3}\mathbf{4}\rbrack 
    \right)}
    {s-M_W^2}.
\end{align}

We begin with two charged lepton neutrino pairs of different generations.  The amplitude is
\begin{align}
    \mathcal{M}_{l_1\bar{\nu}_1\bar{l}_2\nu_2} &= 
    -\frac{e^{2}}{2M_W^{2} s_W^{2}} \frac{\left(
    2M_W^{2} \lbrack2\mathbf{3}\rbrack \langle\mathbf{1}4\rangle
    -m_1m_2\langle\mathbf{3}4\rangle\lbrack\mathbf{1}2\rbrack 
    \right)}
    {s-M_W^{2}},
\end{align}
where $m_1$ and $m_2$ are the masses of the charged leptons.
We have validated this amplitude against Feynman diagrams in the processes $\mu \bar{e} \to \bar{\nu}_e \nu_\mu$ and $\mu \nu_e \to e \nu_\mu$ in SPINAS for a variety of masses and collider energies.

We next consider a lepton pair and a quark pair.  We have
\begin{align}
    \mathcal{M}_{q_1\bar{q}_2\bar{l}\nu_l} &=
    -\frac{e^{2}}{2M_W^{2} s_W^{2}} 
    \frac{\left(
    2M_W^{2} \langle\mathbf{1}4\rangle\lbrack\mathbf{2}\mathbf{3}\rbrack 
    +m_lm_2\langle\mathbf{1}\mathbf{2}\rangle \langle\mathbf{3}4\rangle
    -m_1m_l\langle\mathbf{3}4\rangle\lbrack\mathbf{1}\mathbf{2}\rbrack 
    \right)}{s-M_W^{2}}
    \delta^{i_1}_{i_2}.
\end{align}
We have validated this amplitude against Feynman diagrams in the processes $u \bar{d} \to \nu_\tau \bar{\tau}$ and $u \tau \to \nu_\tau d$ in SPINAS for a variety of masses and collider energies.

If we have two pairs of quarks from different generations, we have
\begin{align}
    \mathcal{M}_{q_1\bar{q}_2\bar{q}_3q_4} &=
    \mathcal{M}_{WS}\delta^{i_1}_{i_2}\delta^{i_4}_{i_3}.
\end{align}
We have validated this amplitude against Feynman diagrams in the processes $u \bar{d} \to t \bar{b}$ and $u b \to d t$ in SPINAS for a variety of masses and collider energies.

\section{\label{sec:A and g}Amplitudes with an External Photon or Gluon}
In this section, we consider amplitudes that have an external photon or gluon.  As we will see, they all simplify considerably to a single numerator multiplied by a propagator denominator structure and every term in the propagator denominator structure has at least two propagator denominators.  Moreover, every term has no high-energy-growth terms as the energy growth of the numerator cancels against that of the denominator term for term \cite{Christensen:2024C}.

\subsection{$\mathbf{f \bar{f} \gamma h}$ and $\mathbf{f \bar{f} g h}$}
These amplitudes were also found in \cite{Christensen:2022nja}.
The amplitude can be calculated in either the T or the U channel.  Both give the same amplitude.  For a positive-helicity photon or gluon, the amplitude is
\begin{align}
    \mathcal{M}_{f\bar{f}(\gamma^+/g^+)h} &=
    \frac{e^2m_fQ_f}{\sqrt{2}M_Ws_W}
    \frac{\left(
    m_h^{2} \lbrack\mathbf{1}3\rbrack \lbrack\mathbf{2}3\rbrack 
    +m_f\lbrack\mathbf{2}3\rbrack \lbrack3\lvert p_{4} \rvert \mathbf{1}\rangle 
    +m_f\lbrack\mathbf{1}3\rbrack \lbrack3\lvert p_{4} \rvert \mathbf{2}\rangle 
    +\langle\mathbf{1}\mathbf{2}\rangle \lbrack3\lvert p_{2} p_{4} \rvert 3\rbrack 
    \right)}
    {(t-m_f^2)(u-m_f^2)}.
\end{align}
To obtain the negative helicity amplitude, simply switch angle and square brackets ($\langle\rangle\longleftrightarrow\lbrack\rbrack$), to obtain
\begin{align}
    \mathcal{M}_{f\bar{f}(\gamma^-/g^-)h} &=
    \frac{e^2m_fQ_f}{\sqrt{2}M_Ws_W}
    \frac{\left(
    m_h^{2} \langle\mathbf{1}3\rangle \langle\mathbf{2}3\rangle
    +m_f\langle\mathbf{2}3\rangle \langle3\lvert p_{4} \rvert \mathbf{1}\rbrack
    +m_f\langle\mathbf{1}3\rangle \langle3\lvert p_{4} \rvert \mathbf{2}\rbrack
    +\lbrack\mathbf{1}\mathbf{2}\rbrack \langle3\lvert p_{2} p_{4} \rvert 3\rangle
    \right)}
    {(t-m_f^2)(u-m_f^2)}.
\end{align}
If the amplitude is for a photon, there is also a Kronecker delta ($\delta^{i_1}_{i_2})$ for the colors if the fermions are quarks.  If the amplitude is for a gluon, replace $eQ_f$ with $g_s$ and include a color matrix $T_{a\ i_2}^{\ i_1}$ for the colors.
We have validated this amplitude against Feynman diagrams in the processes $e \bar{e} \to \gamma h$, $e \gamma \to e h$, $u \bar{u} \to \gamma h$, $u \gamma \to u h$, $d \bar{d} \to \gamma h$, $d \gamma \to d h$, $u g \to u h$, $h g \to u \bar{u}$, $d g \to d h$ and $h g \to d \bar{d}$ in SPINAS
for a variety of masses and collider energies.  We have additionally validated it for the individual helicities of the photon and gluon when in the initial state for each of these masses and energies.

\subsection{$\mathbf{f \bar{f} \gamma Z}$ and $\mathbf{f \bar{f} g Z}$}
We get the same result whether we calculate this in the T or the U channel.  For the positive helicity photon or gluon, we have
\begin{align}
    \mathcal{M}_{f\bar{f}(\gamma^+/g^+)Z} &=
    \frac{e^2Q_f}{M_Ws_W}
    \frac{\left(
    g_R\langle\mathbf{2}\mathbf{4}\rangle\left(
    M_Z \lbrack\mathbf{1}3\rbrack \lbrack3\lvert p_{2} \rvert \mathbf{4}\rangle 
    +m_f\lbrack3\mathbf{4}\rbrack  \lbrack3\lvert p_{4} \rvert \mathbf{1}\rangle 
    \right)
    -g_L\langle\mathbf{1}\mathbf{4}\rangle\left(
    M_Z \lbrack\mathbf{2}3\rbrack \lbrack3\lvert p_{1} \rvert \mathbf{4}\rangle 
    +m_f\lbrack3\mathbf{4}\rbrack  \lbrack3\lvert p_{4} \rvert \mathbf{2}\rangle 
    \right)
    \right)}
    {(t-m_f^2)(u-m_f^2)}.
\end{align}
For the negative helicity photon or gluon, switch the angle and square brackets ($\langle\rangle\longleftrightarrow\lbrack\rbrack$) and switch the chiral couplings ($g_L\longleftrightarrow g_R$), to obtain
\begin{align}
    \mathcal{M}_{f\bar{f}(\gamma^-/g^-)Z} &=
    \frac{e^2Q_f}{M_Ws_W}
    \frac{\left(
    g_L\lbrack\mathbf{2}\mathbf{4}\rbrack\left(
    M_Z \langle\mathbf{1}3\rangle \langle3\lvert p_{2} \rvert \mathbf{4}\rbrack
    +m_f\langle3\mathbf{4}\rangle  \langle3\lvert p_{4} \rvert \mathbf{1}\rbrack
    \right)
    -g_R\lbrack\mathbf{1}\mathbf{4}\rbrack\left(
    M_Z \langle\mathbf{2}3\rangle \langle3\lvert p_{1} \rvert \mathbf{4}\rbrack
    +m_f\langle3\mathbf{4}\rangle \langle3\lvert p_{4} \rvert \mathbf{2}\rbrack
    \right)
    \right)}
    {(t-m_f^2)(u-m_f^2)}.
\end{align}
As for the previous amplitude, if the fermion is a quark, there is a $\delta^{i_1}_{i_2}$ for the photon amplitude and a $T_{a\ i_2}^{\ i_1}$ for the gluon amplitude, in addition to replacing $eQ_f$ with $g_s$.  
We have validated this amplitude against Feynman diagrams in the processes $\gamma Z \to \bar{e} e$, $\gamma e \to Z e$, $\gamma Z \to \bar{u} u$, $\gamma u \to Z u$, $\gamma Z \to \bar{d} d$, $\gamma d \to Z d$, $g Z \to \bar{u} u$, $g u \to Z u$, $g Z \to \bar{d} d$ and $g d \to Z d$ in SPINAS
for a variety of masses and collider energies.  We have additionally validated it for the individual helicities of the photon and gluon when in the initial state for each of these masses and energies.

\subsection{\label{sec:ffAW}$\mathbf{q_1 \bar{q}_2 g W}$ and $\mathbf{f_1 \bar{f}_2 \gamma W}$}
The amplitude for the gluon can be calculated in either the T or the U channel, giving the same result, which, for positive-helicity gluons, is
\begin{align}
    \mathcal{M}_{q_1\bar{q}_2g^+W} &=
    -\frac{\sqrt{2}\ e g_s}{M_Ws_W}
    \frac{
    \langle\mathbf{1}\mathbf{4}\rangle 
    \left(
    \lbrack3\mathbf{4}\rbrack \left(
    m_1^{2} \lbrack\mathbf{2}3\rbrack 
    +m_2\lbrack3\lvert p_{1} \rvert \mathbf{2}\rangle 
    \right)
    -M_W\lbrack\mathbf{2}3\rbrack \lbrack3\lvert p_{1} \rvert \mathbf{4}\rangle 
    \right)}
    {(t-m_1^2)(u-m_2^2)}
    T_{a\ i_2}^{\ i_1}.
\end{align}
This expression differs from the left-chiral part of the $Z$ boson amplitude from the previous subsection because the masses of the two fermions are different.  This amplitude doesn't simplify as much as the $Z$-boson vertex did.  
For negative helicity gluons, it is not as simple as interchanging angle and square brackets, because we would also have to interchange the left- and right-chiral couplings for the $W$ boson (see the previous subsection with the $Z$ boson).  But, since the $W$ boson only couples to the left-chiral fermions, we don't have both expression to interchange.  We find
\begin{align}
    \mathcal{M}_{q_1\bar{q}_2g^-W} &=
    \frac{\sqrt{2}\ e g_s}{M_Ws_W}
    \frac{
    \lbrack\mathbf{2}\mathbf{4}\rbrack \left(\langle3\mathbf{4}\rangle \left(m_2^{2} \langle\mathbf{1}3\rangle+m_1\lbrack\mathbf{1}\lvert p_{2} \rvert 3\rangle\right)-M_W\langle\mathbf{1}3\rangle\lbrack\mathbf{4}\lvert p_{2} \rvert 3\rangle\right)
    }
    {(t-m_1^2)(u-m_2^2)}
    T_{a\ i_2}^{\ i_1}.
\end{align}
Although it isn't a simple transformation of the positive-helicity case, we can see a resemblance to the right-chiral part of the $Z$-boson amplitude from the previous subsection, after switching angle and square brackets.  But, once again, the masses of the two fermions are different here, so it is not an exact correspondence.  
We have validated this amplitude against Feynman diagrams in the processes $d W \to g u$ and $g W \to u \bar{d}$ in SPINAS
for a variety of masses and collider energies.  We have additionally validated it for the individual helicities of the gluon when in the initial state for each of these masses and energies.

For the photon process, there are three possible diagrams for the quarks, but two for the leptons.  We will describe the amplitude in the case of three diagrams and then consider the lepton case as a special case.  In the previous photon cases, where there were only two possible diagrams, either diagram could be used to obtain the amplitude and obtain identical results, a final amplitude with both propagator denominators combined.  However, in this case, we have three propagator denominators, but we still only have two propagator denominators combined per term.  Moreover, the electric charge of the particle that the photon connects to is different in each of the diagrams.  This results in a slightly more complicated amplitude.  

The numerator expression with the spinor products is the same for all three diagrams.  It is only the charge and the propagator denominators that is different for each diagram.  We find that we have to combine the propagator denominators with the charge to obtain the correct amplitude, which is
\begin{align}
    \mathcal{M}_{f_1\bar{f}_2\gamma^+W} &=
    \frac{\sqrt{2}\ e^2}{M_Ws_W}
    \frac{\langle\mathbf{1}\mathbf{4}\rangle\left(
    \lbrack3\mathbf{4}\rbrack\left(
    m_1^{2}   \lbrack\mathbf{2}3\rbrack 
    +m_2  \lbrack3\lvert p_{1} \rvert \mathbf{2}\rangle 
    \right)
    -M_W \lbrack\mathbf{2}3\rbrack \lbrack3\lvert p_{1} \rvert \mathbf{4}\rangle 
    \right)}
    {s-M_W^2}
    \left(
        \frac{Q_1}{t-m_1^2}
        +\frac{Q_2}{u-m_2^2}
    \right).
    \label{eq:M_f,fb,gamma,W}
\end{align}
If the fermion is a quark, then there is also a Kronecker delta for the colors.  We have validated this amplitude against Feynman diagrams in the processes $\bar{u} d \to \gamma \bar{W}$ and $\gamma W \to u \bar{d}$ in SPINAS
for a variety of masses and collider energies.  We have additionally validated it for the individual helicities of the gluon when in the initial state for each of these masses and energies.
  
The propagator part is equivalent to the following form
\begin{align}
    \frac{1}{3}\left[
        \frac{Q_W}{s-M_W^2}\left(
        \frac{1}{t-m_1^2}
        -\frac{1}{u-m_2^2}
        \right)
        +\frac{Q_1}{t-m_1^2}\left(
        \frac{1}{u-m_2^2}-\frac{1}{s-M_W^2}
        \right)
        +\frac{Q_2}{u-m_2^2}\left(
        \frac{1}{t-m_1^2}-\frac{1}{s-M_W^2}
        \right)
    \right] =
    \nonumber\\
    \frac{-1}{s-M_W^2}\left(
        \frac{Q_1}{t-m_1^2}
        +\frac{Q_2}{u-m_2^2}
    \right),
\end{align}
where $Q_W=Q_2-Q_1$.
The three terms come from each of the three diagrams.  Each one has the charge of the particle that connects with the photon as well as its propagator denominator.  Additionally, each has one of the other propagator denominators.  However, the two extra propagator denominators are equivalent when the particle attached to the photon is on shell.  For example, for the $W$ S-channel diagram, we have $Q_W/(s-M_W^2)$.  In addition, we have either $1/(t-m_1^2)$ or $-1/(u-m_2^2)$, which are related by $s+t+u=M_W^2+m_1^2+m_2^2$.  This reduces to $t+u=m_1^2+m_2^2$, when $s=M_W^2$.  Rearranging, we have $(t-m_1^2)=-(u-m_2^2)$.  Both of these possibilities are included.  Finally, the entire combination reduces to the simpler form that we present in Eq.~(\ref{eq:M_f,fb,gamma,W}).  The longer form may be useful as we explore higher-multiplicity amplitudes.

For leptons, this amplitude simplifies considerably.  Since one of the diagrams does not contribute (because the neutrino is electrically neutral), we only have two diagrams, and once again, either diagram can be used and they give identical results.  The amplitude is
\begin{align}
    \mathcal{M}_{l\bar{\nu}_l\gamma^+W} &=
    \frac{\sqrt{2}\ e^{2}}{M_Ws_W} 
    \frac{\lbrack23\rbrack \langle\mathbf{1}\mathbf{4}\rangle \left(
    M_W\lbrack3\lvert p_{1} \rvert \mathbf{4}\rangle 
    -m_1^{2} \lbrack3\mathbf{4}\rbrack 
    \right)}
    {\left(s-M_W^{2}\right)\left(t-m_1^{2}\right)} .
\end{align}
We have validated this amplitude against Feynman diagrams in the processes $\bar{e} \nu_e \to \gamma W$ and $\gamma \nu_e \to W e$ in SPINAS
for a variety of masses and collider energies.  We have additionally validated it for the individual helicities of the photon when in the initial state for each of these masses and energies.

\subsection{\label{sec:ffAA}$\mathbf{f \bar{f} \gamma \gamma}$, $\mathbf{f \bar{f} g \gamma}$ and $\mathbf{f \bar{f} g g}$}

The amplitude with two photons or one gluon and one photon has two possible diagrams, one in the T channel and the other in the U channel.  These amplitudes were also found in \cite{Christensen:2022nja}.  Both give identical results, which is
\begin{align}
    \mathcal{M}_{f\bar{f}(\gamma^+/g^+)\gamma^+} &=
    2Q_f^2e^2m_f\frac{\langle\mathbf{1}\mathbf{2}\rangle\lbrack34\rbrack ^{2}  }{(t-m_f^2)(u-m_f^2)},
\end{align}
for the case where both bosons have positive helicity and
\begin{align}
    \mathcal{M}_{f\bar{f}(\gamma^+/g^+)\gamma^-} &=
    -2Q_f^2e^2
    \frac{\left(\langle\mathbf{2}4\rangle\lbrack\mathbf{1}3\rbrack +\langle\mathbf{1}4\rangle\lbrack\mathbf{2}3\rbrack \right)\lbrack3\lvert p_{1} \rvert 4\rangle}
    {(t-m_f^2)(u-m_f^2)},
\end{align}
where the bosons have opposite helicity. For the other helicity combinations, simply interchange angle and square brackets to obtain
\begin{align}
    \mathcal{M}_{f\bar{f}(\gamma^-/g^-)\gamma^-} &=
    2Q_f^2e^2m_f\frac{\lbrack\mathbf{1}\mathbf{2}\rbrack\langle34\rangle ^{2}  }{(t-m_f^2)(u-m_f^2)}
\end{align}
and
\begin{align}
    \mathcal{M}_{f\bar{f}(\gamma^-/g^-)\gamma^+} &=
    -2Q_f^2e^2
    \frac{\left(\lbrack\mathbf{2}4\rbrack\langle\mathbf{1}3\rangle
    +\lbrack\mathbf{1}4\rbrack\langle\mathbf{2}3\rangle \right)\langle3\lvert p_{1} \rvert 4\rbrack}
    {(t-m_f^2)(u-m_f^2)}.
\end{align}
If the fermions are quarks, there is also a $\delta^{i_1}_{i_2}$ for the colors if particles 3 and 4 are photons.  If the the third particle is a gluon, then replace one power of $Q_f e$ with $g_s$ and include $T_{a\ i_2}^{\ i_1}$ for the colors.  We have validated this amplitude against Feynman diagrams in the processes $e \bar{e} \to \gamma \gamma$, $e \gamma \to \gamma e$, $\gamma \gamma \to e \bar{e}$, $u \bar{u} \to \gamma \gamma$, $u \gamma \to \gamma u$, $\gamma \gamma \to u \bar{u}$, $d \bar{d} \to \gamma \gamma$, $d \gamma \to \gamma d$, $\gamma \gamma \to d \bar{d}$, $g \gamma \to u \bar{u}$, $g u \to \gamma u$, $g \gamma \to d \bar{d}$ and $g d \to \gamma d$ in SPINAS
for a variety of masses and collider energies.  We have additionally validated it for the individual helicities of the photons and gluons when in the initial state for each of these masses and energies.

For the case of two gluons, we have three possible diagrams, with the addition of a gluon S-channel diagram.  Although, in this case, unlike $f_1 \bar{f}_2 \gamma W$ from the previous subsection, there aren't different charges to deal with, there is a more complicated color structure, between the diagrams.  Once again, the structure of the numerator with all the spinor products is the same for all three diagrams.  We only need to worry about the propagators and the colors.  

However, the propagator structure for the amplitude is not simply a combination of the denominators and products of the QCD matrices and Kronecker delta functions for the colors.  In order to give the result in the simplest way, we first specify the numerators of the amplitudes, which are
\begin{align}
    \mathcal{N}_{q\bar{q}g^+g^+} &=
    2g_s^2m_f\langle\mathbf{1}\mathbf{2}\rangle\lbrack34\rbrack ^{2},
\end{align}
for the case where both gluons  have positive helicity and
\begin{align}
    \mathcal{N}_{q\bar{q}g^+g^-} &=
    -2g_2^2
    \left(\langle\mathbf{2}4\rangle\lbrack\mathbf{1}3\rbrack +\langle\mathbf{1}4\rangle\lbrack\mathbf{2}3\rbrack \right)\lbrack3\lvert p_{1} \rvert 4\rangle,
\end{align}
where the gluons have opposite helicity.  For the other helicity combinations, simply interchange angle and square brackets.  All the helicity combinations have the same propagator denominators and color structure.  We only have the result with the propagators and colors for the squared amplitude.  We find
\begin{align}
    \lvert\mathcal{M}_{q\bar{q}g^{h_3}g^{h_4}}\rvert^2 &=
    \lvert\mathcal{N}_{q\bar{q}g^{h_3}g^{h_4}}\rvert^2
    \left(
        \frac{6}{s^2(t-m_q^2)^2}
        +\frac{6}{s^2(u-m_q^2)^2}
        +\frac{-2/3}{(t-m_q^2)^2(u-m_q^2)^2}
    \right),
\end{align}
for each of the helicity combinations, after summing over colors.  We have validated this amplitude squared against Feynman diagrams in the processes $g g \to u \bar{u}$, $g u \to g u$, $g g \to d \bar{d}$ and $g d \to g d$ in SPINAS
for a variety of masses and collider energies.  We have additionally validated it for the individual helicities of the gluons when in the initial state for each of these masses and energies.

We do not know a natural way to write the propagator part of this amplitude before squaring and including the color structure that directly gives this result.  In previous processes, we have only had two diagrams, and only obtained a single term.  In those cases, the color structure was simple and its square gave the appropriate factor.  In this case however, if we were to write each combination of two propagator denominators and assign each a color factor, it is difficult to obtain both the correct factors for the terms that are in the final squared amplitude and vanishing cross terms (such as $1/[s^2(t-m_q^2)(u-m_q^2)]$).

There are some hints though, and we know how to produce this final result.  The numerator factors come from traces of the color matrices in the following way:
\begin{align}
    6 &= \sum_{abc}f_{abc}\mbox{Tr}\left(T_aT_cT_b\right)\\
    6 &= -\sum_{abc}f_{abc}\mbox{Tr}\left(T_aT_bT_c\right)\\
    -\frac{2}{3} &= \sum_{abc}\mbox{Tr}\left(T_aT_bT_aT_b\right).
\end{align}
The first is the product of the color part of the S and T diagrams, the second is minus the product of the color part of the S and the U diagrams, and the third is the product of the color part of the T and the U diagrams.  In order to understand the signs, note that if we begin with the S-channel diagram, the denominator is either $1/(s(t-m_q^2))$ or $-1/(s(u-m_q^2))$ since when the S-channel diagram is on shell, $(t-m_q^2)=-(u-m_q^2)$, suggesting a sign difference between these two terms.  The sign between these propagator denominators then cancels the sign in $\sum_{abc}f_{abc}\mbox{Tr}\left(T_aT_bT_c\right)$.  Finally, if we had $-1/(s(u-m_q^2))$ in the U-channel diagram, it would be equivalent to $1/((t-m_q^2)(u-m_q^2))$, giving a positive sign for $\sum_{abc}\mbox{Tr}\left(T_aT_bT_aT_b\right)$.  To say it more succinctly, we can understand the sign in front of the color traces as switching in the order $1/(s(t-m_q^2))\to-1/(s(u-m_q^2))\to1/((t-m_q^2)(u-m_q^2))$, giving a relative minus sign for the S-U diagram.

\subsection{$\mathbf{\gamma h W \bar{W}}$}
This amplitude can be obtained from either the T or the U channel.  They give identical results.  The amplitude for positive photon helicity is
\begin{align}
    \mathcal{M}_{\gamma^+ h W \bar{W}} &=
    \frac{\sqrt{2}\ e^2}{M_Ws_W}
    \frac{\langle\mathbf{3}\mathbf{4}\rangle \left(
    m_h^{2} \lbrack1\mathbf{3}\rbrack \lbrack1\mathbf{4}\rbrack 
    -M_W\left(
    \lbrack1\mathbf{4}\rbrack \lbrack1\lvert p_{2} \rvert \mathbf{3}\rangle 
    +\lbrack1\mathbf{3}\rbrack \lbrack1\lvert p_{2} \rvert \mathbf{4}\rangle 
    \right)
    \right)}
    {(t-M_W^2)(u-M_W^2)}.
\end{align}
The negative-helicity amplitude is obtained by interchanging angle and square brackets and is
\begin{align}
    \mathcal{M}_{\gamma^- h W \bar{W}} &=
    \frac{\sqrt{2}\ e^2}{M_Ws_W}
    \frac{\lbrack\mathbf{3}\mathbf{4}\rbrack \left(
    m_h^{2} \langle1\mathbf{3}\rangle    \langle1\mathbf{4}\rangle 
    -M_W\left(
    \langle1\mathbf{4}\rangle \langle1\lvert p_{2} \rvert \mathbf{3}\rbrack  
    +\langle1\mathbf{3}\rangle \langle1\lvert p_{2} \rvert \mathbf{4}\rbrack
    \right)
    \right)}
    {(t-M_W^2)(u-M_W^2)}.
\end{align}
We have validated this amplitude squared against Feynman diagrams in the processes $\gamma h \to W \bar{W}$ and $\gamma W \to W h$ in SPINAS
for a variety of masses and collider energies.  We have additionally validated it for the individual helicities of the photons when in the initial state for each of these masses and energies.

\subsection{$\mathbf{\gamma Z \bar{W} W}$}
We show that this amplitude does not require a 4-point vertex, unlike Feynman diagrams, and is perturbatively unitary without one.  That is, we find that there is no high-energy growth in this amplitude in the absence of a 4-point vertex \cite{Christensen:2024C}.
We can obtain this amplitude from either the T- or U-channel diagram.  The identical result, for positive helicity, is
\begin{align}
    \mathcal{M}_{\gamma^+h W \bar{W}} =
    \frac{2e^2}{c_Ws_W(t-M_W^2)(u-M_W^2)}
    \Big(&
    c_W^{2} \lbrack1\mathbf{4}\rbrack ^{2} \langle\mathbf{2}\mathbf{3}\rangle ^{2} 
    +\left(c_W^{2} - s_W^{2} \right)\lbrack1\mathbf{3}\rbrack \lbrack1\mathbf{4}\rbrack \langle\mathbf{2}\mathbf{3}\rangle \langle\mathbf{2}\mathbf{4}\rangle 
    +c_W^{2} \lbrack1\mathbf{3}\rbrack ^{2} \langle\mathbf{2}\mathbf{4}\rangle ^{2} 
    \nonumber\\
    &-c_W\lbrack1\mathbf{2}\rbrack \lbrack1\mathbf{4}\rbrack \langle\mathbf{2}\mathbf{3}\rangle \langle\mathbf{3}\mathbf{4}\rangle 
    +c_W\lbrack1\mathbf{2}\rbrack \lbrack1\mathbf{3}\rbrack \langle\mathbf{2}\mathbf{4}\rangle \langle\mathbf{3}\mathbf{4}\rangle 
    +c_W^{2} \lbrack1\mathbf{2}\rbrack ^{2} \langle\mathbf{3}\mathbf{4}\rangle ^{2} \Big).
\end{align}
In fact, we can see by inspection that both the numerator and the denominator grow quartically with energy, resulting in at most a constant at high energy.  Thus this amplitude is perturbatively unitary and has no need for a 4-point vertex to cancel the high energy growth.
The negative-helicity amplitude is obtained by interchanging angle and square brackets.  It is
\begin{align}
    \mathcal{M}_{\gamma^-h W \bar{W}} =
    \frac{2e^2}{c_Ws_W(t-M_W^2)(u-M_W^2)}
    \Big(&
    c_W^{2} \langle1\mathbf{4}\rangle ^{2} \lbrack\mathbf{2}\mathbf{3}\rbrack ^{2} 
    +\left(c_W^{2} - s_W^{2} \right)\langle1\mathbf{3}\rangle \langle1\mathbf{4}\rangle \lbrack\mathbf{2}\mathbf{3}\rbrack \lbrack\mathbf{2}\mathbf{4}\rbrack
    +c_W^{2} \langle1\mathbf{3}\rangle ^{2} \lbrack\mathbf{2}\mathbf{4}\rbrack ^{2} 
    \nonumber\\
    &-c_W\langle1\mathbf{2}\rangle \langle1\mathbf{4}\rangle \lbrack\mathbf{2}\mathbf{3}\rbrack \lbrack\mathbf{3}\mathbf{4}\rbrack
    +c_W\langle1\mathbf{2}\rangle \langle1\mathbf{3}\rangle \lbrack\mathbf{2}\mathbf{4}\rbrack \lbrack\mathbf{3}\mathbf{4}\rbrack
    +c_W^{2} \langle1\mathbf{2}\rangle ^{2} \lbrack\mathbf{3}\mathbf{4}\rbrack ^{2} \Big).
\end{align}
We have validated this amplitude squared against Feynman diagrams in the processes $\gamma Z \to W \bar{W}$ and $\gamma W \to Z W$ in SPINAS
for a variety of masses and collider energies.  We have additionally validated it for the individual helicities of the photons when in the initial state for each of these masses and energies.

\subsection{$\mathbf{\gamma \gamma \bar{W} W}$}
As in the previous subsection, we show that this amplitude does not require a 4-point vertex, unlike Feynman diagrams, and is perturbatively unitary without one in \cite{Christensen:2024C}.
The T- and U-channel diagrams give an identical result.  If both photons have positive helicity, the amplitude is
\begin{align}
    \mathcal{M}_{\gamma^+\gamma^+\bar{W}W} &=
    2e^{2} 
    \frac{\lbrack12\rbrack ^{2} \langle\mathbf{3}\mathbf{4}\rangle ^{2} }
    {(t-M_W^2)(u-M_W^2)}.
\end{align}
If the photons have opposite helicity, we have
\begin{align}
    \mathcal{M}_{\gamma^+\gamma^-\bar{W}W} &=
    2e^2
    \frac{\left(\langle2\mathbf{4}\rangle \lbrack1\mathbf{3}\rbrack 
    +\langle2\mathbf{3}\rangle \lbrack1\mathbf{4}\rbrack \right)^{2} }
    {(t-M_W^2)(u-M_W^2)}.
\end{align}
The amplitudes for the other helicity combinations are given by interchange of the angle and square brackets.  They are
\begin{align}
    \mathcal{M}_{\gamma^-\gamma^-\bar{W}W} &=
    2e^{2} 
    \frac{\langle12\rangle ^{2} \lbrack\mathbf{3}\mathbf{4}\rbrack ^{2} }
    {(t-M_W^2)(u-M_W^2)}
\end{align}
and
\begin{align}
    \mathcal{M}_{\gamma^-\gamma^+\bar{W}W} &=
    2e^2
    \frac{\left(\lbrack2\mathbf{4}\rbrack \langle1\mathbf{3}\rangle 
    +\lbrack2\mathbf{3}\rbrack \langle1\mathbf{4}\rangle \right)^{2} }
    {(t-M_W^2)(u-M_W^2)}.
\end{align}
This last helicity combination can also be obtained from $\mathcal{M}_{\gamma^+\gamma^-\bar{W}W}$ by interchanging particles 1 and 2.
We have validated this amplitude squared against Feynman diagrams in the processes $\gamma \gamma \to W \bar{W}$ and $\gamma W \to \gamma W$ in SPINAS
for a variety of masses and collider energies.  We have additionally validated it for the individual helicities of the photons when in the initial state for each of these masses and energies.

\subsection{$\mathbf{g g g g}$}
The amplitude for this process is well known \cite{Parke:1986gb,Gastmans:1990xh,Dixon:1996wi,Britto:2005fq}, but we give it here for completeness.  
As for the $f \bar{f} g g$ process above, each diagram contributes the same to the numerator, but the color structure is different for each propagator denominator.  We don't have a simple form of the amplitude with propagator denominators and color matrices before squaring, so we give the numerators first and then follow them with the full squared amplitude.

We have six different non-zero color combinations
\begin{align}
    \mathcal{N}_{g^+g^+g^-g^-} &=
    2g_s^2
    \lbrack12\rbrack^2\langle34\rangle^2,
    \nonumber\\
    \mathcal{N}_{g^+g^-g^+g^-} &=
    2g_s^2
    \lbrack13\rbrack^2\langle24\rangle^2,
    \nonumber\\
    \mathcal{N}_{g^+g^-g^-g^+} &=
    2g_s^2
    \lbrack14\rbrack^2\langle23\rangle^2,
\end{align}
and the other helicity combinations related by interchange of the angle and square brackets are
\begin{align}
    \mathcal{N}_{g^-g^-g^+g^+} &=
    2g_s^2
    \langle12\rangle^2\lbrack34\rbrack^2,
    \nonumber\\
    \mathcal{N}_{g^-g^+g^-g^+} &=
    2g_s^2
    \langle13\rangle^2\lbrack24\rbrack^2,
    \nonumber\\
    \mathcal{N}_{g^-g^+g^+g^-} &=
    2g_s^2
    \langle14\rangle^2\lbrack23\rbrack^2.
\end{align}
All of these can also be obtained from the first by interchanging the position of the particles.  Furthermore, all of these spinor products are completely determined by the helicity of the gluons.
All the helicity combinations have the same propagator denominators and color structure.  After squaring and summing over colors, we have
\begin{align}
    \lvert\mathcal{M}_{g^{h_1}g^{h_2}g^{h_3}g^{h_4}}\rvert^2 &=
    \lvert\mathcal{N}_{g^{h_1}g^{h_2}g^{h_3}g^{h_4}}\rvert^2
    \left(
        \frac{36}{s^2t^2}
        +\frac{36}{s^2u^2}
        +\frac{36}{t^2u^2}
    \right),
\end{align}
for each helicity combination.  We have validated this amplitude squared against Feynman diagrams in the processes $g g \to g g$ in SPINAS
for all the combinations of initial helicities in addition to the case where all the helicities are summed over.

Just as with $f \bar{f} g g$, the factor $36$ comes from summing over the gluons, as in
\begin{align}
    \sum_{abcdeg}f_{bac}f_{cde}f_{gda}f_{gbe} &= 36
    \nonumber\\
    \sum_{abcdeg}f_{bac}f_{cde}f_{gea}f_{bdg} &= -36
    \nonumber\\
    \sum_{abcdeg}f_{cad}f_{ceb}f_{gea}f_{gbd} &= 36,
\end{align}
where the first is the S-T color combination, the second is the S-U color combination and the third is the T-U color combination.  Once again, we find the sign flip as we go from $1/(st)\to -1/(su) \to 1/(tu)$.

\section{\label{sec:other with no 4-point vertex}Other Amplitudes without a 4-Point Vertex}
In this section, we describe all the remaining 4-point amplitudes that do not involve a 4-point vertex.

\subsection{$\mathbf{f \bar{f} h h}$}
We begin with a fermion-anti-fermion pair and two Higgs bosons.  There is a contribution from the Higgs in the S channel, given by
\begin{align}
    \mathcal{M}_{hS} &=
    -\frac{3e^2m_fm_h^2}{4 M_W^{2} s_W^{2}} 
    \frac{\left(\langle\mathbf{1}\mathbf{2}\rangle +\lbrack\mathbf{1}\mathbf{2}\rbrack \right)}
    {s-m_h^2}.
\end{align}
There are also contributions from the fermion in the T and U channels, given by
\begin{align}
    \mathcal{M}_{fT} &=
    -\frac{e^2m_f^2}{4M_W^2s_W^2}
    \frac{\left(
    2m_f\left(\langle\mathbf{1}\mathbf{2}\rangle +\lbrack\mathbf{1}\mathbf{2}\rbrack \right)
    +\lbrack\mathbf{1}\lvert p_{3} \rvert \mathbf{2}\rangle +\lbrack\mathbf{2}\lvert p_{3} \rvert \mathbf{1}\rangle \right)}
    {t-m_f^2}
\end{align}
and
\begin{align}
    \mathcal{M}_{fU} &=
    -\frac{e^2m_f^2}{4M_W^2s_W^2}
    \frac{\left(
    2m_f\left(\langle\mathbf{1}\mathbf{2}\rangle +\lbrack\mathbf{1}\mathbf{2}\rbrack \right)
    +\lbrack\mathbf{1}\lvert p_{4} \rvert \mathbf{2}\rangle +\lbrack\mathbf{2}\lvert p_{4} \rvert \mathbf{1}\rangle \right)}
    {u-m_f^2}.
\end{align}
The total amplitude is
\begin{align}
    \mathcal{M}_{f\bar{f}hh} &= \mathcal{M}_{hS} + \mathcal{M}_{fT} + \mathcal{M}_{fU}.
\end{align}
If the fermion is a quark, then there is also a color delta function, ensuring the colors are the same.
We have validated this amplitude against Feynman diagrams in the processes $e \bar{e} \to h h$, $e h \to e h$, $u \bar{u} \to h h$, $u h \to u h$, $d \bar{d} \to h h$ and $d h \to d h$ in SPINAS
for a variety of masses and collider energies.

\subsection{$\mathbf{f \bar{f} Z h}$}
Our next amplitude involves a fermion-anti-fermion pair, a $Z$ boson and a Higgs boson.  There is a contribution from the $Z$ boson in the S channel, given by
\begin{align}
    \mathcal{M}_{ZS} &=
    \frac{e^2}{2\sqrt{2}M_W^2s_W^2}
    \frac{\left(
    2M_Z^{2} \left(g_R\langle\mathbf{2}\mathbf{3}\rangle \lbrack\mathbf{1}\mathbf{3}\rbrack +g_L\langle\mathbf{1}\mathbf{3}\rangle \lbrack\mathbf{2}\mathbf{3}\rbrack \right)
    +m_f\left(g_L-g_R\right)
    \left(\langle\mathbf{1}\mathbf{2}\rangle -\lbrack\mathbf{1}\mathbf{2}\rbrack \right)\lbrack\mathbf{3}\lvert p_{4} \rvert \mathbf{3}\rangle 
    \right)}
    {s-M_Z^2}.
\end{align}
There is also a contribution by the fermion in the T and U channels, given by
\begin{align}
    \mathcal{M}_{fT} &=
    \frac{e^2m_f}{2\sqrt{2}M_W^2s_W^2}
    \frac{\left(
    g_R\lbrack\mathbf{1}\mathbf{3}\rbrack 
    \left(2m_f\langle\mathbf{2}\mathbf{3}\rangle 
    +\lbrack\mathbf{2}\lvert p_{4} \rvert \mathbf{3}\rangle \right)
    +g_L\langle\mathbf{1}\mathbf{3}\rangle 
    \left(2m_f\lbrack\mathbf{2}\mathbf{3}\rbrack 
    +\lbrack\mathbf{3}\lvert p_{4} \rvert \mathbf{2}\rangle \right)
    \right)}
    {t-m_f^2}
\end{align}
and
\begin{align}
    \mathcal{M}_{fU} &=
    \frac{e^2m_f}{2\sqrt{2}M_W^2s_W^2}
    \frac{\left(
    g_L\lbrack\mathbf{2}\mathbf{3}\rbrack 
    \left(2m_f\langle\mathbf{1}\mathbf{3}\rangle 
    +\lbrack\mathbf{1}\lvert p_{4} \rvert \mathbf{3}\rangle \right)
    +g_R\langle\mathbf{2}\mathbf{3}\rangle 
    \left(2m_f\lbrack\mathbf{1}\mathbf{3}\rbrack 
    +\lbrack\mathbf{3}\lvert p_{4} \rvert \mathbf{1}\rangle \right)
    \right)}
    {u-m_f^2}.
\end{align}

The neutrino amplitude is particularly simple.  It only has a contribution from the $Z$ boson in the S channel, and is
\begin{align}
    \mathcal{M}_{\nu\bar{\nu}Zh} &=
    \frac{e^{2}}{\sqrt{2}c_W^{2} s_W^{2}}
    \frac{\langle1\mathbf{3}\rangle \lbrack2\mathbf{3}\rbrack }{ \left(s-M_Z^{2}\right)} .
\end{align}
We have validated this amplitude against Feynman diagrams in the processes $\nu_e \bar{\nu}_e \to Z h$ and $\nu_e Z \to \nu_e h$ in SPINAS
for a variety of masses and collider energies.

For charged leptons and quarks, we have
\begin{align}
    \mathcal{M}_{f\bar{f}Zh} &=
    \mathcal{M}_{ZS} + \mathcal{M}_{fT} + \mathcal{M}_{fU},
\end{align}
with a Kronecker delta for the color if the fermion is a quark.  We have validated this amplitude against Feynman diagrams in the processes $e \bar{e} \to Z h$, $e Z \to e h$, $u \bar{u} \to Z h$, $u Z \to u h$, $d \bar{d} \to Z h$ and $d Z \to d h$ in SPINAS
for a variety of masses and collider energies.

\subsection{$\mathbf{f_1 \bar{f}_2 \bar{W} h}$ and $\mathbf{\bar{f}_1 f_2 W h}$}
We now consider a fermion and the antifermion of its isospin partner along with an anti-$W$ boson and a Higgs boson.  There is a contribution from a $W$ boson in the S channel, given by
\begin{align}
    \mathcal{M}_{WS} &=
    -\frac{e^2}{2M_W^2s_W^2}
    \frac{\left(
    2M_W^{2} \langle\mathbf{13}\rangle \lbrack\mathbf{23}\rbrack 
    + 
    \left(m_2\langle\mathbf{12}\rangle-m_1\lbrack\mathbf{12}\rbrack \right)
    \lbrack\mathbf{3}\lvert p_{4} \rvert \mathbf{3}\rangle
    \right)}
    {s-M_W^2}.
    \label{eq:f1f2Wh:MWS}
\end{align}
There are also contributions from the fermions in the T and U channels, given by
\begin{align}
    \mathcal{M}_{f_2T} &= 
    -\frac{e^2m_2}{2M_W^2s_W^2}
    \frac{
    \langle\mathbf{13}\rangle 
    \left(2m_2\lbrack\mathbf{23}\rbrack 
    +\lbrack\mathbf{3}\lvert p_{4} \rvert \mathbf{2}\rangle\right)
    }
    {t-m_2^2}
\end{align}
and
\begin{align}
    \mathcal{M}_{f_1U} &= 
    -\frac{e^2m_1}{2M_W^2s_W^2}
    \frac{\lbrack\mathbf{23}\rbrack \left(2m_1\langle\mathbf{13}\rangle +\lbrack\mathbf{1}\lvert p_{4} \rvert \mathbf{3}\rangle \right)}
    {u-m_1^2}.
\end{align}

For the leptons, we have the S and T channels, giving us
\begin{align}
    \mathcal{M}_{\nu_l\bar{l},\bar{W},h} &=
    \mathcal{M}_{WS} + \mathcal{M}_{lT},
\end{align}
where $m_1=0$.  We have validated this amplitude against Feynman diagrams in the processes $\bar{e} \nu_e \to W h$ and $h \nu_e \to W e$ in SPINAS
for a variety of masses and collider energies.

For the quarks, we have all three channels, giving
\begin{align}
    \mathcal{M}_{q_1\bar{q}_2\bar{W}h} &=
    \left(\mathcal{M}_{WS} + \mathcal{M}_{q_2T} + \mathcal{M}_{q_1U}\right)\delta^{i_1}_{i_2}.
    \label{eq:f1f2Wh:MqqWh}
\end{align}
We have validated this amplitude against Feynman diagrams in the processes $u \bar{d} \to W h$ and $u h \to W d$ in SPINAS
for a variety of masses and collider energies.

The amplitude for $\bar{f}_1 f_2 W h$, on the other hand, is obtained from Eqs.~(\ref{eq:f1f2Wh:MWS}) through (\ref{eq:f1f2Wh:MqqWh}), by simply interchanging angle and square brackets ($\langle\rangle\longleftrightarrow\lbrack\rbrack$).

\subsection{$\mathbf{f \bar{f} \bar{W} W}$}
For this amplitude, we have a contribution from the Higgs in the S channel, which is
\begin{equation}
    \mathcal{M}_{hS} = 
    -\frac{e^{2} m_f}{2M_W^{2} s_W^{2}}
    \frac{ \left(\langle\mathbf{1}\mathbf{2}\rangle +\lbrack\mathbf{1}\mathbf{2}\rbrack \right)\langle\mathbf{3}\mathbf{4}\rangle\lbrack\mathbf{3}\mathbf{4}\rbrack }{ \left(s-m_h^{2}\right)} .
\end{equation}

The contribution from the isospin partner in the T channel contributes if the isospin of $f$ is $+\frac{1}{2}$.  It is
\begin{equation}
    \mathcal{M}_{f'T} = \frac{e^{2} }{M_W^{2} s_W^{2}} 
    \frac{\langle\mathbf{1}\mathbf{3}\rangle \lbrack\mathbf{2}\mathbf{4}\rbrack \left(M_W\langle\mathbf{3}\mathbf{4}\rangle +\lbrack\mathbf{3}\lvert p_{1} \rvert \mathbf{4}\rangle \right)}{ \left(t-m_{f'}^{2}\right)} .
    \label{eq:ffWW:f' T}
\end{equation}
If the isospin of $f$ is $-\frac{1}{2}$, the contribution from the isospin partner is in the U channel, and is
\begin{equation}
    \mathcal{M}_{f'U} = -\frac{e^{2} }{M_W^{2} s_W^{2}} 
    \frac{\langle\mathbf{1}\mathbf{4}\rangle \lbrack\mathbf{2}\mathbf{3}\rbrack \left(M_W\langle\mathbf{3}\mathbf{4}\rangle -\lbrack\mathbf{4}\lvert p_{1} \rvert \mathbf{3}\rangle \right)}{ \left(u-m_{f'}^{2}\right)} .
    \label{eq:ffWW:f' U}
\end{equation}

The contribution from the photon in the S channel is
\begin{align}
    \mathcal{M}_{\gamma S} = 
    \frac{2e^2Q_f}{M_W^2s}
    \Big(&
    M_W\left(
    \langle\mathbf{2}\mathbf{4}\rangle \lbrack\mathbf{1}\mathbf{3}\rbrack 
    +\langle\mathbf{2}\mathbf{3}\rangle \lbrack\mathbf{1}\mathbf{4}\rbrack 
    +\langle\mathbf{1}\mathbf{4}\rangle \lbrack\mathbf{2}\mathbf{3}\rbrack 
    +\langle\mathbf{1}\mathbf{3}\rangle \lbrack\mathbf{2}\mathbf{4}\rbrack \right)\left(\langle\mathbf{3}\mathbf{4}\rangle +\lbrack\mathbf{3}\mathbf{4}\rbrack \right)
    \nonumber\\
    &+\langle\mathbf{3}\mathbf{4}\rangle \lbrack\mathbf{3}\mathbf{4}\rbrack \left(
    \lbrack\mathbf{1}\lvert p_{3} \rvert \mathbf{2}\rangle + \lbrack\mathbf{2}\lvert p_{3} \rvert \mathbf{1}\rangle 
    \right)
    \Big).
    \label{eq:ffWW:A S}
\end{align}
We found this by using a massive photon and taking the massless limit at the end of the calculation.  We have also confirmed that this result can be obtained by use of the $x$ factor (and a massless photon) and show this derivation in App.~\ref{app:internal photons}.

The contribution from the $Z$ boson in the S channel is
\begin{align}
\mathcal{M}_{ZS} &= 
    \frac{e^2}{2 M_W^2 s_W^2\left(s-M_Z^{2}\right)}
    \Bigg(
    2\langle\mathbf{3}\mathbf{4}\rangle \lbrack\mathbf{3}\mathbf{4}\rbrack 
    \left(g_R\lbrack\mathbf{1}\lvert p_{3} \rvert \mathbf{2}\rangle + g_L\lbrack\mathbf{2}\lvert p_{3} \rvert \mathbf{1}\rangle \right)
    - m_f \langle\mathbf{3}\mathbf{4}\rangle \lbrack\mathbf{3}\mathbf{4}\rbrack
    \left(\langle\mathbf{1}\mathbf{2}\rangle  - \lbrack\mathbf{1}\mathbf{2}\rbrack \right)
    \nonumber\\
    &+2M_W    \left(\langle\mathbf{3}\mathbf{4}\rangle+\lbrack\mathbf{3}\mathbf{4}\rbrack\right)
    \Big(g_R
    \left(\langle\mathbf{2}\mathbf{4}\rangle \lbrack\mathbf{1}\mathbf{3}\rbrack +\langle\mathbf{2}\mathbf{3}\rangle \lbrack\mathbf{1}\mathbf{4}\rbrack 
    \right)
    +g_L \left(\langle\mathbf{1}\mathbf{4}\rangle  \lbrack\mathbf{2}\mathbf{3}\rbrack +\langle\mathbf{1}\mathbf{3}\rangle  \lbrack\mathbf{2}\mathbf{4}\rbrack  \right)
    \Big)
    \Bigg) .
    \label{eq:ffWW:Z S}
\end{align}

If the fermion is a neutrino, we only have the electron T-channel diagram and the $Z$ boson S-channel diagram.  All together, the amplitude is
\begin{align}
    \mathcal{M}_{\nu\bar{\nu}\bar{W}W} &=
    \frac{e^2}{M_W^2s_W^2}
    \left(
    \frac{\langle1\mathbf{3}\rangle \lbrack2\mathbf{4}\rbrack \left(M_W\langle\mathbf{3}\mathbf{4}\rangle +\lbrack\mathbf{3}\lvert p_{1} \rvert \mathbf{4}\rangle \right)}
    {t-m_l^2}
    +\frac{M_W\left(\langle1\mathbf{4}\rangle \lbrack2\mathbf{3}\rbrack +\langle1\mathbf{3}\rangle \lbrack2\mathbf{4}\rbrack \right)\left(\langle\mathbf{3}\mathbf{4}\rangle +\lbrack\mathbf{3}\mathbf{4}\rbrack \right)+\langle\mathbf{3}\mathbf{4}\rangle \lbrack\mathbf{3}\mathbf{4}\rbrack \lbrack2\lvert p_{3} \rvert 1\rangle}
    {s-M_Z^2}
    \right),
\end{align}
where $l$ represents the charged lepton partner of the neutrino.  We have validated this amplitude against Feynman diagrams in the processes $\nu_e \bar{\nu}_e \to 
W \bar{W}$ and $\nu_e W \to W \nu_e$ in SPINAS
for a variety of masses and collider energies.

The amplitude for up-type quarks doesn't simplify, but also includes the T-channel diagram, giving
\begin{align}
    \mathcal{M}_{u\bar{u}\bar{W}W} &=
    \left(
    \mathcal{M}_{hS} + \mathcal{M}_{dT} + \mathcal{M}_{\gamma S} + \mathcal{M}_{Z S}
    \right)\delta^{i_1}_{i_2},
\end{align}
where $u$ stands for any of $u, c$ or $t$ and $d$ stands for $d, s$ or $b$, respectively.  We have validated this amplitude against Feynman diagrams in the processes $u \bar{u} \to 
W \bar{W}$ and $u W \to W u$ in SPINAS
for a variety of masses and collider energies.

For the charged leptons and down-type quarks, we have
\begin{align}
    \mathcal{M}_{f\bar{f}\bar{W}W} &=
    \left(
    \mathcal{M}_{hS} + \mathcal{M}_{f'U} + \mathcal{M}_{\gamma S} + \mathcal{M}_{Z S}
    \right),
\end{align}
where $f$ stands for any of $e, \mu, \tau, d, s$ or $b$ and $f'$ stands for $\nu_e, \nu_\mu, \nu_\tau, u, c$ or $t$, respectively.  There is also a Kronecker delta for the colors if $f$ is a quark.  We have validated this amplitude against Feynman diagrams in the processes $e \bar{e} \to 
W \bar{W}$, $e W \to W e$, $d \bar{d} \to 
W \bar{W}$ and $d W \to W d$ in SPINAS
for a variety of masses and collider energies.

\subsection{$\mathbf{f \bar{f} Z Z}$}
The contribution from the Higgs in the S channel is 
\begin{equation}
\mathcal{M}_{hS} =
    -\frac{e^{2} m_f}{2M_W^{2} s_W^{2}} \frac{\langle\mathbf{3}\mathbf{4}\rangle \lbrack\mathbf{3}\mathbf{4}\rbrack
    \left(\langle\mathbf{1}\mathbf{2}\rangle 
    + \lbrack\mathbf{1}\mathbf{2}\rbrack \right)}{ \left(s-m_h^{2}\right)} ,
\end{equation}
where there is also a color Kronecker delta function when the fermion is a quark.

The contributions from the fermion in the T and U channels are
\begin{align}
\mathcal{M}_{fT} &=
    \frac{e^{2} }{2 M_W^{2} s_W^{2}(t-m_f^2)}
    \Bigg(
    g_L^{2} \left(M_Z\langle\mathbf{3}\mathbf{4}\rangle +\lbrack\mathbf{3}\lvert p_{1} \rvert \mathbf{4}\rangle \right) \langle\mathbf{1}\mathbf{3}\rangle \lbrack\mathbf{2}\mathbf{4}\rbrack  
    +g_R^{2} \left(M_Z\lbrack\mathbf{3}\mathbf{4}\rbrack +\lbrack\mathbf{4}\lvert p_{1} \rvert \mathbf{3}\rangle \right)\langle\mathbf{2}\mathbf{4}\rangle \lbrack\mathbf{1}\mathbf{3}\rbrack 
    \nonumber\\
    &+g_Lg_Rm_f\left(\langle\mathbf{3}\mathbf{4}\rangle \lbrack\mathbf{1}\mathbf{3}\rbrack \lbrack\mathbf{2}\mathbf{4}\rbrack +\langle\mathbf{1}\mathbf{3}\rangle \langle\mathbf{2}\mathbf{4}\rangle \lbrack\mathbf{3}\mathbf{4}\rbrack \right)
    \Bigg)
    \label{eq:ffZZ:f T}
\end{align}
and
\begin{align}
\mathcal{M}_{fU} &=
    \frac{-e^{2} }{2 M_W^{2} s_W^{2}(u-m_f^2)}
    \Bigg(
    g_R^{2}\left(M_Z\lbrack\mathbf{3}\mathbf{4}\rbrack -\lbrack\mathbf{3}\lvert p_{1} \rvert \mathbf{4}\rangle \right)  \langle\mathbf{2}\mathbf{3}\rangle \lbrack\mathbf{1}\mathbf{4}\rbrack  
    +g_L^{2}\left(M_Z\langle\mathbf{3}\mathbf{4}\rangle -\lbrack\mathbf{4}\lvert p_{1} \rvert \mathbf{3}\rangle \right)  \langle\mathbf{1}\mathbf{4}\rangle \lbrack\mathbf{2}\mathbf{3}\rbrack  
    \nonumber\\
    &+g_Lg_Rm_f\left(\langle\mathbf{3}\mathbf{4}\rangle \lbrack\mathbf{1}\mathbf{4}\rbrack \lbrack\mathbf{2}\mathbf{3}\rbrack +\langle\mathbf{1}\mathbf{4}\rangle \langle\mathbf{2}\mathbf{3}\rangle \lbrack\mathbf{3}\mathbf{4}\rbrack \right)
    \Bigg) ,
    \label{eq:ffZZ:f U}
\end{align}
where these contributions also have a color Kronecker delta function when the fermion is a quark.

The neutrino amplitude is particularly simple, given by
\begin{align}
    \mathcal{M}_{\nu\bar{nu}ZZ} &=
    \frac{e^2}{2M_W^2s_W^2}
    \left(
    \frac{\langle1\mathbf{3}\rangle \lbrack2\mathbf{4}\rbrack \left(\lbrack\mathbf{3}\lvert p_{1} \rvert \mathbf{4}\rangle + M_Z\langle\mathbf{3}\mathbf{4}\rangle \right)}
    {t}
    +\frac{\langle1\mathbf{4}\rangle \lbrack2\mathbf{3}\rbrack \left( \lbrack\mathbf{4}\lvert p_{1} \rvert \mathbf{3}\rangle -M_Z\langle\mathbf{3}\mathbf{4}\rangle\right)}
    {u}
    \right).
\end{align}
We have validated this amplitude against Feynman diagrams in the processes $\nu_e \bar{\nu}_e \to Z Z$ and $\nu_e Z \to Z \nu_e$ in SPINAS
for a variety of masses and collider energies.

For the other fermions, the amplitude is given by
\begin{align}
    \mathcal{M}_{f\bar{f}ZZ} &=
    \mathcal{M}_{hS} + \mathcal{M}_{fT} + \mathcal{M}_{fU},
\end{align}
with a Kronecker delta for colors if $f$ is a quark.  We have validated this amplitude against Feynman diagrams in the processes $e \bar{e} \to Z Z$, $e Z \to Z e$, $u \bar{u} \to Z Z$, $u Z \to Z u$, $d \bar{d} \to Z Z$ and $d Z \to Z d$ in SPINAS
for a variety of masses and collider energies.

\subsection{$\mathbf{\bar{f}_1 f_2 Z W}$ and $\mathbf{\bar{f}_1 f_2 Z \bar{W}}$}
We begin with $\bar{f}_1 f_2 Z W$.
The contribution from the $W$ boson in the S channel is
\begin{align}
\mathcal{M}_{WS} &= 
    \frac{e^2}{\sqrt{2}M_W^{2} M_Z^{2} s_W^{2} \left(s-M_W^{2}\right)}
    \Bigg(
    2M_W^{3} 
    \left(
    \langle\mathbf{2}\mathbf{4}\rangle \langle\mathbf{3}\mathbf{4}\rangle \lbrack\mathbf{1}\mathbf{3}\rbrack 
    + \langle\mathbf{2}\mathbf{3}\rangle \lbrack\mathbf{1}\mathbf{4}\rbrack \lbrack\mathbf{3}\mathbf{4}\rbrack
    \right)
    +2M_W^{2} M_Z \left(
    \langle\mathbf{2}\mathbf{3}\rangle \langle\mathbf{3}\mathbf{4}\rangle \lbrack\mathbf{1}\mathbf{4}\rbrack
    + \langle\mathbf{2}\mathbf{4}\rangle \lbrack\mathbf{1}\mathbf{3}\rbrack \lbrack\mathbf{3}\mathbf{4}\rbrack
    \right)
    \nonumber\\
    &+2M_W^{2} \langle\mathbf{3}\mathbf{4}\rangle \lbrack\mathbf{3}\mathbf{4}\rbrack \lbrack\mathbf{1}\lvert p_{3} \rvert \mathbf{2}\rangle 
    +\left(m_{f_2}\lbrack\mathbf{1}\mathbf{2}\rbrack
    -m_{f_1}\langle\mathbf{1}\mathbf{2}\rangle\right)
    \left(M_Z^2-2M_W^{2}\right) \langle\mathbf{3}\mathbf{4}\rangle \lbrack\mathbf{3}\mathbf{4}\rbrack 
    \Bigg),
    \label{eq:ffZW:W S}
\end{align}
with a Kronecker delta for quarks.

The contribution from $f_1$ in the T channel is
\begin{align}
\mathcal{M}_{f_1T} &= 
    \frac{e^2}{\sqrt{2}M_W^{2} s_W^{2}}
    \frac{\langle\mathbf{2}\mathbf{4}\rangle \left(g_{Rf_1}m_{f_1}\langle\mathbf{1}\mathbf{3}\rangle \lbrack\mathbf{3}\mathbf{4}\rbrack +g_{Lf_1}\lbrack\mathbf{1}\mathbf{3}\rbrack\left(M_Z\lbrack\mathbf{3}\mathbf{4}\rbrack +\lbrack\mathbf{4}\lvert p_{1} \rvert \mathbf{3}\rangle \right) \right)}
    {\left(t-m_{f_1}^{2}\right)},
    \label{eq:ffZW:f1 T}
\end{align}
with a Kronecker delta for quarks.

The contribution from the isospin partner in the U channel is
\begin{align}
\mathcal{M}_{f_2 U} &= 
    -\frac{e^2}{\sqrt{2}M_W^{2} s_W^{2}}
    \frac{\lbrack\mathbf{1}\mathbf{4}\rbrack \left(g_{Rf_2}m_{f_2}\langle\mathbf{3}\mathbf{4}\rangle \lbrack\mathbf{2}\mathbf{3}\rbrack +g_{Lf_2}\langle\mathbf{2}\mathbf{3}\rangle\left(M_W\lbrack\mathbf{3}\mathbf{4}\rbrack -\lbrack\mathbf{3}\lvert p_{1} \rvert \mathbf{4}\rangle \right) \right)}
    {(u-m_{f_2}^2)},
    \label{eq:ffZW:f2 U}
\end{align}
with a Kronecker delta for quarks.  

All together, we have
\begin{align}
    \mathcal{M}_{\bar{f}_1f_2ZW} &=
    \mathcal{M}_{WS} + \mathcal{M}_{f_1T} + \mathcal{M}_{f_2U},
\end{align}
times a color Kronecker delta for quarks.  We have validated this amplitude against Feynman diagrams in the processes $\bar{u} d \to Z W$ and $W d \to Z u$ in SPINAS
for a variety of masses and collider energies.

We now turn to $\bar{f}_1 f_2 Z \bar{W}$.  $\mathcal{M}_{WS}$ flips sign and interchanges $f_1$ and $f_2$.  We obtain
\begin{align}
\mathcal{M}_{WS} &= 
    -\frac{e^2}{\sqrt{2}M_W^{2} M_Z^{2} s_W^{2} \left(s-M_W^{2}\right)}
    \Bigg(
    2M_W^{3} 
    \left(
    \langle\mathbf{2}\mathbf{4}\rangle \langle\mathbf{3}\mathbf{4}\rangle \lbrack\mathbf{1}\mathbf{3}\rbrack 
    + \langle\mathbf{2}\mathbf{3}\rangle \lbrack\mathbf{1}\mathbf{4}\rbrack \lbrack\mathbf{3}\mathbf{4}\rbrack
    \right)
    +2M_W^{2} M_Z \left(
    \langle\mathbf{2}\mathbf{3}\rangle \langle\mathbf{3}\mathbf{4}\rangle \lbrack\mathbf{1}\mathbf{4}\rbrack
    + \langle\mathbf{2}\mathbf{4}\rangle \lbrack\mathbf{1}\mathbf{3}\rbrack \lbrack\mathbf{3}\mathbf{4}\rbrack
    \right)
    \nonumber\\
    &+2M_W^{2} \langle\mathbf{3}\mathbf{4}\rangle \lbrack\mathbf{3}\mathbf{4}\rbrack \lbrack\mathbf{1}\lvert p_{3} \rvert \mathbf{2}\rangle 
    +\left(m_{f_1}\lbrack\mathbf{1}\mathbf{2}\rbrack
    -m_{f_2}\langle\mathbf{1}\mathbf{2}\rangle\right)
    \left(M_Z^2-2M_W^{2}\right) \langle\mathbf{3}\mathbf{4}\rangle \lbrack\mathbf{3}\mathbf{4}\rbrack 
    \Bigg),
    \label{eq:ffZW:W2 S}
\end{align}
with a Kronecker delta for quarks.  For the T- and U-channel diagrams, the result is related to the previous results by $f_1\longleftrightarrow f_2$.  They are
\begin{align}
    \mathcal{M}_{f_2T} &= \frac{e^2}{\sqrt{2}M_W^{2} s_W^{2}}
    \frac{\langle\mathbf{2}\mathbf{4}\rangle \left(g_{Rf_2}m_{f_2}\langle\mathbf{1}\mathbf{3}\rangle \lbrack\mathbf{3}\mathbf{4}\rbrack +g_{Lf_2}\lbrack\mathbf{1}\mathbf{3}\rbrack\left(M_Z\lbrack\mathbf{3}\mathbf{4}\rbrack +\lbrack\mathbf{4}\lvert p_{1} \rvert \mathbf{3}\rangle \right) \right)}
    {\left(t-m_{f_2}^{2}\right)}
    \label{eq:ffZW:f2 T}
\end{align}
and 
\begin{align}
\mathcal{M}_{f_1 U} &= 
    -\frac{e^2}{\sqrt{2}M_W^{2} s_W^{2}}
    \frac{\lbrack\mathbf{1}\mathbf{4}\rbrack \left(g_{Rf_1}m_{f_1}\langle\mathbf{3}\mathbf{4}\rangle \lbrack\mathbf{2}\mathbf{3}\rbrack +g_{Lf_1}\langle\mathbf{2}\mathbf{3}\rangle\left(M_W\lbrack\mathbf{3}\mathbf{4}\rbrack -\lbrack\mathbf{3}\lvert p_{1} \rvert \mathbf{4}\rangle \right) \right)}
    {(u-m_{f_1}^2)},
    \label{eq:ffZW:f1 U}
\end{align}
both with a Kronecker delta for quarks.  All together, we have
\begin{align}
    \mathcal{M}_{\bar{f}_1f_2Z\bar{W}} &=
    \mathcal{M}_{WS} + \mathcal{M}_{f_2T} + \mathcal{M}_{f_1U},
\end{align}
times a color Kronecker delta for quarks.  We have validated this amplitude against Feynman diagrams in the processes $\bar{e} \nu_e \to Z W$ and $Z \nu_e \to W e$ in SPINAS
for a variety of masses and collider energies.

\subsection{$\mathbf{Z h \bar{W} W}$}
The contribution from the $Z$ boson in the S channel is
\begin{align}
\mathcal{M}_{ZS} &=
    \frac{e^{2} }{\sqrt{2}M_W^{2} s_W^{2}}
    \frac{
    2M_W
    \left(\langle\mathbf{1}\mathbf{4}\rangle \lbrack\mathbf{1}\mathbf{3}\rbrack
    +\langle\mathbf{1}\mathbf{3}\rangle \lbrack\mathbf{1}\mathbf{4}\rbrack\right)\left(\langle\mathbf{3}\mathbf{4}\rangle +\lbrack\mathbf{3}\mathbf{4}\rbrack \right) 
    +\left(\lbrack\mathbf{1}\lvert p_{3} \rvert \mathbf{1}\rangle -\lbrack\mathbf{1}\lvert p_{4} \rvert \mathbf{1}\rangle \right)\langle\mathbf{3}\mathbf{4}\rangle \lbrack\mathbf{3}\mathbf{4}\rbrack }{ \left(s-M_Z^{2}\right)} .
    \label{eq:ZhWW:Z S}
\end{align}
The contribution from a $W$ boson in the T channel is
\begin{align}
\mathcal{M}_{WT} &=
    \frac{e^{2} }{\sqrt{2}M_W^{2} M_Z^{2} s_W^{2} \left(t-M_W^{2}\right)}
    \Bigg(
    2M_W^{3} \langle\mathbf{1}\mathbf{3}\rangle \langle\mathbf{3}\mathbf{4}\rangle \lbrack\mathbf{1}\mathbf{4}\rbrack 
    +2M_W^{2} M_Z \left(
    \langle\mathbf{3}\mathbf{4}\rangle \lbrack\mathbf{1}\mathbf{3}\rbrack \lbrack\mathbf{1}\mathbf{4}\rbrack 
    + \langle\mathbf{1}\mathbf{3}\rangle \langle\mathbf{1}\mathbf{4}\rangle \lbrack\mathbf{3}\mathbf{4}\rbrack 
    \right)
    \nonumber\\
    &-M_Z^{2} \langle\mathbf{1}\mathbf{3}\rangle \lbrack\mathbf{1}\mathbf{3}\rbrack \lbrack\mathbf{4}\lvert p_{2} \rvert \mathbf{4}\rangle 
    +2M_W^{2} \langle\mathbf{3}\mathbf{4}\rangle \lbrack\mathbf{1}\mathbf{3}\rbrack \lbrack\mathbf{4}\lvert p_{3} \rvert \mathbf{1}\rangle 
    \Bigg).
    \label{eq:ZhWW:W T}
\end{align}
The contribution from a $W$ boson in the U channel is
\begin{align}
\mathcal{M}_{WU} &=
    \frac{e^{2}}{\sqrt{2}M_W^{2} M_Z^{2} s_W^{2} \left(u-M_W^{2}\right)}  
    \Bigg(
    2M_W^{3} \langle\mathbf{1}\mathbf{4}\rangle \langle\mathbf{3}\mathbf{4}\rangle \lbrack\mathbf{1}\mathbf{3}\rbrack 
    +2M_W^{2} M_Z \left(
    \langle\mathbf{3}\mathbf{4}\rangle \lbrack\mathbf{1}\mathbf{3}\rbrack \lbrack\mathbf{1}\mathbf{4}\rbrack 
    + \langle\mathbf{1}\mathbf{3}\rangle \langle\mathbf{1}\mathbf{4}\rangle \lbrack\mathbf{3}\mathbf{4}\rbrack
    \right)
    \nonumber\\
    &+M_Z^{2} \langle\mathbf{1}\mathbf{4}\rangle \lbrack\mathbf{1}\mathbf{4}\rbrack \lbrack\mathbf{3}\lvert p_{2} \rvert \mathbf{3}\rangle 
    +2M_W^{2} \langle\mathbf{3}\mathbf{4}\rangle \lbrack\mathbf{1}\mathbf{4}\rbrack \lbrack\mathbf{3}\lvert p_{4} \rvert \mathbf{1}\rangle 
    \Bigg).
    \label{eq:ZhWW:W U}
\end{align}
The combination of these gives the amplitude,
\begin{align}
    \mathcal{M}_{Zh\bar{W}W} &=
    \mathcal{M}_{ZS} + \mathcal{M}_{WT} + \mathcal{M}_{WU}.
\end{align}
We have validated this amplitude against Feynman diagrams in the processes $Z h \to W \bar{W}$ and $Z W \to W h$ in SPINAS
for a variety of masses and collider energies.

\subsection{$\mathbf{Z Z Z Z}$}
This amplitude has contributions from the Higgs boson in the S, T and U channels.  It is
\begin{align}
    \mathcal{M}_{ZZZZ} &= 
    \frac{e^2}{M_W^2s_W^2}
    \left(
    \frac{\langle\mathbf{1}\mathbf{2}\rangle \langle\mathbf{3}\mathbf{4}\rangle \lbrack\mathbf{1}\mathbf{2}\rbrack \lbrack\mathbf{3}\mathbf{4}\rbrack }{ \left(s-m_h^{2}\right)} 
    +\frac{\langle\mathbf{1}\mathbf{3}\rangle \langle\mathbf{2}\mathbf{4}\rangle \lbrack\mathbf{1}\mathbf{3}\rbrack \lbrack\mathbf{2}\mathbf{4}\rbrack }{ \left(t-m_h^{2}\right)} 
    +\frac{\langle\mathbf{1}\mathbf{4}\rangle \langle\mathbf{2}\mathbf{3}\rangle \lbrack\mathbf{1}\mathbf{4}\rbrack \lbrack\mathbf{2}\mathbf{3}\rbrack }{ \left(u-m_h^{2}\right)} 
    \right).
\end{align}
We have validated this amplitude against Feynman diagrams in the processes $Z Z \to Z Z$ in SPINAS
for a variety of masses and collider energies.

\subsection{$\mathbf{Z Z \bar{W} W}$}

The amplitude in this subsection is also found in \cite{Christensen:2024B}, but with particles 3 and 4 outgoing, where we show that it satisfies perturbative unitarity.

The contribution to the amplitude coming from an S-channel Higgs is
\begin{equation}
    \mathcal{M}_{hS} =
    -\frac{e^{2}}{M_W^{2} s_W^{2}}
    \frac{\langle\mathbf{1}\mathbf{2}\rangle \langle\mathbf{3}\mathbf{4}\rangle \lbrack\mathbf{1}\mathbf{2}\rbrack \lbrack\mathbf{3}\mathbf{4}\rbrack }{ \left(s-m_h^{2}\right)} .
    \label{eq:ZZWW:Higgs Amp}
\end{equation}

The contribution to the amplitude coming from a T-channel $W$ boson is
\begin{align}
\mathcal{M}_{WT} &= -\frac{e^2}{2M_W^3s_W^2\left(t-M_W^2\right)}
\Bigg(
M_W\Big(
    4c_W^{3} \langle\mathbf{2}\mathbf{4}\rangle \langle\mathbf{3}\mathbf{4}\rangle \lbrack\mathbf{1}\mathbf{2}\rbrack \lbrack\mathbf{1}\mathbf{3}\rbrack 
    -3c_W^{3} \langle\mathbf{1}\mathbf{3}\rangle \langle\mathbf{2}\mathbf{3}\rangle \langle\mathbf{2}\mathbf{4}\rangle \lbrack\mathbf{1}\mathbf{4}\rbrack 
    -4c_W^{2} \langle\mathbf{2}\mathbf{3}\rangle \langle\mathbf{2}\mathbf{4}\rangle \lbrack\mathbf{1}\mathbf{3}\rbrack \lbrack\mathbf{1}\mathbf{4}\rbrack 
    \nonumber\\
    &-3c_W^{4} \langle\mathbf{1}\mathbf{4}\rangle \langle\mathbf{2}\mathbf{4}\rangle \lbrack\mathbf{1}\mathbf{3}\rbrack \lbrack\mathbf{2}\mathbf{3}\rbrack 
    -3c_W^{2} \langle\mathbf{1}\mathbf{3}\rangle \langle\mathbf{2}\mathbf{4}\rangle \lbrack\mathbf{1}\mathbf{4}\rbrack \lbrack\mathbf{2}\mathbf{3}\rbrack 
    -c_W^{4} \langle\mathbf{1}\mathbf{3}\rangle \langle\mathbf{2}\mathbf{4}\rangle \lbrack\mathbf{1}\mathbf{4}\rbrack \lbrack\mathbf{2}\mathbf{3}\rbrack 
    -c_W^{3} \langle\mathbf{2}\mathbf{4}\rangle \lbrack\mathbf{1}\mathbf{3}\rbrack \lbrack\mathbf{1}\mathbf{4}\rbrack \lbrack\mathbf{2}\mathbf{3}\rbrack 
    \nonumber\\
    &-3c_W^{3} \langle\mathbf{1}\mathbf{3}\rangle \langle\mathbf{1}\mathbf{4}\rangle \langle\mathbf{2}\mathbf{3}\rangle \lbrack\mathbf{2}\mathbf{4}\rbrack 
    -3c_W^{2} \langle\mathbf{1}\mathbf{4}\rangle \langle\mathbf{2}\mathbf{3}\rangle \lbrack\mathbf{1}\mathbf{3}\rbrack \lbrack\mathbf{2}\mathbf{4}\rbrack 
    -c_W^{4} \langle\mathbf{1}\mathbf{4}\rangle \langle\mathbf{2}\mathbf{3}\rangle \lbrack\mathbf{1}\mathbf{3}\rbrack \lbrack\mathbf{2}\mathbf{4}\rbrack 
    -2\langle\mathbf{1}\mathbf{3}\rangle \langle\mathbf{2}\mathbf{4}\rangle \lbrack\mathbf{1}\mathbf{3}\rbrack \lbrack\mathbf{2}\mathbf{4}\rbrack 
    \nonumber\\
    &+6c_W^{2} \langle\mathbf{1}\mathbf{3}\rangle \langle\mathbf{2}\mathbf{4}\rangle \lbrack\mathbf{1}\mathbf{3}\rbrack \lbrack\mathbf{2}\mathbf{4}\rbrack 
    -2c_W^{4} \langle\mathbf{1}\mathbf{3}\rangle \langle\mathbf{2}\mathbf{4}\rangle \lbrack\mathbf{1}\mathbf{3}\rbrack \lbrack\mathbf{2}\mathbf{4}\rbrack 
    -2c_W^{2} \langle\mathbf{1}\mathbf{2}\rangle \langle\mathbf{3}\mathbf{4}\rangle \lbrack\mathbf{1}\mathbf{3}\rbrack \lbrack\mathbf{2}\mathbf{4}\rbrack 
    +2c_W^{4} \langle\mathbf{1}\mathbf{2}\rangle \langle\mathbf{3}\mathbf{4}\rangle \lbrack\mathbf{1}\mathbf{3}\rbrack \lbrack\mathbf{2}\mathbf{4}\rbrack 
    \nonumber\\
    &-c_W^{4} \langle\mathbf{1}\mathbf{3}\rangle \langle\mathbf{2}\mathbf{3}\rangle \lbrack\mathbf{1}\mathbf{4}\rbrack \lbrack\mathbf{2}\mathbf{4}\rbrack 
    -4c_W^{2} \langle\mathbf{1}\mathbf{3}\rangle \langle\mathbf{1}\mathbf{4}\rangle \lbrack\mathbf{2}\mathbf{3}\rbrack \lbrack\mathbf{2}\mathbf{4}\rbrack 
    -c_W^{3} \langle\mathbf{1}\mathbf{4}\rangle \lbrack\mathbf{1}\mathbf{3}\rbrack \lbrack\mathbf{2}\mathbf{3}\rbrack \lbrack\mathbf{2}\mathbf{4}\rbrack 
    +4c_W^{3} \langle\mathbf{1}\mathbf{2}\rangle \langle\mathbf{1}\mathbf{3}\rangle \lbrack\mathbf{2}\mathbf{4}\rbrack \lbrack\mathbf{3}\mathbf{4}\rbrack 
    \Big)
    \nonumber\\
    &-c_W^3\Big(
    3c_W \langle\mathbf{1}\mathbf{3}\rangle \langle\mathbf{2}\mathbf{4}\rangle \lbrack\mathbf{2}\mathbf{3}\rbrack 
    +3c_W \langle\mathbf{1}\mathbf{3}\rangle \langle\mathbf{2}\mathbf{3}\rangle \lbrack\mathbf{2}\mathbf{4}\rbrack 
    + \langle\mathbf{2}\mathbf{3}\rangle \lbrack\mathbf{1}\mathbf{3}\rbrack \lbrack\mathbf{2}\mathbf{4}\rbrack 
    + \langle\mathbf{1}\mathbf{3}\rangle \lbrack\mathbf{2}\mathbf{3}\rbrack \lbrack\mathbf{2}\mathbf{4}\rbrack 
    \Big)\lbrack\mathbf{1}\lvert p_{3} \rvert \mathbf{4}\rangle 
    \nonumber\\
    &-c_W^3\Big(
    3 \langle\mathbf{2}\mathbf{3}\rangle \langle\mathbf{2}\mathbf{4}\rangle \lbrack\mathbf{1}\mathbf{3}\rbrack 
    +3 \langle\mathbf{1}\mathbf{3}\rangle \langle\mathbf{2}\mathbf{4}\rangle \lbrack\mathbf{2}\mathbf{3}\rbrack 
    +c_W \langle\mathbf{2}\mathbf{4}\rangle \lbrack\mathbf{1}\mathbf{3}\rbrack \lbrack\mathbf{2}\mathbf{3}\rbrack 
    +c_W \langle\mathbf{2}\mathbf{3}\rangle \lbrack\mathbf{1}\mathbf{3}\rbrack \lbrack\mathbf{2}\mathbf{4}\rbrack 
    \Big)\lbrack\mathbf{4}\lvert p_{3} \rvert \mathbf{1}\rangle
\Bigg).
\end{align}

The contribution from the U-channel $W$ boson is
\begin{align}
\mathcal{M}_{WU} &= -\frac{e^2}{2M_W^3s_W^2\left(u-M_W^2\right)}
\Bigg(
    \Big(
    -3c_W^{3} \langle\mathbf{1}\mathbf{4}\rangle \langle\mathbf{2}\mathbf{3}\rangle \langle\mathbf{2}\mathbf{4}\rangle \lbrack\mathbf{1}\mathbf{3}\rbrack 
    -4c_W^{3} \langle\mathbf{2}\mathbf{3}\rangle \langle\mathbf{3}\mathbf{4}\rangle \lbrack\mathbf{1}\mathbf{2}\rbrack \lbrack\mathbf{1}\mathbf{4}\rbrack 
    -4c_W^{2} \langle\mathbf{2}\mathbf{3}\rangle \langle\mathbf{2}\mathbf{4}\rangle \lbrack\mathbf{1}\mathbf{3}\rbrack \lbrack\mathbf{1}\mathbf{4}\rbrack 
    \nonumber\\
    &-3c_W^{3} \langle\mathbf{1}\mathbf{3}\rangle \langle\mathbf{1}\mathbf{4}\rangle \langle\mathbf{2}\mathbf{4}\rangle \lbrack\mathbf{2}\mathbf{3}\rbrack 
    -c_W^{4} \langle\mathbf{1}\mathbf{4}\rangle \langle\mathbf{2}\mathbf{4}\rangle \lbrack\mathbf{1}\mathbf{3}\rbrack \lbrack\mathbf{2}\mathbf{3}\rbrack 
    -2\langle\mathbf{1}\mathbf{4}\rangle \langle\mathbf{2}\mathbf{3}\rangle \lbrack\mathbf{1}\mathbf{4}\rbrack \lbrack\mathbf{2}\mathbf{3}\rbrack 
    +6c_W^{2} \langle\mathbf{1}\mathbf{4}\rangle \langle\mathbf{2}\mathbf{3}\rangle \lbrack\mathbf{1}\mathbf{4}\rbrack \lbrack\mathbf{2}\mathbf{3}\rbrack 
    \nonumber\\
    &-2c_W^{4} \langle\mathbf{1}\mathbf{4}\rangle \langle\mathbf{2}\mathbf{3}\rangle \lbrack\mathbf{1}\mathbf{4}\rbrack \lbrack\mathbf{2}\mathbf{3}\rbrack 
    -3c_W^{2} \langle\mathbf{1}\mathbf{3}\rangle \langle\mathbf{2}\mathbf{4}\rangle \lbrack\mathbf{1}\mathbf{4}\rbrack \lbrack\mathbf{2}\mathbf{3}\rbrack 
    -c_W^{4} \langle\mathbf{1}\mathbf{3}\rangle \langle\mathbf{2}\mathbf{4}\rangle \lbrack\mathbf{1}\mathbf{4}\rbrack \lbrack\mathbf{2}\mathbf{3}\rbrack 
    +2c_W^{2} \langle\mathbf{1}\mathbf{2}\rangle \langle\mathbf{3}\mathbf{4}\rangle \lbrack\mathbf{1}\mathbf{4}\rbrack \lbrack\mathbf{2}\mathbf{3}\rbrack 
    \nonumber\\
    &-2c_W^{4} \langle\mathbf{1}\mathbf{2}\rangle \langle\mathbf{3}\mathbf{4}\rangle \lbrack\mathbf{1}\mathbf{4}\rbrack \lbrack\mathbf{2}\mathbf{3}\rbrack 
    -3c_W^{2} \langle\mathbf{1}\mathbf{4}\rangle \langle\mathbf{2}\mathbf{3}\rangle \lbrack\mathbf{1}\mathbf{3}\rbrack \lbrack\mathbf{2}\mathbf{4}\rbrack 
    -c_W^{4} \langle\mathbf{1}\mathbf{4}\rangle \langle\mathbf{2}\mathbf{3}\rangle \lbrack\mathbf{1}\mathbf{3}\rbrack \lbrack\mathbf{2}\mathbf{4}\rbrack 
    -3c_W^{4} \langle\mathbf{1}\mathbf{3}\rangle \langle\mathbf{2}\mathbf{3}\rangle \lbrack\mathbf{1}\mathbf{4}\rbrack \lbrack\mathbf{2}\mathbf{4}\rbrack 
    \nonumber\\
    &-c_W^{3} \langle\mathbf{2}\mathbf{3}\rangle \lbrack\mathbf{1}\mathbf{3}\rbrack \lbrack\mathbf{1}\mathbf{4}\rbrack \lbrack\mathbf{2}\mathbf{4}\rbrack 
    -4c_W^{2} \langle\mathbf{1}\mathbf{3}\rangle \langle\mathbf{1}\mathbf{4}\rangle \lbrack\mathbf{2}\mathbf{3}\rbrack \lbrack\mathbf{2}\mathbf{4}\rbrack 
    -c_W^{3} \langle\mathbf{1}\mathbf{3}\rangle \lbrack\mathbf{1}\mathbf{4}\rbrack \lbrack\mathbf{2}\mathbf{3}\rbrack \lbrack\mathbf{2}\mathbf{4}\rbrack 
    -4c_W^{3} \langle\mathbf{1}\mathbf{2}\rangle \langle\mathbf{1}\mathbf{4}\rangle \lbrack\mathbf{2}\mathbf{3}\rbrack \lbrack\mathbf{3}\mathbf{4}\rbrack 
    \Big)M_W
    \nonumber\\
    &+\Big(
    -3c_W^{4} \langle\mathbf{1}\mathbf{4}\rangle \langle\mathbf{2}\mathbf{4}\rangle \lbrack\mathbf{2}\mathbf{3}\rbrack 
    -c_W^{3} \langle\mathbf{2}\mathbf{4}\rangle \lbrack\mathbf{1}\mathbf{4}\rbrack \lbrack\mathbf{2}\mathbf{3}\rbrack 
    -3c_W^{4} \langle\mathbf{1}\mathbf{4}\rangle \langle\mathbf{2}\mathbf{3}\rangle \lbrack\mathbf{2}\mathbf{4}\rbrack 
    -c_W^{3} \langle\mathbf{1}\mathbf{4}\rangle \lbrack\mathbf{2}\mathbf{3}\rbrack \lbrack\mathbf{2}\mathbf{4}\rbrack 
    \Big)\lbrack\mathbf{1}\lvert p_{4} \rvert \mathbf{3}\rangle 
    \nonumber\\
    &+\Big(
    -3c_W^{3} \langle\mathbf{2}\mathbf{3}\rangle \langle\mathbf{2}\mathbf{4}\rangle \lbrack\mathbf{1}\mathbf{4}\rbrack 
    -c_W^{4} \langle\mathbf{2}\mathbf{4}\rangle \lbrack\mathbf{1}\mathbf{4}\rbrack \lbrack\mathbf{2}\mathbf{3}\rbrack 
    -3c_W^{3} \langle\mathbf{1}\mathbf{4}\rangle \langle\mathbf{2}\mathbf{3}\rangle \lbrack\mathbf{2}\mathbf{4}\rbrack 
    -c_W^{4} \langle\mathbf{2}\mathbf{3}\rangle \lbrack\mathbf{1}\mathbf{4}\rbrack \lbrack\mathbf{2}\mathbf{4}\rbrack 
    \Big)\lbrack\mathbf{3}\lvert p_{4} \rvert \mathbf{1}\rangle
\Bigg).
\end{align}
The combined amplitude is
\begin{align}
    \mathcal{M}_{ZZ\bar{W}W} &= 
    \mathcal{M}_{hS} + \mathcal{M}_{WT} + \mathcal{M}_{WU}.
\end{align}
We have validated this amplitude against Feynman diagrams in the processes $Z Z \to W \bar{W}$ and $Z W \to Z W$ in SPINAS
for a variety of masses and collider energies.

\subsection{$\mathbf{W W \bar{W} \bar{W}}$}
The amplitude in this subsection is also found in \cite{Christensen:2024B}, but with particles 3 and 4 outgoing, where we show that it satisfies perturbative unitarity.

The Higgs boson contributes in both the T and the U channels.  The amplitudes with all particles incoming is
\begin{align}
\mathcal{M}_{Th} &= 
    -\frac{e^2}{M_W^2s_W^2}
    \frac{\langle\mathbf{1}\mathbf{3}\rangle \langle\mathbf{2}\mathbf{4}\rangle \lbrack\mathbf{1}\mathbf{3}\rbrack \lbrack\mathbf{2}\mathbf{4}\rbrack }{ \left(t-m_h^{2}\right)} 
    \\
\mathcal{M}_{Uh} &= 
    -\frac{e^2}{M_W^2s_W^2}
    \frac{\langle\mathbf{1}\mathbf{4}\rangle \langle\mathbf{2}\mathbf{3}\rangle \lbrack\mathbf{1}\mathbf{4}\rbrack \lbrack\mathbf{2}\mathbf{3}\rbrack }{ \left(u-m_h^{2}\right)} .
\end{align}

The contribution coming from the photon in the T and U channels are
\begin{align}
    \mathcal{M}_{\gamma T} &= 
    \frac{e^2}{M_W^3 t}
    \Bigg(
        \Big(
        -2\langle\mathbf{2}\mathbf{4}\rangle \langle\mathbf{3}\mathbf{4}\rangle \lbrack\mathbf{1}\mathbf{2}\rbrack \lbrack\mathbf{1}\mathbf{3}\rbrack 
        +\langle\mathbf{1}\mathbf{3}\rangle \langle\mathbf{2}\mathbf{3}\rangle \langle\mathbf{2}\mathbf{4}\rangle \lbrack\mathbf{1}\mathbf{4}\rbrack 
        +2\langle\mathbf{2}\mathbf{3}\rangle \langle\mathbf{2}\mathbf{4}\rangle \lbrack\mathbf{1}\mathbf{3}\rbrack \lbrack\mathbf{1}\mathbf{4}\rbrack 
        +\langle\mathbf{1}\mathbf{4}\rangle \langle\mathbf{2}\mathbf{4}\rangle \lbrack\mathbf{1}\mathbf{3}\rbrack \lbrack\mathbf{2}\mathbf{3}\rbrack 
        +2\langle\mathbf{1}\mathbf{3}\rangle \langle\mathbf{2}\mathbf{4}\rangle \lbrack\mathbf{1}\mathbf{4}\rbrack \lbrack\mathbf{2}\mathbf{3}\rbrack 
        \nonumber\\
        &+\langle\mathbf{2}\mathbf{4}\rangle \lbrack\mathbf{1}\mathbf{3}\rbrack \lbrack\mathbf{1}\mathbf{4}\rbrack \lbrack\mathbf{2}\mathbf{3}\rbrack 
        +\langle\mathbf{1}\mathbf{3}\rangle \langle\mathbf{1}\mathbf{4}\rangle \langle\mathbf{2}\mathbf{3}\rangle \lbrack\mathbf{2}\mathbf{4}\rbrack 
        +2\langle\mathbf{1}\mathbf{4}\rangle \langle\mathbf{2}\mathbf{3}\rangle \lbrack\mathbf{1}\mathbf{3}\rbrack \lbrack\mathbf{2}\mathbf{4}\rbrack 
        -2\langle\mathbf{1}\mathbf{3}\rangle \langle\mathbf{2}\mathbf{4}\rangle \lbrack\mathbf{1}\mathbf{3}\rbrack \lbrack\mathbf{2}\mathbf{4}\rbrack 
        +\langle\mathbf{1}\mathbf{2}\rangle \langle\mathbf{3}\mathbf{4}\rangle \lbrack\mathbf{1}\mathbf{3}\rbrack \lbrack\mathbf{2}\mathbf{4}\rbrack 
        \nonumber\\
        &+\langle\mathbf{1}\mathbf{3}\rangle \langle\mathbf{2}\mathbf{3}\rangle \lbrack\mathbf{1}\mathbf{4}\rbrack \lbrack\mathbf{2}\mathbf{4}\rbrack 
        +2\langle\mathbf{1}\mathbf{3}\rangle \langle\mathbf{1}\mathbf{4}\rangle \lbrack\mathbf{2}\mathbf{3}\rbrack \lbrack\mathbf{2}\mathbf{4}\rbrack 
        +\langle\mathbf{1}\mathbf{4}\rangle \lbrack\mathbf{1}\mathbf{3}\rbrack \lbrack\mathbf{2}\mathbf{3}\rbrack \lbrack\mathbf{2}\mathbf{4}\rbrack 
        +\langle\mathbf{1}\mathbf{3}\rangle \langle\mathbf{2}\mathbf{4}\rangle \lbrack\mathbf{1}\mathbf{2}\rbrack \lbrack\mathbf{3}\mathbf{4}\rbrack -2\langle\mathbf{1}\mathbf{2}\rangle \langle\mathbf{1}\mathbf{3}\rangle \lbrack\mathbf{2}\mathbf{4}\rbrack \lbrack\mathbf{3}\mathbf{4}\rbrack 
    \Big)M_W
    \nonumber\\
    &
    +\Big(
        \langle\mathbf{1}\mathbf{3}\rangle \langle\mathbf{2}\mathbf{4}\rangle \lbrack\mathbf{2}\mathbf{3}\rbrack 
        +\langle\mathbf{1}\mathbf{3}\rangle \langle\mathbf{2}\mathbf{3}\rangle \lbrack\mathbf{2}\mathbf{4}\rbrack 
        +\langle\mathbf{2}\mathbf{3}\rangle \lbrack\mathbf{1}\mathbf{3}\rbrack \lbrack\mathbf{2}\mathbf{4}\rbrack 
        +\langle\mathbf{1}\mathbf{3}\rangle \lbrack\mathbf{2}\mathbf{3}\rbrack \lbrack\mathbf{2}\mathbf{4}\rbrack 
    \Big)\lbrack\mathbf{1}\lvert p_{3} \rvert \mathbf{4}\rangle 
    \nonumber\\
    &
    +\Big(
        \langle\mathbf{2}\mathbf{3}\rangle \langle\mathbf{2}\mathbf{4}\rangle \lbrack\mathbf{1}\mathbf{3}\rbrack 
        +\langle\mathbf{1}\mathbf{3}\rangle \langle\mathbf{2}\mathbf{4}\rangle \lbrack\mathbf{2}\mathbf{3}\rbrack 
        +\langle\mathbf{2}\mathbf{4}\rangle \lbrack\mathbf{1}\mathbf{3}\rbrack \lbrack\mathbf{2}\mathbf{3}\rbrack 
        +\langle\mathbf{2}\mathbf{3}\rangle \lbrack\mathbf{1}\mathbf{3}\rbrack \lbrack\mathbf{2}\mathbf{4}\rbrack 
    \Big)\lbrack\mathbf{4}\lvert p_{3} \rvert \mathbf{1}\rangle
    \Bigg)
    \label{eq:WWWW:TA full}
\end{align}
and
\begin{align}
    \mathcal{M}_{\gamma U} &=
    \frac{e^2}{M_W^3 u}\Bigg(
        \Big(
            \langle\mathbf{1}\mathbf{4}\rangle \langle\mathbf{2}\mathbf{3}\rangle \langle\mathbf{2}\mathbf{4}\rangle \lbrack\mathbf{1}\mathbf{3}\rbrack 
            +2\langle\mathbf{2}\mathbf{3}\rangle \langle\mathbf{3}\mathbf{4}\rangle \lbrack\mathbf{1}\mathbf{2}\rbrack \lbrack\mathbf{1}\mathbf{4}\rbrack 
            +2\langle\mathbf{2}\mathbf{3}\rangle \langle\mathbf{2}\mathbf{4}\rangle \lbrack\mathbf{1}\mathbf{3}\rbrack \lbrack\mathbf{1}\mathbf{4}\rbrack 
            +\langle\mathbf{1}\mathbf{3}\rangle \langle\mathbf{1}\mathbf{4}\rangle \langle\mathbf{2}\mathbf{4}\rangle \lbrack\mathbf{2}\mathbf{3}\rbrack 
            +\langle\mathbf{1}\mathbf{4}\rangle \langle\mathbf{2}\mathbf{4}\rangle \lbrack\mathbf{1}\mathbf{3}\rbrack \lbrack\mathbf{2}\mathbf{3}\rbrack 
            \nonumber\\
            &-2\langle\mathbf{1}\mathbf{4}\rangle \langle\mathbf{2}\mathbf{3}\rangle \lbrack\mathbf{1}\mathbf{4}\rbrack \lbrack\mathbf{2}\mathbf{3}\rbrack 
            +2\langle\mathbf{1}\mathbf{3}\rangle \langle\mathbf{2}\mathbf{4}\rangle \lbrack\mathbf{1}\mathbf{4}\rbrack \lbrack\mathbf{2}\mathbf{3}\rbrack 
            -\langle\mathbf{1}\mathbf{2}\rangle \langle\mathbf{3}\mathbf{4}\rangle \lbrack\mathbf{1}\mathbf{4}\rbrack \lbrack\mathbf{2}\mathbf{3}\rbrack +2\langle\mathbf{1}\mathbf{4}\rangle \langle\mathbf{2}\mathbf{3}\rangle \lbrack\mathbf{1}\mathbf{3}\rbrack \lbrack\mathbf{2}\mathbf{4}\rbrack 
            +\langle\mathbf{1}\mathbf{3}\rangle \langle\mathbf{2}\mathbf{3}\rangle \lbrack\mathbf{1}\mathbf{4}\rbrack \lbrack\mathbf{2}\mathbf{4}\rbrack 
            \nonumber\\
            &+\langle\mathbf{2}\mathbf{3}\rangle \lbrack\mathbf{1}\mathbf{3}\rbrack \lbrack\mathbf{1}\mathbf{4}\rbrack \lbrack\mathbf{2}\mathbf{4}\rbrack 
            +2\langle\mathbf{1}\mathbf{3}\rangle \langle\mathbf{1}\mathbf{4}\rangle \lbrack\mathbf{2}\mathbf{3}\rbrack \lbrack\mathbf{2}\mathbf{4}\rbrack 
            +\langle\mathbf{1}\mathbf{3}\rangle \lbrack\mathbf{1}\mathbf{4}\rbrack \lbrack\mathbf{2}\mathbf{3}\rbrack \lbrack\mathbf{2}\mathbf{4}\rbrack 
            -\langle\mathbf{1}\mathbf{4}\rangle \langle\mathbf{2}\mathbf{3}\rangle \lbrack\mathbf{1}\mathbf{2}\rbrack \lbrack\mathbf{3}\mathbf{4}\rbrack 
            +2\langle\mathbf{1}\mathbf{2}\rangle \langle\mathbf{1}\mathbf{4}\rangle \lbrack\mathbf{2}\mathbf{3}\rbrack \lbrack\mathbf{3}\mathbf{4}\rbrack 
        \Big)M_W
        \nonumber\\
        &+\Big(\langle\mathbf{1}\mathbf{4}\rangle \langle\mathbf{2}\mathbf{4}\rangle \lbrack\mathbf{2}\mathbf{3}\rbrack 
        +\langle\mathbf{2}\mathbf{4}\rangle \lbrack\mathbf{1}\mathbf{4}\rbrack \lbrack\mathbf{2}\mathbf{3}\rbrack 
        +\langle\mathbf{1}\mathbf{4}\rangle \langle\mathbf{2}\mathbf{3}\rangle \lbrack\mathbf{2}\mathbf{4}\rbrack 
        +\langle\mathbf{1}\mathbf{4}\rangle \lbrack\mathbf{2}\mathbf{3}\rbrack \lbrack\mathbf{2}\mathbf{4}\rbrack 
        \Big)\lbrack\mathbf{1}\lvert p_{4} \rvert \mathbf{3}\rangle 
        \nonumber\\
        &+\Big(
            \langle\mathbf{2}\mathbf{3}\rangle \langle\mathbf{2}\mathbf{4}\rangle \lbrack\mathbf{1}\mathbf{4}\rbrack 
            +\langle\mathbf{2}\mathbf{4}\rangle \lbrack\mathbf{1}\mathbf{4}\rbrack \lbrack\mathbf{2}\mathbf{3}\rbrack 
            +\langle\mathbf{1}\mathbf{4}\rangle \langle\mathbf{2}\mathbf{3}\rangle \lbrack\mathbf{2}\mathbf{4}\rbrack 
            +\langle\mathbf{2}\mathbf{3}\rangle \lbrack\mathbf{1}\mathbf{4}\rbrack \lbrack\mathbf{2}\mathbf{4}\rbrack 
        \Big)\lbrack\mathbf{3}\lvert p_{4} \rvert \mathbf{1}\rangle
    \bigg).
    \label{eq:WWWW:UA full}
\end{align}
Both of these diagrams were originally calculated using a massive photon and taking the massless limit at the end.  However, we have also calculated it using the $x$ factor (and a massless photon) and outline this derivation in App.~\ref{app:internal photons}.

The contribution coming from the Z boson in the T and U channels are
\begin{align}
\mathcal{M}_{ZT} &= -\frac{e^2}{2M_W^2M_Zs_W^2\left(t-M_Z^2\right)}
    \Bigg(
        \Big(
            -4c_W^{2} \langle\mathbf{2}\mathbf{4}\rangle \langle\mathbf{3}\mathbf{4}\rangle \lbrack\mathbf{1}\mathbf{2}\rbrack \lbrack\mathbf{1}\mathbf{3}\rbrack 
            +3c_W^{2} \langle\mathbf{1}\mathbf{3}\rangle \langle\mathbf{2}\mathbf{3}\rangle \langle\mathbf{2}\mathbf{4}\rangle \lbrack\mathbf{1}\mathbf{4}\rbrack 
            +4c_W^{2} \langle\mathbf{2}\mathbf{3}\rangle \langle\mathbf{2}\mathbf{4}\rangle \lbrack\mathbf{1}\mathbf{3}\rbrack \lbrack\mathbf{1}\mathbf{4}\rbrack 
            \nonumber\\
            &+3c_W^{2} \langle\mathbf{1}\mathbf{4}\rangle \langle\mathbf{2}\mathbf{4}\rangle \lbrack\mathbf{1}\mathbf{3}\rbrack \lbrack\mathbf{2}\mathbf{3}\rbrack 
            +\langle\mathbf{1}\mathbf{3}\rangle \langle\mathbf{2}\mathbf{4}\rangle \lbrack\mathbf{1}\mathbf{4}\rbrack \lbrack\mathbf{2}\mathbf{3}\rbrack 
            +3c_W^{2} \langle\mathbf{1}\mathbf{3}\rangle \langle\mathbf{2}\mathbf{4}\rangle \lbrack\mathbf{1}\mathbf{4}\rbrack \lbrack\mathbf{2}\mathbf{3}\rbrack 
            +c_W^{2} \langle\mathbf{2}\mathbf{4}\rangle \lbrack\mathbf{1}\mathbf{3}\rbrack \lbrack\mathbf{1}\mathbf{4}\rbrack \lbrack\mathbf{2}\mathbf{3}\rbrack 
            \nonumber\\
            &+3c_W^{2} \langle\mathbf{1}\mathbf{3}\rangle \langle\mathbf{1}\mathbf{4}\rangle \langle\mathbf{2}\mathbf{3}\rangle \lbrack\mathbf{2}\mathbf{4}\rbrack 
            +\langle\mathbf{1}\mathbf{4}\rangle \langle\mathbf{2}\mathbf{3}\rangle \lbrack\mathbf{1}\mathbf{3}\rbrack \lbrack\mathbf{2}\mathbf{4}\rbrack 
            +3c_W^{2} \langle\mathbf{1}\mathbf{4}\rangle \langle\mathbf{2}\mathbf{3}\rangle \lbrack\mathbf{1}\mathbf{3}\rbrack \lbrack\mathbf{2}\mathbf{4}\rbrack 
            -2c_W^{2} \langle\mathbf{1}\mathbf{3}\rangle \langle\mathbf{2}\mathbf{4}\rangle \lbrack\mathbf{1}\mathbf{3}\rbrack \lbrack\mathbf{2}\mathbf{4}\rbrack \nonumber\\
            &-2\langle\mathbf{1}\mathbf{2}\rangle \langle\mathbf{3}\mathbf{4}\rangle \lbrack\mathbf{1}\mathbf{3}\rbrack \lbrack\mathbf{2}\mathbf{4}\rbrack 
            +2c_W^{2} \langle\mathbf{1}\mathbf{2}\rangle \langle\mathbf{3}\mathbf{4}\rangle \lbrack\mathbf{1}\mathbf{3}\rbrack \lbrack\mathbf{2}\mathbf{4}\rbrack 
            +c_W^{2} \langle\mathbf{1}\mathbf{3}\rangle \langle\mathbf{2}\mathbf{3}\rangle \lbrack\mathbf{1}\mathbf{4}\rbrack \lbrack\mathbf{2}\mathbf{4}\rbrack 
            +4c_W^{2} \langle\mathbf{1}\mathbf{3}\rangle \langle\mathbf{1}\mathbf{4}\rangle \lbrack\mathbf{2}\mathbf{3}\rbrack \lbrack\mathbf{2}\mathbf{4}\rbrack 
            \nonumber\\
            &+c_W^{2} \langle\mathbf{1}\mathbf{4}\rangle \lbrack\mathbf{1}\mathbf{3}\rbrack \lbrack\mathbf{2}\mathbf{3}\rbrack \lbrack\mathbf{2}\mathbf{4}\rbrack 
            -4c_W^{2} \langle\mathbf{1}\mathbf{2}\rangle \langle\mathbf{1}\mathbf{3}\rangle \lbrack\mathbf{2}\mathbf{4}\rbrack \lbrack\mathbf{3}\mathbf{4}\rbrack 
        \Big)MZ
        \nonumber\\
        &+\Big(
            3c_W\langle\mathbf{1}\mathbf{3}\rangle \langle\mathbf{2}\mathbf{4}\rangle \lbrack\mathbf{2}\mathbf{3}\rbrack 
            +3c_W\langle\mathbf{1}\mathbf{3}\rangle \langle\mathbf{2}\mathbf{3}\rangle \lbrack\mathbf{2}\mathbf{4}\rbrack 
            +c_W\langle\mathbf{2}\mathbf{3}\rangle \lbrack\mathbf{1}\mathbf{3}\rbrack \lbrack\mathbf{2}\mathbf{4}\rbrack 
            +c_W\langle\mathbf{1}\mathbf{3}\rangle \lbrack\mathbf{2}\mathbf{3}\rbrack \lbrack\mathbf{2}\mathbf{4}\rbrack 
        \Big)\lbrack\mathbf{1}\lvert p_{3} \rvert \mathbf{4}\rangle 
        \nonumber\\
        &+\Big(
            3c_W\langle\mathbf{2}\mathbf{3}\rangle \langle\mathbf{2}\mathbf{4}\rangle \lbrack\mathbf{1}\mathbf{3}\rbrack 
            +3c_W\langle\mathbf{1}\mathbf{3}\rangle \langle\mathbf{2}\mathbf{4}\rangle \lbrack\mathbf{2}\mathbf{3}\rbrack 
            +c_W\langle\mathbf{2}\mathbf{4}\rangle \lbrack\mathbf{1}\mathbf{3}\rbrack \lbrack\mathbf{2}\mathbf{3}\rbrack 
            +c_W\langle\mathbf{2}\mathbf{3}\rangle \lbrack\mathbf{1}\mathbf{3}\rbrack \lbrack\mathbf{2}\mathbf{4}\rbrack 
        \Big)\lbrack\mathbf{4}\lvert p_{3} \rvert \mathbf{1}\rangle
    \Bigg)
    \label{eq:WWWW:TZ full}
\end{align}

\begin{align}
\mathcal{M}_{ZU} &= -\frac{e^2}{2M_W^2M_Zs_W^2\left(u-M_Z^2\right)}
    \Bigg(
        \Big(
            3c_W^{2} \langle\mathbf{1}\mathbf{4}\rangle \langle\mathbf{2}\mathbf{3}\rangle \langle\mathbf{2}\mathbf{4}\rangle \lbrack\mathbf{1}\mathbf{3}\rbrack 
            +4c_W^{2} \langle\mathbf{2}\mathbf{3}\rangle \langle\mathbf{3}\mathbf{4}\rangle \lbrack\mathbf{1}\mathbf{2}\rbrack \lbrack\mathbf{1}\mathbf{4}\rbrack 
            +4c_W^{2} \langle\mathbf{2}\mathbf{3}\rangle \langle\mathbf{2}\mathbf{4}\rangle \lbrack\mathbf{1}\mathbf{3}\rbrack \lbrack\mathbf{1}\mathbf{4}\rbrack 
            \nonumber\\
            &+3c_W^{2} \langle\mathbf{1}\mathbf{3}\rangle \langle\mathbf{1}\mathbf{4}\rangle \langle\mathbf{2}\mathbf{4}\rangle \lbrack\mathbf{2}\mathbf{3}\rbrack 
            +c_W^{2} \langle\mathbf{1}\mathbf{4}\rangle \langle\mathbf{2}\mathbf{4}\rangle \lbrack\mathbf{1}\mathbf{3}\rbrack \lbrack\mathbf{2}\mathbf{3}\rbrack 
            -2c_W^{2} \langle\mathbf{1}\mathbf{4}\rangle \langle\mathbf{2}\mathbf{3}\rangle \lbrack\mathbf{1}\mathbf{4}\rbrack \lbrack\mathbf{2}\mathbf{3}\rbrack 
            +\langle\mathbf{1}\mathbf{3}\rangle \langle\mathbf{2}\mathbf{4}\rangle \lbrack\mathbf{1}\mathbf{4}\rbrack \lbrack\mathbf{2}\mathbf{3}\rbrack 
            \nonumber\\
            &+3c_W^{2} \langle\mathbf{1}\mathbf{3}\rangle \langle\mathbf{2}\mathbf{4}\rangle \lbrack\mathbf{1}\mathbf{4}\rbrack \lbrack\mathbf{2}\mathbf{3}\rbrack 
            +2\langle\mathbf{1}\mathbf{2}\rangle \langle\mathbf{3}\mathbf{4}\rangle \lbrack\mathbf{1}\mathbf{4}\rbrack \lbrack\mathbf{2}\mathbf{3}\rbrack 
            -2c_W^{2} \langle\mathbf{1}\mathbf{2}\rangle \langle\mathbf{3}\mathbf{4}\rangle \lbrack\mathbf{1}\mathbf{4}\rbrack \lbrack\mathbf{2}\mathbf{3}\rbrack 
            +\langle\mathbf{1}\mathbf{4}\rangle \langle\mathbf{2}\mathbf{3}\rangle \lbrack\mathbf{1}\mathbf{3}\rbrack \lbrack\mathbf{2}\mathbf{4}\rbrack 
            \nonumber\\
            &+3c_W^{2} \langle\mathbf{1}\mathbf{4}\rangle \langle\mathbf{2}\mathbf{3}\rangle \lbrack\mathbf{1}\mathbf{3}\rbrack \lbrack\mathbf{2}\mathbf{4}\rbrack 
            +3c_W^{2} \langle\mathbf{1}\mathbf{3}\rangle \langle\mathbf{2}\mathbf{3}\rangle \lbrack\mathbf{1}\mathbf{4}\rbrack \lbrack\mathbf{2}\mathbf{4}\rbrack 
            +c_W^{2} \langle\mathbf{2}\mathbf{3}\rangle \lbrack\mathbf{1}\mathbf{3}\rbrack \lbrack\mathbf{1}\mathbf{4}\rbrack \lbrack\mathbf{2}\mathbf{4}\rbrack 
            +4c_W^{2} \langle\mathbf{1}\mathbf{3}\rangle \langle\mathbf{1}\mathbf{4}\rangle \lbrack\mathbf{2}\mathbf{3}\rbrack \lbrack\mathbf{2}\mathbf{4}\rbrack 
            \nonumber\\
            &+c_W^{2} \langle\mathbf{1}\mathbf{3}\rangle \lbrack\mathbf{1}\mathbf{4}\rbrack \lbrack\mathbf{2}\mathbf{3}\rbrack \lbrack\mathbf{2}\mathbf{4}\rbrack 
            +4c_W^{2} \langle\mathbf{1}\mathbf{2}\rangle \langle\mathbf{1}\mathbf{4}\rangle \lbrack\mathbf{2}\mathbf{3}\rbrack \lbrack\mathbf{3}\mathbf{4}\rbrack 
        \Big)M_Z
        \nonumber\\
        &+\Big(
            3c_W\langle\mathbf{1}\mathbf{4}\rangle \langle\mathbf{2}\mathbf{4}\rangle \lbrack\mathbf{2}\mathbf{3}\rbrack 
            +c_W\langle\mathbf{2}\mathbf{4}\rangle \lbrack\mathbf{1}\mathbf{4}\rbrack \lbrack\mathbf{2}\mathbf{3}\rbrack 
            +3c_W\langle\mathbf{1}\mathbf{4}\rangle \langle\mathbf{2}\mathbf{3}\rangle \lbrack\mathbf{2}\mathbf{4}\rbrack 
            +c_W\langle\mathbf{1}\mathbf{4}\rangle \lbrack\mathbf{2}\mathbf{3}\rbrack \lbrack\mathbf{2}\mathbf{4}\rbrack 
        \Big)\lbrack\mathbf{1}\lvert p_{4} \rvert \mathbf{3}\rangle 
        \nonumber\\
        &+\Big(
            3c_W\langle\mathbf{2}\mathbf{3}\rangle \langle\mathbf{2}\mathbf{4}\rangle \lbrack\mathbf{1}\mathbf{4}\rbrack 
            +c_W\langle\mathbf{2}\mathbf{4}\rangle \lbrack\mathbf{1}\mathbf{4}\rbrack \lbrack\mathbf{2}\mathbf{3}\rbrack 
            +3c_W\langle\mathbf{1}\mathbf{4}\rangle \langle\mathbf{2}\mathbf{3}\rangle \lbrack\mathbf{2}\mathbf{4}\rbrack 
            +c_W\langle\mathbf{2}\mathbf{3}\rangle \lbrack\mathbf{1}\mathbf{4}\rbrack \lbrack\mathbf{2}\mathbf{4}\rbrack 
        \Big)\lbrack\mathbf{3}\lvert p_{4} \rvert \mathbf{1}\rangle
    \Bigg).
    \label{eq:WWWW:UZ full}
\end{align}

The total amplitude is
\begin{align}
    \mathcal{M}_{WW\bar{W}\bar{W}} &=
    \mathcal{M}_{hT} + \mathcal{M}_{hU} + \mathcal{M}_{\gamma T} + \mathcal{M}_{\gamma U} + \mathcal{M}_{Z T} + \mathcal{M}_{Z U}.
\end{align}
We have validated this amplitude against Feynman diagrams in the processes $W \bar{W} \to W \bar{W}$ and $W W \to W W$ in SPINAS
for a variety of masses and collider energies.

\section{\label{sec:4-point vertex}Amplitudes with a 4-Point Vertex}
In this section, we consider amplitudes that require a 4-point vertex.  The Higgs 4-point vertex is the same as in Feynman diagrams.  We found the other 4-point vertices by considering perturbative unitarity \cite{Christensen:2024B}.  The amplitudes in this section, except for $\mathcal{M}_{hhhh}$, are also found in that paper, but with particles 3 and 4 outgoing.  We include them for completeness.

\subsection{$\mathbf{h h h h}$}
This amplitude is the same as with Feynman diagrams.  We include it for completeness.  It has contribution from a 4-point vertex and a Higgs in the S-, T- and U-channel diagrams.  All together, we have
\begin{align}
    \mathcal{M}_{hhhh} &= 
    -\frac{9e^2m_h^4}{4M_W^2s_W^2}
    \left(
    \frac{1}{3m_h^2}
    +\frac{1}{s-m_h^2}
    +\frac{1}{t-m_h^2}
    +\frac{1}{u-m_h^2}
    \right).
\end{align}
We have validated this amplitude against Feynman diagrams in the processes $h h \to h h$ in SPINAS
for a variety of masses and collider energies.

\subsection{$\mathbf{h h Z Z}$ and $\mathbf{h h \bar{W} W}$}
Both these processes have very similar details.  In fact, only the masses are changed between them.  We will describe them together.  

The contribution from the 4-point vertex, in both processes, is
\begin{equation}
    \mathcal{M}_4 = \frac{e^2}{2M_W^2s_W^2}\lbrack\mathbf{34}\rbrack\langle\mathbf{34}\rangle .
\end{equation}

The contribution to both amplitudes coming from an S-channel Higgs, with all momenta incoming, is
\begin{equation}
    \mathcal{M}_{hS} = -\frac{3e^2m_h^2}{2M_W^2s_W^2}\frac{\lbrack\mathbf{34}\rbrack\langle\mathbf{34}\rangle}{(s-m_h^2)} .
\end{equation}

The contribution coming from a $Z$ boson in the T and U channels are
\begin{align}
\mathcal{M}_{TZ} &= -\frac{e^{2}}{2M_W^{2} s_W^{2}} \frac{\left(2M_Z^{2} \langle\mathbf{3}\mathbf{4}\rangle \lbrack\mathbf{3}\mathbf{4}\rbrack +\lbrack\mathbf{3}\lvert p_{1} \rvert \mathbf{4}\rangle \lbrack\mathbf{4}\lvert p_{1} \rvert \mathbf{3}\rangle +M_Z\left(\lbrack\mathbf{3}\mathbf{4}\rbrack \lbrack\mathbf{3}\lvert p_{1} \rvert \mathbf{4}\rangle +\langle\mathbf{3}\mathbf{4}\rangle \lbrack\mathbf{4}\lvert p_{1} \rvert \mathbf{3}\rangle \right)\right)}{ \left(t-M_Z^{2}\right)} 
    \label{eq:hhZZ:TZ diagram}
\end{align}
and
\begin{align}
\mathcal{M}_{UZ} &= 
    -\frac{e^{2}}{2M_W^{2} s_W^{2}} 
    \frac{\left(2M_Z^{2} \langle\mathbf{3}\mathbf{4}\rangle \lbrack\mathbf{3}\mathbf{4}\rbrack +\lbrack\mathbf{3}\lvert p_{1} \rvert \mathbf{4}\rangle \lbrack\mathbf{4}\lvert p_{1} \rvert \mathbf{3}\rangle -M_Z\left(\langle\mathbf{3}\mathbf{4}\rangle \lbrack\mathbf{3}\lvert p_{1} \rvert \mathbf{4}\rangle +\lbrack\mathbf{3}\mathbf{4}\rbrack \lbrack\mathbf{4}\lvert p_{1} \rvert \mathbf{3}\rangle \right)\right)}{ \left(u-M_Z^{2}\right)} .
    \label{eq:hhZZ:UZ diagram}
\end{align}

The contribution coming from a $W$ boson in the T and U channels are
\begin{align}
\mathcal{M}_{TW} &= -\frac{e^{2}}{2M_W^{2} s_W^{2}} \frac{\left(2M_W^{2} \langle\mathbf{3}\mathbf{4}\rangle \lbrack\mathbf{3}\mathbf{4}\rbrack +\lbrack\mathbf{3}\lvert p_{1} \rvert \mathbf{4}\rangle \lbrack\mathbf{4}\lvert p_{1} \rvert \mathbf{3}\rangle +M_W\left(\lbrack\mathbf{3}\mathbf{4}\rbrack \lbrack\mathbf{3}\lvert p_{1} \rvert \mathbf{4}\rangle +\langle\mathbf{3}\mathbf{4}\rangle \lbrack\mathbf{4}\lvert p_{1} \rvert \mathbf{3}\rangle \right)\right)}{ \left(t-M_W^{2}\right)} 
    \label{eq:hhWW:TW diagram}
\end{align}
and
\begin{align}
\mathcal{M}_{UW} &= 
    -\frac{e^{2}}{2M_W^{2} s_W^{2}} 
    \frac{\left(2M_W^{2} \langle\mathbf{3}\mathbf{4}\rangle \lbrack\mathbf{3}\mathbf{4}\rbrack +\lbrack\mathbf{3}\lvert p_{1} \rvert \mathbf{4}\rangle \lbrack\mathbf{4}\lvert p_{1} \rvert \mathbf{3}\rangle -M_W\left(\langle\mathbf{3}\mathbf{4}\rangle \lbrack\mathbf{3}\lvert p_{1} \rvert \mathbf{4}\rangle +\lbrack\mathbf{3}\mathbf{4}\rbrack \lbrack\mathbf{4}\lvert p_{1} \rvert \mathbf{3}\rangle \right)\right)}{ \left(u-M_W^{2}\right)} .
    \label{eq:hhWW:UW diagram}
\end{align}

Putting these together, we have
\begin{align}
    \mathcal{M}_{hhZZ} &= 
    \mathcal{M}_4 + \mathcal{M}_{hS} + \mathcal{M}_{ZT} + \mathcal{M}_{ZU}
\end{align}
and
\begin{align}
    \mathcal{M}_{hh\bar{W}W} &= 
    \mathcal{M}_4 + \mathcal{M}_{hS} + \mathcal{M}_{WT} + \mathcal{M}_{WU}.
\end{align}
We have validated these amplitudes against Feynman diagrams in the processes $h h \to Z Z$, $h Z \to Z h$, $h h \to W \bar{W}$ and $h W \to W h$ in SPINAS
for a variety of masses and collider energies.

\section{Acknowledgements}
We would like to thank Ishmam Mahbub and Zhen Liu for helpful discussions during the completion of this work.

\section{\label{sec:conclusions}Conclusion}
In this paper, we give a comprehensive set of 4-point amplitudes in the constructive Standard Model (CSM).  Some of these amplitudes had previously been found but, here, we introduce the new results for the full amplitudes, including all the contributions, for $f_1\bar{f}_1f_2\bar{f}_2$ (only the photon contribution was previously known), $f \bar{f} (\gamma/g) Z$, $f_1 \bar{f}_2(\gamma/g)W$, $q\bar{q}gg$, $\gamma hW\bar{W}$, $\gamma Z\bar{W}W$, $\gamma \gamma \bar{W}W$, $f \bar{f} h h$, $f\bar{f}Zh$, $f_1\bar{f}_2\bar{W}h$, $f\bar{f}\bar{W}W$, $f\bar{f}ZZ$, $\bar{f}_1f_2Z(W/\bar{W})$, $Zh\bar{W}W$, $ZZZZ$, $hhZZ$, $hh\bar{W}W$, $ZZ\bar{W}W$, and $W\bar{W}\bar{W}W$.  The processes $\gamma Z\bar{W}W$, $\gamma\gamma\bar{W}W$, $hhZZ$, $hh\bar{W}W$, $ZZ\bar{W}W$ and $W\bar{W}\bar{W}W$ are presented simultaneously in a companion to this paper in \cite{Christensen:2024C}, where we use perturbative unitarity to determine the absence of a 4-point vertex in the first two and the last two and we also find the nature of the 4-point vertex for $hhZZ$ and $hh\bar{W}W$.  Any 4-point amplitude in the SM can be obtained from these by a suitable choice of masses and chiral couplings, a permutation of particles (crossing symmetry) and an inversion of the direction of the momentum of the outgoing particles (all the amplitudes presented in this paper are for all incoming particles).  

Furthermore, we validated all these amplitudes in a large number of $2\to 2$ processes by comparing them to Feynman rules at the squared amplitude level using our new computational package SPINAS, which numerically calculates phase-space points for constructive amplitudes, simultaneously published in a companion to this paper \cite{Christensen:2024B}.   Our validations include a comparison with Feynman diagrams over a variety of masses, scattering energies and angles.  In fact, downloading the SPINAS package includes a SM directory containing all the code for our validations.  Moreover, for every process, we compare with Feynman diagrams in at least two configurations related by crossing symmetry and for the separate helicities of the photons and gluons when they are in the initial state.

Most of our calculations followed the standard rules of constructive amplitude calculations \cite{Britto:2005fq,Arkani-Hamed:2017jhn,Christensen:2018zcq,Christensen:2019mch,Christensen:2022nja,Lai:2023upa,Christensen:2024C}, although the simplification procedure was complicated in some cases.  However, there were some new features that emerged when we calculated all the 4-point amplitudes with external photons and gluons, regarding the propagator structure, when more than 2 diagrams are possible.  In particular, in Sec.~\ref{sec:ffAW}, we showed that for the amplitude $f_1\bar{f}_2\gamma W$, although the three diagrams gave the same spinor product expressions in the numerators, the propagator denominators and charges were different and resulted in a slightly more complicated propagator structure that is not obtained by only considering one diagram.  Similarly, when considering the amplitude $f\bar{f}gg$ in Sec.~\ref{sec:ffAA}, we find that the spinor product expression is the same between the three diagrams, but the propagator structure is more complicated.  In fact, we don't have a simple way to write it for the amplitude that gives the correct squared amplitude, but we do have the correct squared propagator structure, including the contributions from the colors.  Although we don't have a simple form for the amplitude propagator structure (before squaring), we do have a systematic way of obtaining its square, summed over colors, as we describe in that section.  These statements also apply to $gggg$, although we assume the full structure was already well known in this case.  We hope that these findings will support the calculation of higher-multiplicity amplitudes and their validation with Feynman rules at higher multiplicity.

In the future, we plan to extend these calculations to 5-point and 6-point amplitudes and to complete a general algorithm for generating any N-point amplitude in the CSM, and encapsulate it in a computational package and validate it against Feynman rules.  Although most of the algorithm is expected to already be known, we think there are still some details to work out with regard to photons and gluons in the external states at higher multiplicity, especially when these photons and gluons are in an amplitude with more than one fermion line.  Moreover, as we have discussed, we do not understand the propagator and color structure of amplitudes with more than one gluon in the external state before squaring.  We hope that by analyzing further examples at 5-point, we will be able to elucidate this structure, and we will make this a matter of focus in upcoming projects.

Beyond this, we would like to explore complete sets of explicit CSM amplitudes at one loop.  In principle, it should be possible to connect two legs of a diagram and integrate over the resulting loop momentum.  For example, very briefly, if a massive fermion is present on two external legs of a 4-point amplitude, we could reverse the momentum of one and set it equal to the momentum of the other leg.  We would also lower the spin index on one and contract it with the other, summing over it.  This would result in either a momentum or a mass, depending on whether the spinors were different types (angle-square or square-angle) or the same type (angle-angle or square-square), respectively.  With this, we will have a trace over momenta and mass in the loop, analogous to Feynman diagrams.  Finally, the integration can be done, with the usual regularization for infinite integrals.  We conjecture that the number of such regularized integrals requiring renormalization will be finite, and the CSM will be renormalizable.  However, this must be proved.  As always, we will report on our methodology and any subtelties and we will compare our results with Feynman diagrams to ensure our calculations are correct.

With regard to both higher-multiplicity and higher-loop amplitudes, we note in the companion to this paper \cite{Christensen:2024B} and in this paper that we did not need any contact terms at 4-points in the CSM, beyond the 4-point vertices we already had in Feynman rules and, indeed, we need fewer 4-point vertices in the CSM than in Feynman rules.  In particular, unlike in Feynman rules, it was already well known that we do not need a 4-point gluon vertex, but we also showed that we do not need a $\gamma Z\bar{W}W$ vertex, a $\gamma\gamma\bar{W}W$ vertex, a $ZZ\bar{W}W$ vertex or a $WW\bar{W}\bar{W}$ vertex.  We expect this property to hold at all multiplicities and all orders in perturbation theory in a renormalizable theory such as the SM.  We do not think this is any different in the CSM and, in fact, we think the CSM is better behaved, with the smaller set of 4-point vertices.  Nevertheless, this must be proved.  

As we build higher-multiplicity amplitudes in the future, we would also like to analyze how the computational efficiency of the phase-space calculations scales and compare this with Feynman diagrams.  It appears pretty clear that they will be much more efficient when external photons or gluons are part of the amplitude and we are hopeful that this will remain true for amplitudes without external photons or gluons, although less profound.  We did not analyze this at 4-point because we think it is more significant to compare the scaling behavior more than the efficiency at 4 points, since 4-point amplitudes are trivial for computers to calculate with either method.  The real test will be whether their computational load will grow less quickly as the number of particles increases.  

Finally, as we complete the algorithm for the CSM and the computational package to generate any amplitude, we would like to open it up to any constructive theory, whether renormalizable or effective.  As our skills and tools improve, we would also like to incorporate gravity in future calculations.

\appendix

\section{\label{app:internal photons}Internal Photons}
In \cite{Lai:2023upa}, the authors showed that the amplitude for $e \bar{e} \bar{\mu} \mu$ could be obtained from the $x$ factor and that there was a factor of $s$ that had not been taken into account in earlier work.  In this appendix, we will reproduce their result and then attempt to apply it to $f \bar{f} \bar{W} W$.  We then will find an alternate way of calculating $f \bar{f} \bar{W} W$ by factoring it in a way that we can reuse the $f_1 \bar{f}_1 \bar{f}_2 f_2$ result.  This second (factoring) method will be more straight forward for the amplitude of $W W \bar{W} \bar{W}$ and we have achieved success for that amplitude as well.  We initially obtained all the internal photon results in this article by use of a massive photon and taking the massless limit at the end, as described in \cite{Christensen:2022nja}.  We have found that this method works in all cases and we have validated our results with SPINAS.  This is important because it gives us a validated form to compare with results using the $x$ factor.

We begin with a representation of the $x$ factor \cite{Arkani-Hamed:2017jhn,Christensen:2018zcq}, given by
\begin{align}
    x_{i,j} &= -\frac{\langle q\lvert p_j\rvert k\rbrack}{m\langle q k\rangle},
    \\
    \tilde{x}_{i,j} &= -\frac{\lbrack q\lvert p_j\rvert k\rangle}{m\lbrack q k\rbrack},
\end{align}
where $q=\xi$ is a massless reference momenta and $k=(p_i+p_j)$ is minus the massless momentum coming into the vertex.  We have also used the convention $\lvert -k\rangle=-\lvert k\rangle$ and $\lvert -k\rbrack = \lvert k\rbrack$ and momentum conservation to simplify the form on the right.  

Multiplying by the angle or square bracket, and applying the Schouten and mass identities, we have
\begin{align}
    x_{i,j}\langle\mathbf{ij}\rangle &= 
    \frac{-m\lbrack\mathbf{j}k\rbrack\langle\mathbf{i}q\rangle
    +\langle\mathbf{j}q\rangle\langle\mathbf{i}\lvert p_j\rvert k\rbrack}
    {m\langle qk\rangle},
    \\
    \tilde{x}_{i,j}\lbrack\mathbf{ij}\rbrack &= 
    \frac{-m\langle\mathbf{j} k\rangle\lbrack\mathbf{i}q\rbrack
    +\lbrack\mathbf{j}q\rbrack\lbrack\mathbf{i}\lvert p_j\rvert k\rangle}
    {m\lbrack qk\rbrack}.
\end{align}
In each of these, we can add and subtract $p_i$ to the 2nd term and use the on-shell condition $k\lvert k\rangle = 0$ and $k\lvert k\rbrack = 0$, where $k=(p_i+p_j)$.  We also use the mass identity, giving us
\begin{align}
    x_{i,j}\langle\mathbf{ij}\rangle &= -
    \frac{\lbrack\mathbf{j}k\rbrack\langle\mathbf{i}q\rangle
    +\langle\mathbf{j}q\rangle\lbrack\mathbf{i} k\rbrack}
    {\langle qk\rangle},
    \label{eq:app:x<ij>}
    \\
    \tilde{x}_{i,j}\lbrack\mathbf{ij}\rbrack &= 
    -\frac{\langle\mathbf{j} k\rangle\lbrack\mathbf{i}q\rbrack
    +\lbrack\mathbf{j}q\rbrack\langle\mathbf{i} k\rangle}
    {\lbrack qk\rbrack}.
    \label{eq:app:xt[ij]}
\end{align}
We note that this expression is symmetric in $i$ and $j$.
By starting at this point, we can simplify some of the algebra in the amplitudes.

\subsection{$f_1 \bar{f}_1 \bar{f}_2 f_2$}
We begin by reproducing the result of \cite{Lai:2023upa} for $f_1 \bar{f}_1 \bar{f}_2 f_2$.
We do this by plugging in the expressions from Eqs.~(\ref{eq:app:x<ij>}) and (\ref{eq:app:xt[ij]}) (and remember that $k=(p_1+p_2)=-(p_3+p_4)$), giving
\begin{align}
    \mathcal{M}_{f_1\bar{f}_1\bar{f}_2f_2}^\gamma &= -\frac{2e^2Q_1Q_2}{s}\left(
    \langle\mathbf{4}\mathbf{3}\rangle \lbrack\mathbf{1}\mathbf{2}\rbrack x_{43} \tilde{x}_{12} 
    +\langle\mathbf{1}\mathbf{2}\rangle \lbrack\mathbf{4}\mathbf{3}\rbrack x_{12} \tilde{x}_{43}
    \right)
    \nonumber\\
    &= \frac{-e^2Q_1Q_2}{s\ q\cdot k}\big(
    \left(\lbrack\mathbf{3}k\rbrack\langle\mathbf{4}q\rangle
    +\langle\mathbf{3}q\rangle\lbrack\mathbf{4} k\rbrack\right)
    \left(\langle\mathbf{2} k\rangle\lbrack\mathbf{1}q\rbrack
    +\lbrack\mathbf{2}q\rbrack\langle\mathbf{1} k\rangle\right)
    +\left(\lbrack\mathbf{2}k\rbrack\langle\mathbf{1}q\rangle
    +\langle\mathbf{2}q\rangle\lbrack\mathbf{1} k\rbrack\right)
    \left(\langle\mathbf{4} k\rangle\lbrack\mathbf{3}q\rbrack
    +\lbrack\mathbf{4}q\rbrack\langle\mathbf{3} k\rangle\right)
    \big),
    \label{eq:app:ffff: initial}
\end{align}
where we have used the same reference momentum $q$ for both vertices and $\langle qk\rangle\lbrack qk\rbrack = -2q\cdot k$.  Expanding and using $\lvert k\rangle\lbrack k\rvert = k$ and similarly for $q$, we have
\begin{align}
    \mathcal{M}_{f_1\bar{f}_1\bar{f}_2f_2}^\gamma =  
    \frac{-e^2Q_1Q_2}{s\ q\cdot k}\big(&
    \lbrack\mathbf{3}\lvert k\rvert \mathbf{2}\rangle\lbrack\mathbf{1}\lvert q\rvert \mathbf{4}\rangle
    +\lbrack\mathbf{4}\lvert k\rvert \mathbf{2}\rangle\lbrack\mathbf{1}\lvert q\rvert \mathbf{3}\rangle
    +\lbrack\mathbf{3}\lvert k\rvert\mathbf{1}\rangle\lbrack\mathbf{2}\lvert q\rvert \mathbf{4}\rangle
    +\lbrack\mathbf{4}\lvert k\rvert\mathbf{1}\rangle\lbrack\mathbf{2}\lvert q\rvert\mathbf{3}\rangle
    \nonumber\\
    &+
    \lbrack\mathbf{4}\lvert q\rvert \mathbf{1}\rangle\lbrack\mathbf{2}\lvert k\rvert \mathbf{3}\rangle
    +\lbrack\mathbf{3}\lvert q\rvert \mathbf{1}\rangle\lbrack\mathbf{2}\lvert k\rvert \mathbf{4}\rangle
    +\lbrack\mathbf{4}\lvert q\rvert \mathbf{2}\rangle\lbrack\mathbf{1}\lvert k\rvert\mathbf{3}\rangle
    +\lbrack\mathbf{3}\lvert q\rvert\mathbf{2}\rangle\lbrack\mathbf{1}\lvert k\rvert\mathbf{4}\rangle
    \big).
    \label{eq:app:ffff:A8}
\end{align}
Our next step is to apply another Schouten identity, giving us
\begin{align}
    \mathcal{M}_{f_1\bar{f}_1\bar{f}_2f_2}^\gamma =  
    \frac{e^2Q_1Q_2}{s\ q\cdot k}\big(&
    \langle\mathbf{42}\rangle\lbrack\mathbf{1}\lvert qk\rvert\mathbf{3}\rbrack
    +\langle\mathbf{32}\rangle\lbrack\mathbf{1}\lvert qk\rvert\mathbf{4}\rbrack
    +\langle\mathbf{41}\rangle\lbrack\mathbf{2}\lvert qk\rvert\mathbf{3}\rbrack
    +\langle\mathbf{31}\rangle\lbrack\mathbf{2}\lvert qk\rvert\mathbf{4}\rbrack
    \nonumber\\
    &+\langle\mathbf{31}\rangle\lbrack\mathbf{2}\lvert kq\rvert\mathbf{4}\rbrack
    +\langle\mathbf{41}\rangle\lbrack\mathbf{2}\lvert kq\rvert\mathbf{3}\rbrack
    +\langle\mathbf{32}\rangle\lbrack\mathbf{1}\lvert kq\rvert\mathbf{4}\rbrack
    +\langle\mathbf{42}\rangle\lbrack\mathbf{1}\lvert kq\rvert\mathbf{3}\rbrack
    \big),
    \label{eq:app:ffff:A9}
\end{align}
where we have discarded terms of the form 
\begin{align}
    \left(\langle\mathbf{4}\lvert k\rvert\mathbf{3}\rbrack+\langle\mathbf{3}\lvert k\rvert\mathbf{4}\rbrack\right) &=
    m_3\left(
        \langle\mathbf{43}\rangle-\lbrack\mathbf{43}\rbrack 
        + \langle\mathbf{34}\rangle-\lbrack\mathbf{34}\rbrack 
    \right)
    = 0.
\end{align}
Our final step is to interchange the order of the momenta in the bottom line using $kq = 2q\cdot k - qk$.  The terms with $qk$ all cancel, and we are left with
\begin{align}
    \mathcal{M}_{f_1\bar{f}_1\bar{f}_2f_2}^\gamma =  
    \frac{-2e^2Q_1Q_2}{s}\big(&
    \langle\mathbf{13}\rangle\lbrack\mathbf{24}\rbrack
    +\langle\mathbf{14}\rangle\lbrack\mathbf{23}\rbrack
    +\langle\mathbf{23}\rangle\lbrack\mathbf{14}\rbrack
    +\langle\mathbf{24}\rangle\lbrack\mathbf{13}\rbrack
    \big),
    \label{eq:app:ffff:A S}
\end{align}
in agreement with Eq.~(\ref{eq:4-f:A S}).

\subsection{$f\bar{f}\bar{W}W$}
Now, what we would like is to do the same for $f \bar{f} \bar{W} W$.  After we finish this calculation, we will redo it another way that will translate better to $W W \bar{W} \bar{W}$.  To begin this calculation, it's amplitude begins as
\begin{align}
    \mathcal{M}_{f\bar{f}\bar{W}W}^\gamma &=
    -\frac{2e^{2}Q_f}{M_W s} 
    \left(
    \langle\mathbf{43}\rangle^{2} \lbrack\mathbf{12}\rbrack x_{43} \tilde{x}_{12} 
    +\langle\mathbf{12}\rangle \lbrack\mathbf{43}\rbrack^{2} x_{12} \tilde{x}_{43} 
    \right)
    \nonumber\\
    &=
    -\frac{e^{2}Q_f}{M_W s\ q\cdot k} 
    \big(
    \langle\mathbf{43}\rangle\left(
    \lbrack\mathbf{3}\lvert k\rvert \mathbf{2}\rangle\lbrack\mathbf{1}\lvert q\rvert \mathbf{4}\rangle
    +\lbrack\mathbf{4}\lvert k\rvert \mathbf{2}\rangle\lbrack\mathbf{1}\lvert q\rvert \mathbf{3}\rangle
    +\lbrack\mathbf{3}\lvert k\rvert\mathbf{1}\rangle\lbrack\mathbf{2}\lvert q\rvert \mathbf{4}\rangle
    +\lbrack\mathbf{4}\lvert k\rvert\mathbf{1}\rangle\lbrack\mathbf{2}\lvert q\rvert\mathbf{3}\rangle
    \right)
    \nonumber\\
    &\hspace{0.75in}
    +\lbrack\mathbf{43}\rbrack\left(
    \lbrack\mathbf{4}\lvert q\rvert \mathbf{1}\rangle\lbrack\mathbf{2}\lvert k\rvert \mathbf{3}\rangle
    +\lbrack\mathbf{3}\lvert q\rvert \mathbf{1}\rangle\lbrack\mathbf{2}\lvert k\rvert \mathbf{4}\rangle
    +\lbrack\mathbf{4}\lvert q\rvert \mathbf{2}\rangle\lbrack\mathbf{1}\lvert k\rvert\mathbf{3}\rangle
    +\lbrack\mathbf{3}\lvert q\rvert\mathbf{2}\rangle\lbrack\mathbf{1}\lvert k\rvert\mathbf{4}\rangle
    \right)
    \big),
\end{align}
after inserting the $x$ factors and expanding, as we did to obtain Eq.~(\ref{eq:app:ffff:A8}).  We don't follow the same steps afterwards because we won't get the same cancellation at the end and we need to keep our results symmetric between angle and square brackets to obtain the final result.  We again perform a Schouten identity, but this time we keep a square bracket product without momenta on the top line and an angle bracket product without momenta on the bottom line, to obtain
\begin{align}
    \mathcal{M}_{f\bar{f}\bar{W}W}^\gamma 
    &=
    \frac{e^{2}Q_f}{M_W s\ q\cdot k} 
    \big(
    \langle\mathbf{34}\rangle\left(
    \lbrack\mathbf{13}\rbrack\langle\mathbf{2}\lvert kq\rvert\mathbf{4}\rangle
    +\lbrack\mathbf{14}\rbrack\langle\mathbf{2}\lvert kq\rvert \mathbf{3}\rangle
    +\lbrack\mathbf{23}\rbrack\langle\mathbf{1}\lvert kq\rvert \mathbf{4}\rangle
    +\lbrack\mathbf{24}\rbrack\langle\mathbf{1}\lvert kq\rvert \mathbf{3}\rangle
    \right)
    \nonumber\\
    &\hspace{0.75in}
    +\lbrack\mathbf{34}\rbrack\left(
    \langle\mathbf{13}\rangle\lbrack\mathbf{2}\lvert kq\rvert\mathbf{4}\rbrack
    +\langle\mathbf{14}\rangle\lbrack\mathbf{2}\lvert kq\rvert\mathbf{3}\rbrack
    +\langle\mathbf{23}\rangle\lbrack\mathbf{1}\lvert kq\rvert\mathbf{4}\rbrack
    +\langle\mathbf{24}\rangle\lbrack\mathbf{1}\lvert kq\rvert\mathbf{3}\rbrack
    \right)
    \big),
\end{align}
where we have again thrown away vanishing terms of the form $\lbrack\mathbf{1}\lvert k\rvert\mathbf{2}\rangle+\lbrack\mathbf{2}\lvert k\rvert\mathbf{1}\rangle = 0$ and $\lbrack\mathbf{3}\lvert k\rvert\mathbf{4}\rangle+\lbrack\mathbf{4}\lvert k\rvert\mathbf{3}\rangle = 0$.  At this point, recognizing the result we need from Eq.~(\ref{eq:ffWW:A S}), we interchange the order of $kq$ on both lines to obtain,
\begin{align}
    \mathcal{M}_{f\bar{f}\bar{W}W}^\gamma 
    &=
    \frac{2e^{2}Q_f}{M_W s} 
    \left(\langle\mathbf{34}\rangle+\lbrack\mathbf{34}\rbrack\right)
    \left(
    \lbrack\mathbf{13}\rbrack\langle\mathbf{24}\rangle
    +\lbrack\mathbf{14}\rbrack\langle\mathbf{23}\rangle
    +\lbrack\mathbf{23}\rbrack\langle\mathbf{14}\rangle
    +\lbrack\mathbf{24}\rbrack\langle\mathbf{13}\rangle
    \right)
    \nonumber\\
    &+\frac{-e^{2}Q_f}{M_W s\ q\cdot k} 
    \big(
    \langle\mathbf{34}\rangle\left(
    \lbrack\mathbf{13}\rbrack\langle\mathbf{2}\lvert qk\rvert\mathbf{4}\rangle
    +\lbrack\mathbf{14}\rbrack\langle\mathbf{2}\lvert qk\rvert \mathbf{3}\rangle
    +\lbrack\mathbf{23}\rbrack\langle\mathbf{1}\lvert qk\rvert \mathbf{4}\rangle
    +\lbrack\mathbf{24}\rbrack\langle\mathbf{1}\lvert qk\rvert \mathbf{3}\rangle
    \right)
    \nonumber\\
    &\hspace{0.75in}
    +\lbrack\mathbf{34}\rbrack\left(
    \langle\mathbf{13}\rangle\lbrack\mathbf{2}\lvert qk\rvert\mathbf{4}\rbrack
    +\langle\mathbf{14}\rangle\lbrack\mathbf{2}\lvert qk\rvert\mathbf{3}\rbrack
    +\langle\mathbf{23}\rangle\lbrack\mathbf{1}\lvert qk\rvert\mathbf{4}\rbrack
    +\langle\mathbf{24}\rangle\lbrack\mathbf{1}\lvert qk\rvert\mathbf{3}\rbrack
    \right)
    \big).
\end{align}
The first line is part of Eq.~(\ref{eq:ffWW:A S}).  We need to get the other part.  The second part has two powers of $M_W$ in the denominator and a $\langle\mathbf{34}\rangle\lbrack\mathbf{34}\rbrack$ in the numerator.  To enable this, we insert a $p_3$ or $p_4$ in the numerator, obtaining
\begin{align}
    \mathcal{M}_{f\bar{f}\bar{W}W}^\gamma 
    &=
    \frac{2e^{2}Q_f}{M_W s} 
    \left(\langle\mathbf{34}\rangle+\lbrack\mathbf{34}\rbrack\right)
    \left(
    \lbrack\mathbf{13}\rbrack\langle\mathbf{24}\rangle
    +\lbrack\mathbf{14}\rbrack\langle\mathbf{23}\rangle
    +\lbrack\mathbf{23}\rbrack\langle\mathbf{14}\rangle
    +\lbrack\mathbf{24}\rbrack\langle\mathbf{13}\rangle
    \right)
    \nonumber\\
    &+\frac{e^{2}Q_f}{M_W^2 s\ q\cdot k} 
    \big(
    \langle\mathbf{34}\rangle\left(
    \lbrack\mathbf{13}\rbrack\langle\mathbf{2}\lvert qk p_4\rvert\mathbf{4}\rbrack
    +\lbrack\mathbf{14}\rbrack\langle\mathbf{2}\lvert qk p_3\rvert \mathbf{3}\rbrack
    +\lbrack\mathbf{23}\rbrack\langle\mathbf{1}\lvert qk p_4\rvert \mathbf{4}\rbrack
    +\lbrack\mathbf{24}\rbrack\langle\mathbf{1}\lvert qk p_3\rvert \mathbf{3}\rbrack
    \right)
    \nonumber\\
    &\hspace{0.75in}
    +\lbrack\mathbf{34}\rbrack\left(
    \langle\mathbf{13}\rangle\lbrack\mathbf{2}\lvert qk p_4\rvert\mathbf{4}\rangle
    +\langle\mathbf{14}\rangle\lbrack\mathbf{2}\lvert qk p_3\rvert\mathbf{3}\rangle
    +\langle\mathbf{23}\rangle\lbrack\mathbf{1}\lvert qk p_4\rvert\mathbf{4}\rangle
    +\langle\mathbf{24}\rangle\lbrack\mathbf{1}\lvert qk p_3\rvert\mathbf{3}\rangle
    \right)
    \big),
\end{align}
and apply a Schouten identity to the first and third term of the last two rows, giving
\begin{align}
    \mathcal{M}_{f\bar{f}\bar{W}W}^\gamma 
    &=
    \frac{2e^{2}Q_f}{M_W s} 
    \left(\langle\mathbf{34}\rangle+\lbrack\mathbf{34}\rbrack\right)
    \left(
    \lbrack\mathbf{13}\rbrack\langle\mathbf{24}\rangle
    +\lbrack\mathbf{14}\rbrack\langle\mathbf{23}\rangle
    +\lbrack\mathbf{23}\rbrack\langle\mathbf{14}\rangle
    +\lbrack\mathbf{24}\rbrack\langle\mathbf{13}\rangle
    \right)
    \nonumber\\
    &+\frac{e^{2}Q_f}{M_W^2 s\ q\cdot k} 
    \big(
    \langle\mathbf{34}\rangle\left(
    \lbrack\mathbf{43}\rbrack\langle\mathbf{2}\lvert qk p_4\rvert\mathbf{1}\rbrack
    +\lbrack\mathbf{43}\rbrack\langle\mathbf{1}\lvert qk p_4\rvert \mathbf{2}\rbrack
    \right)
    +\lbrack\mathbf{34}\rbrack\left(
    \langle\mathbf{43}\rangle\lbrack\mathbf{2}\lvert qk p_4\rvert\mathbf{1}\rangle
    +\langle\mathbf{43}\rangle\lbrack\mathbf{1}\lvert qk p_4\rvert\mathbf{2}\rangle
    \right)
    \big)
    \nonumber\\
    &+\frac{e^{2}Q_f}{M_W^2 s\ q\cdot k} 
    \big(
    \langle\mathbf{34}\rangle\left(
    -\lbrack\mathbf{41}\rbrack\langle\mathbf{2}\lvert qk p_4\rvert\mathbf{3}\rbrack
    +\lbrack\mathbf{14}\rbrack\langle\mathbf{2}\lvert qk p_3\rvert \mathbf{3}\rbrack
    -\lbrack\mathbf{42}\rbrack\langle\mathbf{1}\lvert qk p_4\rvert \mathbf{3}\rbrack
    +\lbrack\mathbf{24}\rbrack\langle\mathbf{1}\lvert qk p_3\rvert \mathbf{3}\rbrack
    \right)
    \nonumber\\
    &\hspace{0.75in}
    +\lbrack\mathbf{34}\rbrack\left(
    -\langle\mathbf{41}\rangle\lbrack\mathbf{2}\lvert qk p_4\rvert\mathbf{3}\rangle
    +\langle\mathbf{14}\rangle\lbrack\mathbf{2}\lvert qk p_3\rvert\mathbf{3}\rangle
    -\langle\mathbf{42}\rangle\lbrack\mathbf{1}\lvert qk p_4\rvert\mathbf{3}\rangle
    +\langle\mathbf{24}\rangle\lbrack\mathbf{1}\lvert qk p_3\rvert\mathbf{3}\rangle
    \right)
    \big).
\end{align}
Combining and rearranging, we have
\begin{align}
    \mathcal{M}_{f\bar{f}\bar{W}W}^\gamma 
    &=
    \frac{2e^{2}Q_f}{M_W s} 
    \left(\langle\mathbf{34}\rangle+\lbrack\mathbf{34}\rbrack\right)
    \left(
    \lbrack\mathbf{13}\rbrack\langle\mathbf{24}\rangle
    +\lbrack\mathbf{14}\rbrack\langle\mathbf{23}\rangle
    +\lbrack\mathbf{23}\rbrack\langle\mathbf{14}\rangle
    +\lbrack\mathbf{24}\rbrack\langle\mathbf{13}\rangle
    \right)
    \nonumber\\
    &+\frac{-e^{2}Q_f}{M_W^2 s\ q\cdot k} 
    \langle\mathbf{34}\rangle\lbrack\mathbf{34}\rbrack\left(
    \lbrack\mathbf{1}\lvert \left(p_4 kq + qk p_4\right)\rvert\mathbf{2}\rangle
    +\lbrack\mathbf{2}\lvert \left(p_4 kq + qk p_4\right)\rvert\mathbf{1}\rangle
    \right),
\end{align}
where we have used $k(p_3+p_4)=-k^2=0$ to remove the bottom two lines.  At this point, we interchange $kq$, and find
\begin{align}
    \mathcal{M}_{f\bar{f}\bar{W}W}^\gamma 
    &=
    \frac{2e^{2}Q_f}{M_W s} 
    \left(\langle\mathbf{34}\rangle+\lbrack\mathbf{34}\rbrack\right)
    \left(
    \lbrack\mathbf{13}\rbrack\langle\mathbf{24}\rangle
    +\lbrack\mathbf{14}\rbrack\langle\mathbf{23}\rangle
    +\lbrack\mathbf{23}\rbrack\langle\mathbf{14}\rangle
    +\lbrack\mathbf{24}\rbrack\langle\mathbf{13}\rangle
    \right)
    +\frac{-2e^{2}Q_f}{M_W^2 s} 
    \langle\mathbf{34}\rangle\lbrack\mathbf{34}\rbrack\left(
    \lbrack\mathbf{1}\lvert p_4 \rvert\mathbf{2}\rangle
    +\lbrack\mathbf{2}\lvert p_4 \rvert\mathbf{1}\rangle
    \right)
    \nonumber\\
    &+\frac{e^{2}Q_f}{M_W^2 s\ q\cdot k} 
    \langle\mathbf{34}\rangle\lbrack\mathbf{34}\rbrack\left(
    \lbrack\mathbf{1}\lvert \left(p_4 qk - qk p_4\right)\rvert\mathbf{2}\rangle
    +\lbrack\mathbf{2}\lvert \left(p_4 qk - qk p_4\right)\rvert\mathbf{1}\rangle
    \right).
\end{align}
On the first line, we next use momentum conservation $p_4=-p_1-p_2-p_3$ and the mass identity to remove the extra pieces, giving us agreement of the top line with Eq.~(\ref{eq:ffWW:A S}).  We have only to show that the second line vanishes to complete the derivation.  We begin by moving the $p_4$ to the right past the $q$ (and removing vanishing pieces of the form $\lbrack\mathbf{1}\lvert k\rvert\mathbf{2}\rangle+\lbrack\mathbf{2}\lvert k\rvert\mathbf{1}\rangle=0$), giving
\begin{align}
    \mathcal{R}_{f\bar{f}\bar{W}W}^\gamma &= \langle\mathbf{34}\rangle\lbrack\mathbf{34}\rbrack
    \left(
    \lbrack\mathbf{1}\lvert \left(p_4 qk - qk p_4\right)\rvert\mathbf{2}\rangle
    +\lbrack\mathbf{2}\lvert \left(p_4 qk - qk p_4\right)\rvert\mathbf{1}\rangle
    \right)
    \nonumber\\
    &= -\langle\mathbf{34}\rangle\lbrack\mathbf{34}\rbrack
    \left(
    \lbrack\mathbf{1}\lvert q\left(p_4 k + k p_4\right)\rvert\mathbf{2}\rangle
    +\lbrack\mathbf{2}\lvert q\left(p_4 k + k p_4\right)\rvert\mathbf{1}\rangle
    \right)
    \nonumber\\
    &= -2k\cdot p_4
    \langle\mathbf{34}\rangle\lbrack\mathbf{34}\rbrack
    \left(
    \lbrack\mathbf{1}\lvert q\rvert\mathbf{2}\rangle
    +\lbrack\mathbf{2}\lvert q\rvert\mathbf{1}\rangle
    \right).
\end{align}
We next put the $k\cdot p_4$ inside either $\langle\mathbf{34}\rangle$ or $\lbrack\mathbf{34}\rbrack$.  We will show it inside the angle bracket product.  We obtain
\begin{align}
    \mathcal{R}_{f\bar{f}\bar{W}W}^\gamma &=
    -
    \langle\mathbf{3}\lvert\left( k p_4 + p_4 k\right)\rvert\mathbf{4}\rangle
    \lbrack\mathbf{34}\rbrack
    \left(
    \lbrack\mathbf{1}\lvert q\rvert\mathbf{2}\rangle
    +\lbrack\mathbf{2}\lvert q\rvert\mathbf{1}\rangle
    \right).
\end{align}
We can now change the $p_4$ into $p_3$ on the right side by using $p_4 k=-p_3 k + (p_3+p_4)k = -p_3 k$, since $k=-(p_3+p_4)$ and $k^2=0$, giving us
\begin{align}
    \mathcal{R}_{f\bar{f}\bar{W}W}^\gamma &=
    -
    \langle\mathbf{3}\lvert\left( k p_4 - p_3 k\right)\rvert\mathbf{4}\rangle
    \lbrack\mathbf{34}\rbrack
    \left(
    \lbrack\mathbf{1}\lvert q\rvert\mathbf{2}\rangle
    +\lbrack\mathbf{2}\lvert q\rvert\mathbf{1}\rangle
    \right).
\end{align}
Using the mass identities, we find
\begin{align}
    \mathcal{R}_{f\bar{f}\bar{W}W}^\gamma &=
    M_W\left(
    \langle\mathbf{3}\lvert k \rvert\mathbf{4}\rbrack
    +\lbrack\mathbf{3}\lvert k \rvert\mathbf{4}\rangle
    \right)
    \lbrack\mathbf{34}\rbrack
    \left(
    \lbrack\mathbf{1}\lvert q\rvert\mathbf{2}\rangle
    +\lbrack\mathbf{2}\lvert q\rvert\mathbf{1}\rangle
    \right)
    \nonumber\\
    &= 0,
\end{align}
where, in the last line, we again used $\langle\mathbf{3}\lvert k \rvert\mathbf{4}\rbrack+\lbrack\mathbf{3}\lvert k \rvert\mathbf{4}\rangle = 0$.

On the other hand, there is another way to approach this problem.  If we can factor the amplitude with the $x$ factors into a form such that one of the factors is the same as one of the cases we have already solved, we can reuse its result.  In particular, if we can factor it into a form  where one of the factors is $\langle\mathbf{lk}\rangle \lbrack\mathbf{ij}\rbrack x_{lk} \tilde{x}_{ij} +\langle\mathbf{ij}\rangle \lbrack\mathbf{lk}\rbrack x_{ij} \tilde{x}_{lk}$ as in Eq.~(\ref{eq:app:ffff: initial}), then we can reuse its sequence of identities.  In other words, we can use
\begin{align}
    \langle\mathbf{kl}\rangle \lbrack\mathbf{ij}\rbrack x_{kl} \tilde{x}_{ij} +\langle\mathbf{ij}\rangle \lbrack\mathbf{kl}\rbrack x_{ij} \tilde{x}_{kl}
    &=
    \langle\mathbf{il}\rangle\lbrack\mathbf{jk}\rbrack
    +\langle\mathbf{ik}\rangle\lbrack\mathbf{jl}\rbrack
    +\langle\mathbf{jl}\rangle\lbrack\mathbf{ik}\rbrack
    +\langle\mathbf{jk}\rangle\lbrack\mathbf{il}\rbrack ,
    \label{eq:app:<>[]xx+[]<>xx=}
\end{align}
from Eq.~(\ref{eq:app:ffff:A S}).  In order to get the amplitude into this form, we need to make use of the identities from \cite{Christensen:2022nja}
\begin{align}
    x_{ij}\tilde{x}_{kl}\left(
        \lbrack\mathbf{kl}\rbrack - 
        \langle\mathbf{kl}\rangle 
    \right) &=
    \frac{1}{m_jm_l}\langle\mathbf{k}|p_jp_i|\mathbf{l}\rangle+
    \frac{m_j}{m_l}\langle\mathbf{kl}\rangle
    \label{eq:main:xxt<ij>=xxt[ij]}
    \\
    x_{ij}\tilde{x}_{kl}\left(
        \langle\mathbf{ij}\rangle -
        \lbrack\mathbf{ij}\rbrack 
    \right) &=
    \frac{1}{m_jm_l}\lbrack\mathbf{i}|p_lp_k|\mathbf{j}\rbrack +
    \frac{m_l}{m_j}\lbrack\mathbf{ij}\rbrack,
    \label{eq:main:xxt<kl>=xxt[kl]}
\end{align}
to interchange angle and square brackets.  It may be possible to achieve the correct result without using these identities symmetrically between angle and square brackets, but so far, we have found success with the symmetric application.  Therefore, we will always focus on a symmetric use of them.  

We begin, once again, with 
\begin{align}
    \mathcal{M}_{f\bar{f}\bar{W}W}^\gamma &=
    -\frac{2e^{2}Q_f}{M_W s} 
    \left(
    \left(\frac{1}{2}+\frac{1}{2}\right)
    \langle\mathbf{43}\rangle^{2} \lbrack\mathbf{12}\rbrack x_{43} \tilde{x}_{12} 
    +\left(\frac{1}{2}+\frac{1}{2}\right)
    \langle\mathbf{12}\rangle \lbrack\mathbf{43}\rbrack^{2} x_{12} \tilde{x}_{43} 
    \right),
\end{align}
where we have split each side into two equal pieces.  Our next step is to apply the identities from Eqs.~(\ref{eq:main:xxt<ij>=xxt[ij]}) and (\ref{eq:main:xxt<kl>=xxt[kl]}) to half of each side to obtain
\begin{align}
    \mathcal{M}_{f\bar{f}\bar{W}W}^\gamma &=
    -\frac{e^{2}Q_f}{M_W s}
    \Big(
        \left(
            \langle\mathbf{43}\rangle
            +\lbrack\mathbf{43}\rbrack
        \right)
        \left(
            \langle\mathbf{43}\rangle \lbrack\mathbf{12}\rbrack x_{43} \tilde{x}_{12} 
            +\langle\mathbf{12}\rangle \lbrack\mathbf{43}\rbrack x_{12} \tilde{x}_{43} 
        \right)
    \Big)
        \nonumber\\
    &-\frac{e^{2}Q_f}{M_W s}
    \left(
        \langle\mathbf{43}\rangle \lbrack\mathbf{12}\rbrack
        \left(
            \frac{m_f}{M_W}\lbrack\mathbf{43}\rbrack
            +\frac{1}{m_fM_W}\lbrack\mathbf{4}\lvert p_2p_1\rvert\mathbf{3}\rbrack
        \right)
        +\langle\mathbf{12}\rangle \lbrack\mathbf{43}\rbrack
        \left(
            \frac{m_f}{M_W}\langle\mathbf{43}\rangle
            +\frac{1}{m_fM_W}\langle\mathbf{4}\lvert p_2p_1\rvert\mathbf{3}\rangle
        \right)
    \right)
\end{align}
In our next step, we apply Eq.~(\ref{eq:app:<>[]xx+[]<>xx=}) to the top line.  On the bottom line, we apply a Schouten identity on $\lbrack\mathbf{4}\lvert p_2p_1\rvert\mathbf{3}\rbrack$ and $\langle\mathbf{4}\lvert p_2p_1\rvert\mathbf{3}\rangle$, finding
\begin{align}
    \mathcal{M}_{f\bar{f}\bar{W}W}^\gamma &=
    -\frac{e^{2}Q_f}{M_W s}
    \Big(
        \left(
            \langle\mathbf{43}\rangle
            +\lbrack\mathbf{43}\rbrack
        \right)
        \left(
            \langle\mathbf{13}\rangle\lbrack\mathbf{24}\rbrack
            +\langle\mathbf{14}\rangle\lbrack\mathbf{23}\rbrack
            +\langle\mathbf{23}\rangle\lbrack\mathbf{14}\rbrack
            +\langle\mathbf{24}\rangle\lbrack\mathbf{13}\rbrack
        \right)
    \Big)
        \nonumber\\
    &-\frac{e^{2}Q_f}{M_W s}
    \Bigg(
        \langle\mathbf{43}\rangle 
        \left(
            \frac{m_f}{M_W}\lbrack\mathbf{12}\rbrack\lbrack\mathbf{43}\rbrack
            -\frac{1}{M_W}\lbrack\mathbf{32}\rbrack\lbrack\mathbf{4}\lvert p_2\rvert\mathbf{1}\rangle
            -\frac{1}{m_fM_W}\lbrack\mathbf{31}\rbrack\lbrack\mathbf{4}\lvert p_2p_1\rvert\mathbf{2}\rbrack
        \right)
        \nonumber\\
        &\hspace{0.5in}
        + \lbrack\mathbf{43}\rbrack
        \left(
            \frac{m_f}{M_W}\langle\mathbf{12}\rangle\langle\mathbf{43}\rangle
            -\frac{1}{M_W}\langle\mathbf{32}\rangle\langle\mathbf{4}\lvert p_2\rvert\mathbf{1}\rbrack
            -\frac{1}{m_fM_W}\langle\mathbf{31}\rangle\langle\mathbf{4}\lvert p_2p_1\rvert\mathbf{2}\rangle
        \right)
    \Bigg).
\end{align}
We next interchange the order of $p_2p_1$ and use the on-shell property that $s=(2p_1\cdot p_2+2m_f^2)=0$, giving
\begin{align}
    \mathcal{M}_{f\bar{f}\bar{W}W}^\gamma &=
    -\frac{e^{2}Q_f}{M_W s}
    \Big(
        \left(
            \langle\mathbf{43}\rangle
            +\lbrack\mathbf{43}\rbrack
        \right)
        \left(
            \langle\mathbf{13}\rangle\lbrack\mathbf{24}\rbrack
            +\langle\mathbf{14}\rangle\lbrack\mathbf{23}\rbrack
            +\langle\mathbf{23}\rangle\lbrack\mathbf{14}\rbrack
            +\langle\mathbf{24}\rangle\lbrack\mathbf{13}\rbrack
        \right)
    \Big)
        \nonumber\\
    &-\frac{e^{2}Q_f}{M_W s}
    \Bigg(
        \langle\mathbf{43}\rangle 
        \left(
            \frac{m_f}{M_W}\lbrack\mathbf{12}\rbrack\lbrack\mathbf{43}\rbrack
            -\frac{1}{M_W}\lbrack\mathbf{32}\rbrack\lbrack\mathbf{4}\lvert p_2\rvert\mathbf{1}\rangle
            +\frac{2m_f}{M_W}\lbrack\mathbf{31}\rbrack\lbrack\mathbf{42}\rbrack
            -\frac{1}{M_W}\lbrack\mathbf{31}\rbrack\lbrack\mathbf{4}\lvert p_1\rvert\mathbf{2}\rangle
        \right)
        \nonumber\\
        &\hspace{0.5in}
        + \lbrack\mathbf{43}\rbrack
        \left(
            \frac{m_f}{M_W}\langle\mathbf{12}\rangle\langle\mathbf{43}\rangle
            -\frac{1}{M_W}\langle\mathbf{32}\rangle\langle\mathbf{4}\lvert p_2\rvert\mathbf{1}\rbrack
            +\frac{2m_f}{M_W}\langle\mathbf{31}\rangle\langle\mathbf{42}\rangle
            -\frac{1}{M_W}\langle\mathbf{31}\rangle\langle\mathbf{4}\lvert p_1\rvert\mathbf{2}\rbrack
        \right)
    \Bigg).
\end{align}
After this, we use momentum conservation to replace the inserted momenta and then the mass identities, followed by a Schouten identity on the terms with inserted momenta, resulting in
\begin{align}
    \mathcal{M}_{f\bar{f}\bar{W}W}^\gamma &=
    -\frac{e^{2}Q_f}{M_W s}
    \Big(
        \left(
            \langle\mathbf{43}\rangle
            +\lbrack\mathbf{43}\rbrack
        \right)
        \left(
            \langle\mathbf{13}\rangle\lbrack\mathbf{24}\rbrack
            +\langle\mathbf{14}\rangle\lbrack\mathbf{23}\rbrack
            +\langle\mathbf{23}\rangle\lbrack\mathbf{14}\rbrack
            +\langle\mathbf{24}\rangle\lbrack\mathbf{13}\rbrack
        \right)
    \Big)
        \nonumber\\
    &-\frac{e^{2}m_fQ_f}{M_W^2 s}
    \Big(
        \langle\mathbf{43}\rangle 
        \left(
            \lbrack\mathbf{12}\rbrack\lbrack\mathbf{43}\rbrack
            -\lbrack\mathbf{32}\rbrack\lbrack\mathbf{41}\rbrack
            +\lbrack\mathbf{31}\rbrack\lbrack\mathbf{42}\rbrack
        \right)
        + \lbrack\mathbf{43}\rbrack
        \left(
            \langle\mathbf{12}\rangle\langle\mathbf{43}\rangle
            -\langle\mathbf{32}\rangle\langle\mathbf{41}\rangle
            +\langle\mathbf{31}\rangle\langle\mathbf{42}\rangle
        \right)
    \Big)
        \nonumber\\
    &-\frac{e^{2}Q_f}{M_W s}
    \Bigg(
        \langle\mathbf{43}\rangle 
        \left(
            \lbrack\mathbf{32}\rbrack\langle\mathbf{41}\rangle
            +\lbrack\mathbf{31}\rbrack\langle\mathbf{42}\rangle
        \right)
        + \lbrack\mathbf{43}\rbrack
        \left(
            \langle\mathbf{32}\rangle\lbrack\mathbf{41}\rbrack
            +\langle\mathbf{31}\rangle\lbrack\mathbf{42}\rbrack
        \right)
    \Bigg)
        \nonumber\\
    &-\frac{e^{2}Q_f}{M_W^2 s}
    \Bigg(
        \langle\mathbf{43}\rangle 
        \left(
            \left(
                \lbrack\mathbf{32}\rbrack\lbrack\mathbf{4}\lvert p_3\rvert\mathbf{1}\rangle
                +     \lbrack\mathbf{31}\rbrack\lbrack\mathbf{4}\lvert p_3\rvert\mathbf{2}\rangle
            \right)
        \right)
        + \lbrack\mathbf{43}\rbrack
        \left(
            \left(
                \langle\mathbf{32}\rangle\langle\mathbf{4}\lvert p_3\rvert\mathbf{1}\rbrack
                +\langle\mathbf{31}\rangle\langle\mathbf{4}\lvert p_3\rvert\mathbf{2}\rbrack
            \right)
        \right)
    \Bigg).
\end{align}
At this point, we do another Schouten identity on the second row resulting in that row vanishing and we do a Schouten identity on the term with an inserted $p_3$, finding
\begin{align}
    \mathcal{M}_{f\bar{f}\bar{W}W}^\gamma &=
    \frac{2e^{2}Q_f}{M_W s}
    \left(
        \left(
            \langle\mathbf{34}\rangle
            +\lbrack\mathbf{34}\rbrack
        \right)
        \left(
            \langle\mathbf{13}\rangle\lbrack\mathbf{24}\rbrack
            +\langle\mathbf{14}\rangle\lbrack\mathbf{23}\rbrack
            +\langle\mathbf{23}\rangle\lbrack\mathbf{14}\rbrack
            +\langle\mathbf{24}\rangle\lbrack\mathbf{13}\rbrack
        \right)
    \right)
     +\frac{2e^{2}Q_f}{M_W^2 s}
    \langle\mathbf{34}\rangle \lbrack\mathbf{34}\rbrack
    \left(
        \lbrack\mathbf{1}\lvert p_3\rvert\mathbf{2}\rangle 
        +
        \lbrack\mathbf{2}\lvert p_3\rvert\mathbf{1}\rangle 
    \right),
\end{align}
agreeing with Eq.~(\ref{eq:ffWW:A S}).

\subsection{$WW\bar{W}\bar{W}$}
We have not yet had success with the brute-force method of calculating $W W \bar{W} \bar{W}$ using Eqs.~(\ref{eq:app:x<ij>}) and (\ref{eq:app:xt[ij]}), but we think that it should work if the right sequence of identities can be found.  However, we can use the factoring method.  To get started, we write the initial T-channel diagram for $W W \bar{W} \bar{W}$, 
\begin{align}
    \mathcal{M}_{WW\bar{W}\bar{W}}^\gamma &= 
    \frac{-2e^{2}}{M_W^{2} t }  \left(\langle\mathbf{2}\mathbf{4}\rangle ^{2} \lbrack\mathbf{1}\mathbf{3}\rbrack ^{2} x_{24} \tilde{x}_{13} +\langle\mathbf{1}\mathbf{3}\rangle ^{2} \lbrack\mathbf{2}\mathbf{4}\rbrack ^{2} x_{13} \tilde{x}_{24} \right).
\end{align}
In order to factor this expression, we use the identities from Eqs.~(\ref{eq:main:xxt<ij>=xxt[ij]}) and (\ref{eq:main:xxt<kl>=xxt[kl]})
to interchange angle and square brackets.  In order to obtain a symmetric form, we will shift half of both sides to look like the other.  To make this clear, we first write the expression as
\begin{align}
    \mathcal{M}_{WW\bar{W}\bar{W}}^\gamma &= 
    \frac{-e^{2}}{M_W^{2} t }  
    \left(
    \langle\mathbf{2}\mathbf{4}\rangle \lbrack\mathbf{1}\mathbf{3}\rbrack
    \left(
        \langle\mathbf{2}\mathbf{4}\rangle \lbrack\mathbf{1}\mathbf{3}\rbrack x_{24} \tilde{x}_{13}
        +\langle\mathbf{2}\mathbf{4}\rangle \lbrack\mathbf{1}\mathbf{3}\rbrack x_{24} \tilde{x}_{13}
    \right)
    +\langle\mathbf{1}\mathbf{3}\rangle \lbrack\mathbf{2}\mathbf{4}\rbrack 
    \left(
        \langle\mathbf{1}\mathbf{3}\rangle  \lbrack\mathbf{2}\mathbf{4}\rbrack x_{13} \tilde{x}_{24} 
        +\langle\mathbf{1}\mathbf{3}\rangle \lbrack\mathbf{2}\mathbf{4}\rbrack x_{13} \tilde{x}_{24} 
    \right)
    \right).
\end{align}
We now use Eqs.~(\ref{eq:main:xxt<ij>=xxt[ij]}) and (\ref{eq:main:xxt<kl>=xxt[kl]}) on one term from each side.  We have a choice for the path of switching the form of one side to look like the form of the other side.  
For example, if we begin with $\langle\mathbf{2}\mathbf{4}\rangle \lbrack\mathbf{1}\mathbf{3}\rbrack x_{24} \tilde{x}_{13}$, we could first change $\langle\mathbf{24}\rangle$ into a square bracket, as in
\begin{align}
    \langle\mathbf{2}\mathbf{4}\rangle \lbrack\mathbf{1}\mathbf{3}\rbrack x_{24} \tilde{x}_{13}
    &=\lbrack\mathbf{2}\mathbf{4}\rbrack \lbrack\mathbf{1}\mathbf{3}\rbrack x_{24} \tilde{x}_{13}
    +\lbrack\mathbf{1}\mathbf{3}\rbrack
    \left(
        \lbrack\mathbf{2}\mathbf{4}\rbrack
        +\frac{1}{M_W^2}\lbrack\mathbf{2}\lvert p_3p_1\rvert\mathbf{4}\rbrack
    \right),
\end{align}
and follow this with changing the $\lbrack\mathbf{13}\rbrack$ into an angle bracket, obtaining
\begin{align}
    \langle\mathbf{2}\mathbf{4}\rangle \lbrack\mathbf{1}\mathbf{3}\rbrack x_{24} \tilde{x}_{13}
    &=\lbrack\mathbf{2}\mathbf{4}\rbrack \langle\mathbf{1}\mathbf{3}\rangle x_{24} \tilde{x}_{13}
    +\lbrack\mathbf{2}\mathbf{4}\rbrack
    \left(
        \langle\mathbf{1}\mathbf{3}\rangle
        +\frac{1}{M_W^2}\langle\mathbf{1}\lvert p_4p_2\rvert\mathbf{3}\rangle
    \right)
    +\lbrack\mathbf{1}\mathbf{3}\rbrack
    \left(
        \lbrack\mathbf{2}\mathbf{4}\rbrack
        +\frac{1}{M_W^2}\lbrack\mathbf{2}\lvert p_3p_1\rvert\mathbf{4}\rbrack
    \right).
\end{align}
However, we could instead begin by changing the $\lbrack\mathbf{13}\rbrack$ into an angle bracket and then, afterwards, changing the $\langle\mathbf{24}\rangle$ into a square bracket.  If we used the identities in this order, we would have found
\begin{align}
    \langle\mathbf{2}\mathbf{4}\rangle \lbrack\mathbf{1}\mathbf{3}\rbrack x_{24} \tilde{x}_{13}
    &=\lbrack\mathbf{2}\mathbf{4}\rbrack \langle\mathbf{1}\mathbf{3}\rangle x_{24} \tilde{x}_{13}
    +\langle\mathbf{2}\mathbf{4}\rangle
    \left(
        \langle\mathbf{1}\mathbf{3}\rangle
        +\frac{1}{M_W^2}\langle\mathbf{1}\lvert p_4p_2\rvert\mathbf{3}\rangle
    \right)
    +\langle\mathbf{1}\mathbf{3}\rangle
    \left(
        \lbrack\mathbf{2}\mathbf{4}\rbrack
        +\frac{1}{M_W^2}\lbrack\mathbf{2}\lvert p_3p_1\rvert\mathbf{4}\rbrack
    \right).
\end{align}
In order to find a highly symmetric form, we will include both paths.  We do the same on $\langle\mathbf{1}\mathbf{3}\rangle  \lbrack\mathbf{2}\mathbf{4}\rbrack x_{13} \tilde{x}_{24}$.  This results in
\begin{align}
    \mathcal{M}_{WW\bar{W}\bar{W}}^\gamma &= 
    \frac{-e^{2}}{M_W^{2} t }  
    \left(
        \langle\mathbf{2}\mathbf{4}\rangle \lbrack\mathbf{1}\mathbf{3}\rbrack
        +\lbrack\mathbf{2}\mathbf{4}\rbrack \langle\mathbf{1}\mathbf{3}\rangle
    \right)
    \left(
    \langle\mathbf{2}\mathbf{4}\rangle \lbrack\mathbf{1}\mathbf{3}\rbrack
     x_{24} \tilde{x}_{13}
    +\langle\mathbf{1}\mathbf{3}\rangle \lbrack\mathbf{2}\mathbf{4}\rbrack 
     x_{13} \tilde{x}_{24}
    \right)
    \nonumber\\
    &+\frac{-e^{2}}{2M_W^{2} t }  
    \Big(
    \left(
        \langle\mathbf{2}\mathbf{4}\rangle \lbrack\mathbf{1}\mathbf{3}\rbrack
        +\langle\mathbf{1}\mathbf{3}\rangle \lbrack\mathbf{2}\mathbf{4}\rbrack
    \right)
    \left(
        \lbrack\mathbf{13}\rbrack\lbrack\mathbf{24}\rbrack
        +\langle\mathbf{13}\rangle\langle\mathbf{24}\rangle
    \right)
    +4\langle\mathbf{24}\rangle \lbrack\mathbf{13}\rbrack
    \lbrack\mathbf{24}\rbrack
    \langle\mathbf{13}\rangle
    \Big)
    \nonumber\\
    &+\frac{-e^{2}}{2M_W^4 t }  
    \Bigg(
        \langle\mathbf{2}\mathbf{4}\rangle \lbrack\mathbf{13}\rbrack^2
        \lbrack\mathbf{2}\lvert p_3p_1\rvert \mathbf{4}\rbrack 
        +\lbrack\mathbf{1}\mathbf{3}\rbrack
        \langle\mathbf{24}\rangle^2
        \langle\mathbf{1}\lvert p_4p_2\rvert\mathbf{3}\rangle
        +
        \langle\mathbf{1}\mathbf{3}\rangle \lbrack\mathbf{2}\mathbf{4}\rbrack^2 
        \lbrack\mathbf{1}\lvert p_4p_2\rvert\mathbf{3}\rbrack 
        +\lbrack\mathbf{2}\mathbf{4}\rbrack 
        \langle\mathbf{13}\rangle^2
        \langle\mathbf{2}\lvert p_3p_1\rvert\mathbf{4}\rangle 
    \Bigg)
    \nonumber\\
    &\hspace{-0.5in}
    +\frac{-e^{2}}{2M_W^4 t }  
    \Bigg(
        \langle\mathbf{2}\mathbf{4}\rangle \lbrack\mathbf{1}\mathbf{3}\rbrack
        \lbrack\mathbf{24}\rbrack\langle\mathbf{1}\lvert p_4p_2\rvert\mathbf{3}\rangle
        +\langle\mathbf{2}\mathbf{4}\rangle \lbrack\mathbf{1}\mathbf{3}\rbrack
        \langle\mathbf{13}\rangle\lbrack\mathbf{2}\lvert p_3p_1\rvert \mathbf{4}\rbrack 
        +
        \langle\mathbf{1}\mathbf{3}\rangle \lbrack\mathbf{2}\mathbf{4}\rbrack 
        \lbrack\mathbf{13}\rbrack\langle\mathbf{2}\lvert p_3p_1\rvert\mathbf{4}\rangle 
        +\langle\mathbf{1}\mathbf{3}\rangle \lbrack\mathbf{2}\mathbf{4}\rbrack 
        \langle\mathbf{24}\rangle\lbrack\mathbf{1}\lvert p_4p_2\rvert\mathbf{3}\rbrack 
    \Bigg).
    \label{eq:app:WWWW: intermediate B}
\end{align}
On the first line, we can use Eq.~(\ref{eq:app:<>[]xx+[]<>xx=}).  On the third line, we make use of the spinor product that is squared to do two Schouten identities and apply the mass and on-shell ($t=2p_1\cdot p_3+2M_W^2=0)$ identities.  For example, for the first term on the third line,
\begin{align}
    \lbrack\mathbf{13}\rbrack^2\lbrack\mathbf{2}\lvert p_3p_1\rvert \mathbf{4}\rbrack 
    &= \lbrack\mathbf{43}\rbrack\lbrack\mathbf{13}\rbrack\lbrack\mathbf{2}\lvert p_3p_1\rvert \mathbf{1}\rbrack
    -\lbrack\mathbf{41}\rbrack\lbrack\mathbf{13}\rbrack\lbrack\mathbf{2}\lvert p_3p_1\rvert \mathbf{3}\rbrack
    \nonumber\\
    &=\lbrack\mathbf{43}\rbrack
    \left(
        \lbrack\mathbf{1}\lvert p_1p_3\rvert\mathbf{3}\rbrack \lbrack\mathbf{21}\rbrack
        -
        \lbrack\mathbf{1}\lvert p_1p_3\rvert\mathbf{1}\rbrack \lbrack\mathbf{23}\rbrack
    \right)
    -
    \lbrack\mathbf{41}\rbrack
    \left(
        \lbrack\mathbf{3}\lvert p_1p_3\rvert\mathbf{3}\rbrack\lbrack\mathbf{21}\rbrack
        -
        \lbrack\mathbf{3}\lvert p_1p_3\rvert\mathbf{1}\rbrack\lbrack\mathbf{23}\rbrack
    \right)
    \nonumber\\
    &=\lbrack\mathbf{43}\rbrack
    \left(
        -M_W^2\langle\mathbf{13}\rangle \lbrack\mathbf{21}\rbrack
        -
        M_W\langle\mathbf{1}\lvert p_3\rvert\mathbf{1}\rbrack \lbrack\mathbf{23}\rbrack
    \right)
    -
    \lbrack\mathbf{41}\rbrack
    \left(
        -M_W\lbrack\mathbf{3}\lvert p_1\rvert\mathbf{3}\rangle\lbrack\mathbf{21}\rbrack
        -2p_1\cdot p_3
        \lbrack\mathbf{31}\rbrack\lbrack\mathbf{23}\rbrack
        -M_W^2
        \langle\mathbf{31}\rangle\lbrack\mathbf{23}\rbrack
    \right)
    \nonumber\\
    &=\lbrack\mathbf{43}\rbrack
    \left(
        -M_W^2\langle\mathbf{13}\rangle \lbrack\mathbf{21}\rbrack
        -
        M_W\langle\mathbf{1}\lvert p_3\rvert\mathbf{1}\rbrack \lbrack\mathbf{23}\rbrack
    \right)
    -
    \lbrack\mathbf{41}\rbrack
    \left(
        -M_W\lbrack\mathbf{3}\lvert p_1\rvert\mathbf{3}\rangle\lbrack\mathbf{21}\rbrack
        +2M_W^2
        \lbrack\mathbf{31}\rbrack\lbrack\mathbf{23}\rbrack
        -M_W^2
        \langle\mathbf{31}\rangle\lbrack\mathbf{23}\rbrack
    \right).
\end{align}
We can see that, with this step, we have reduced the inserted momenta to one or less per term for the third row of Eq.~(\ref{eq:app:WWWW: intermediate B}).  On the fourth row, we don't have any spinor product squared, but we can still reduce the momenta to one per term at most.  For example, for the first term of the fourth line, we begin with a Schouten identity, followed by mass identities and interchanging the order of the momenta.  We find
\begin{align}
    \langle\mathbf{24}\rangle\langle\mathbf{1}\lvert p_4p_2\rvert\mathbf{3}\rangle &=
    \langle\mathbf{34}\rangle\langle\mathbf{1}\lvert p_4p_2\rvert\mathbf{2}\rangle
    -2p_2\cdot p_4\langle\mathbf{32}\rangle\langle\mathbf{14}\rangle
    +\langle\mathbf{32}\rangle\langle\mathbf{1}\lvert p_2p_4\rvert\mathbf{4}\rangle
    \nonumber\\
    &=
    -M_W\langle\mathbf{34}\rangle\langle\mathbf{1}\lvert p_4\rvert\mathbf{2}\rbrack
    +2M_W^2\langle\mathbf{32}\rangle\langle\mathbf{14}\rangle
    -M_W\langle\mathbf{32}\rangle\langle\mathbf{1}\lvert p_2\rvert\mathbf{4}\rbrack.
\end{align}

Following the application of these identities on all the terms of the third and fourth rows of Eq.~(\ref{eq:app:WWWW: intermediate B}) and using Eq.~(\ref{eq:app:<>[]xx+[]<>xx=}) on the top row, we obtain
\begin{align}
    \mathcal{M}_{WW\bar{W}\bar{W}}^\gamma &= 
    \frac{e^2}{2M_W^3 t}
    \Big(
        \Big(
            2\langle\mathbf{1}\mathbf{4}\rangle \langle\mathbf{2}\mathbf{3}\rangle \langle\mathbf{2}\mathbf{4}\rangle \lbrack\mathbf{1}\mathbf{3}\rbrack 
            -\langle\mathbf{1}\mathbf{3}\rangle \langle\mathbf{2}\mathbf{4}\rangle ^{2} \lbrack\mathbf{1}\mathbf{3}\rbrack 
            -2\langle\mathbf{2}\mathbf{4}\rangle \langle\mathbf{3}\mathbf{4}\rangle \lbrack\mathbf{1}\mathbf{2}\rbrack \lbrack\mathbf{1}\mathbf{3}\rbrack 
            +2\langle\mathbf{2}\mathbf{3}\rangle \langle\mathbf{2}\mathbf{4}\rangle \lbrack\mathbf{1}\mathbf{3}\rbrack \lbrack\mathbf{1}\mathbf{4}\rbrack 
            \nonumber\\
            &+2\langle\mathbf{1}\mathbf{4}\rangle \langle\mathbf{2}\mathbf{4}\rangle \lbrack\mathbf{1}\mathbf{3}\rbrack \lbrack\mathbf{2}\mathbf{3}\rbrack 
            +4\langle\mathbf{1}\mathbf{3}\rangle \langle\mathbf{2}\mathbf{4}\rangle \lbrack\mathbf{1}\mathbf{4}\rbrack \lbrack\mathbf{2}\mathbf{3}\rbrack 
            +2\langle\mathbf{2}\mathbf{4}\rangle \lbrack\mathbf{1}\mathbf{3}\rbrack \lbrack\mathbf{1}\mathbf{4}\rbrack \lbrack\mathbf{2}\mathbf{3}\rbrack 
            +2\langle\mathbf{1}\mathbf{3}\rangle \langle\mathbf{1}\mathbf{4}\rangle \langle\mathbf{2}\mathbf{3}\rangle \lbrack\mathbf{2}\mathbf{4}\rbrack 
            -\langle\mathbf{1}\mathbf{3}\rangle ^{2} \langle\mathbf{2}\mathbf{4}\rangle \lbrack\mathbf{2}\mathbf{4}\rbrack 
            \nonumber\\
            &-2\langle\mathbf{1}\mathbf{3}\rangle \langle\mathbf{3}\mathbf{4}\rangle \lbrack\mathbf{1}\mathbf{2}\rbrack \lbrack\mathbf{2}\mathbf{4}\rbrack 
            +4\langle\mathbf{1}\mathbf{4}\rangle \langle\mathbf{2}\mathbf{3}\rangle \lbrack\mathbf{1}\mathbf{3}\rbrack \lbrack\mathbf{2}\mathbf{4}\rbrack 
            -4\langle\mathbf{1}\mathbf{3}\rangle \langle\mathbf{2}\mathbf{4}\rangle \lbrack\mathbf{1}\mathbf{3}\rbrack \lbrack\mathbf{2}\mathbf{4}\rbrack 
            +2\langle\mathbf{1}\mathbf{2}\rangle \langle\mathbf{3}\mathbf{4}\rangle \lbrack\mathbf{1}\mathbf{3}\rbrack \lbrack\mathbf{2}\mathbf{4}\rbrack 
            -\langle\mathbf{2}\mathbf{4}\rangle \lbrack\mathbf{1}\mathbf{3}\rbrack ^{2} \lbrack\mathbf{2}\mathbf{4}\rbrack 
            \nonumber\\
            &+2\langle\mathbf{1}\mathbf{3}\rangle \langle\mathbf{2}\mathbf{3}\rangle \lbrack\mathbf{1}\mathbf{4}\rbrack \lbrack\mathbf{2}\mathbf{4}\rbrack 
            +2\langle\mathbf{1}\mathbf{3}\rangle \langle\mathbf{1}\mathbf{4}\rangle \lbrack\mathbf{2}\mathbf{3}\rbrack \lbrack\mathbf{2}\mathbf{4}\rbrack 
            +2\langle\mathbf{1}\mathbf{3}\rangle \lbrack\mathbf{1}\mathbf{4}\rbrack \lbrack\mathbf{2}\mathbf{3}\rbrack \lbrack\mathbf{2}\mathbf{4}\rbrack 
            -\langle\mathbf{1}\mathbf{3}\rangle \lbrack\mathbf{1}\mathbf{3}\rbrack \lbrack\mathbf{2}\mathbf{4}\rbrack ^{2} 
            +2\langle\mathbf{1}\mathbf{3}\rangle \langle\mathbf{2}\mathbf{4}\rangle \lbrack\mathbf{1}\mathbf{2}\rbrack \lbrack\mathbf{3}\mathbf{4}\rbrack 
            \nonumber\\
            &-2\langle\mathbf{1}\mathbf{2}\rangle \langle\mathbf{2}\mathbf{4}\rangle \lbrack\mathbf{1}\mathbf{3}\rbrack \lbrack\mathbf{3}\mathbf{4}\rbrack 
            -2\langle\mathbf{1}\mathbf{2}\rangle \langle\mathbf{1}\mathbf{3}\rangle \lbrack\mathbf{2}\mathbf{4}\rbrack \lbrack\mathbf{3}\mathbf{4}\rbrack 
        \Big)M_W
        -\langle\mathbf{1}\mathbf{3}\rangle \langle\mathbf{2}\mathbf{4}\rangle \lbrack\mathbf{2}\mathbf{3}\rbrack \lbrack\mathbf{1}\lvert p_{2} \rvert \mathbf{4}\rangle 
        -\langle\mathbf{1}\mathbf{3}\rangle \lbrack\mathbf{2}\mathbf{3}\rbrack \lbrack\mathbf{2}\mathbf{4}\rbrack \lbrack\mathbf{1}\lvert p_{2} \rvert \mathbf{4}\rangle 
        \nonumber\\
        &-\langle\mathbf{2}\mathbf{3}\rangle \langle\mathbf{3}\mathbf{4}\rangle \lbrack\mathbf{2}\mathbf{4}\rbrack \lbrack\mathbf{1}\lvert p_{3} \rvert \mathbf{1}\rangle 
        -\langle\mathbf{2}\mathbf{4}\rangle \lbrack\mathbf{2}\mathbf{3}\rbrack \lbrack\mathbf{3}\mathbf{4}\rbrack \lbrack\mathbf{1}\lvert p_{3} \rvert \mathbf{1}\rangle 
        -\langle\mathbf{3}\mathbf{4}\rangle \lbrack\mathbf{1}\mathbf{3}\rbrack \lbrack\mathbf{2}\mathbf{4}\rbrack \lbrack\mathbf{1}\lvert p_{3} \rvert \mathbf{2}\rangle 
        +\langle\mathbf{1}\mathbf{3}\rangle \langle\mathbf{2}\mathbf{4}\rangle \lbrack\mathbf{3}\mathbf{4}\rbrack \lbrack\mathbf{1}\lvert p_{4} \rvert \mathbf{2}\rangle 
        \nonumber\\
        &-\langle\mathbf{1}\mathbf{3}\rangle \langle\mathbf{2}\mathbf{4}\rangle \lbrack\mathbf{1}\mathbf{4}\rbrack \lbrack\mathbf{2}\lvert p_{1} \rvert \mathbf{3}\rangle 
        -\langle\mathbf{2}\mathbf{4}\rangle \lbrack\mathbf{1}\mathbf{3}\rbrack \lbrack\mathbf{1}\mathbf{4}\rbrack \lbrack\mathbf{2}\lvert p_{1} \rvert \mathbf{3}\rangle 
        -\langle\mathbf{1}\mathbf{3}\rangle \langle\mathbf{2}\mathbf{4}\rangle \lbrack\mathbf{3}\mathbf{4}\rbrack \lbrack\mathbf{2}\lvert p_{3} \rvert \mathbf{1}\rangle 
        +\langle\mathbf{3}\mathbf{4}\rangle \lbrack\mathbf{1}\mathbf{3}\rbrack \lbrack\mathbf{2}\mathbf{4}\rbrack \lbrack\mathbf{2}\lvert p_{4} \rvert \mathbf{1}\rangle 
        \nonumber\\
        &+\langle\mathbf{1}\mathbf{4}\rangle \langle\mathbf{3}\mathbf{4}\rangle \lbrack\mathbf{1}\mathbf{3}\rbrack \lbrack\mathbf{2}\lvert p_{4} \rvert \mathbf{2}\rangle 
        +\langle\mathbf{1}\mathbf{3}\rangle \lbrack\mathbf{1}\mathbf{4}\rbrack \lbrack\mathbf{3}\mathbf{4}\rbrack \lbrack\mathbf{2}\lvert p_{4} \rvert \mathbf{2}\rangle 
        -\langle\mathbf{1}\mathbf{3}\rangle \langle\mathbf{1}\mathbf{4}\rangle \lbrack\mathbf{2}\mathbf{4}\rbrack \lbrack\mathbf{3}\lvert p_{1} \rvert \mathbf{2}\rangle 
        -\langle\mathbf{1}\mathbf{4}\rangle \lbrack\mathbf{1}\mathbf{3}\rbrack \lbrack\mathbf{2}\mathbf{4}\rbrack \lbrack\mathbf{3}\lvert p_{1} \rvert \mathbf{2}\rangle 
        \nonumber\\
        &-\langle\mathbf{2}\mathbf{3}\rangle \langle\mathbf{2}\mathbf{4}\rangle \lbrack\mathbf{1}\mathbf{3}\rbrack \lbrack\mathbf{4}\lvert p_{2} \rvert \mathbf{1}\rangle 
        -\langle\mathbf{2}\mathbf{3}\rangle \lbrack\mathbf{1}\mathbf{3}\rbrack \lbrack\mathbf{2}\mathbf{4}\rbrack \lbrack\mathbf{4}\lvert p_{2} \rvert \mathbf{1}\rangle
    \Big).
\end{align}
Our next step is to massage the terms with an inserted momentum into either $\lbrack\mathbf{1}\lvert p_2\rvert\mathbf{4}\rangle$ or $\lbrack\mathbf{4}\lvert p_2\rvert\mathbf{1}\rangle$.  We do this with a lengthy set of steps involving Schouten identities, momentum conservation and mass identities.  Once we get all the momentum-insertion terms into this form, we remove squares of spinor products by applying a few more Schouten identities.  This leaves us with Eq.~(\ref{eq:WWWW:TA full}).  Although we don't include the details here, we have performed the analogous steps for the U-channel diagram and succeeded in obtaining agreement with the validated form found in Eq.~(\ref{eq:WWWW:UA full}).

\end{document}